\documentclass[a4paper,12pt]{article}

\usepackage{amsmath}
\usepackage[dvips]{graphicx}
\usepackage{color}

\def\be{\begin{equation}}
\def\eeq{\end{equation}}
\newcommand{\en}{\end{equation}}
\def\ba{\begin{eqnarray}}
\def\ea{\end{eqnarray}}

\def\tr{{\rm tr}}

\def\D{\nabla}
\def\tD{\tilde{\nabla}}
\def\U{\mathcal{U}}

\def\Dc{\mathcal{D}}

\def\D{\nabla}
\def\tD{\tilde{\nabla}}
\def\Dm{\nabla_{\mu}}
\def\Dn{\nabla_{\nu}}
\def\Drs{\nabla_{\rho\sigma}}

\def\tDm{\tilde{\nabla}_{\mu}}

\def\ta{\tilde{a}}

\def\ta{\tilde{a}}

\makeatletter
\@addtoreset{equation}{section}
\makeatother

\begin{document}

\begin{flushright}
EPHOU-24-019
\\
December, 2024
\end{flushright}

\vfil

\begin{center}

{\Large 
Gauge Covariant Link Formulation of  \\
\vspace{0.3cm}
Twisted $N=D=4$ and  $N=4\ D=5$
 Super Yang-Mills \\
 \vspace{0.3cm}
on a Lattice
}

\vspace{1cm}

{\sc Alessandro D'Adda}\footnote{
This article is dedicated to Alessandro D'Adda who passed away on April 23, 2022.
}$^{a}$, 
{\sc Noboru Kawamoto}\footnote{kawamoto@particle.sci.hokudai.ac.jp}$^{b}$,
{\sc Jun Saito}\footnote{jsaito@obihiro.ac.jp}$^{c}$
 and 
{\sc Kazuhiro Nagata}\footnote{khnagata@yahoo.co.jp}\\

\vspace{0.5cm}
$^{a}${\it{ INFN sezione di Torino, and Dipartimento di Fisica Teorica, Universita 
di Torino, I-10125 Torino, Italy }}\\
$^{b}$
{\it{ Department of Physics, Hokkaido University }}\\
{\it{ Sapporo, 060-0810, Japan}}\\
$^{c}$
{\it{ Department of Human Science, Obihiro University of Agriculture and Veterinary Medicine }}\\
{\it{ Obihiro, 080-8555, Japan}}\\
\end{center}

\vfil

\begin{abstract}

We propose a lattice formulation of four dimensional super Yang-Mills  
model with a twisted $N=4$ supersymmetry in a manifestly gauge covariant manner. 
The formulation we employ here is a four dimensional extension of
the manifestly gauge covariant method which was developed in our 
proposals of Dirac-K\"ahler
twisted $N=D=2$ and $N=4$  $D=3$ super Yang-Mills on a lattice. 
Twisted $N=4$ supersymmetry algebra is geometrically realized on a 
four dimensional lattice with link supercharges and the use of
link (anti-)commutators.
Employing Grassmann parameters with link nature,
we explicitly show that 
the resulting super Yang-Mills action is  invariant under all the supercharges 
on a lattice
without chiral fermion problems.
As a group and algebraic interpretation of the link approach, we show 
that promoting bosonic supercovariant derivatives to their exponentials consistently with the lattice Leibniz rule
naturally gives rise to the notion of gauge covariant link (anti-)commutators.
This can be regarded as a fermionic decomposition of a Lie group element,
which may provide a new methodology in super Lie group and super Lie algebra.
We also provide a five dimensional lift-up of the formulation with  
exact $N=4$ SUSY invariance on a five dimensional lattice. 

\end{abstract}


\newpage

\setcounter{footnote}{0}

\section{Introduction }

\indent

Although supersymmetry (SUSY) has not been observed experimentally so far in
the real  world, 
there are numerous theoretical reasons 
for the necessity of SUSY. 
Investigation of lattice regularization of SUSY has its own interest with possible numerical applications, and there have been a number of papers related to this subject
\cite{Dondi-Nicolai, Kramer, Fujikawa, old-SUSY-works1,old-SUSY-works2}.
See also \cite{Ref_added} and references therein. 
In particular,
 it is interesting to ask if we can formulate super Yang-Mills (SYM) theory 
on the lattice by keeping SUSY exact together with gauge covariance. 
In this paper 
we show that specific
types
of extended super Yang-Mills theories can be formulated 
on the lattice keeping all the supercharges exactly conserved
without chiral fermion problems.

In order to explain our way of thinking and the history of lattice SUSY,
we would like to give 
a short summary of the development of our formulations and related works.

To realize exact lattice SUSY, there are essentially two fundamental obstacles A) and B)
as follows.

A) Difference operator breaks Leibniz rule. 

To see 
this first obstacle, let us 
consider the simplest possible SUSY algebra in a 1-dimensional 
theory which can be represented as\cite{DFKKS}: 
\begin{eqnarray}
Q^2=P,
\label{simp-susy-alg1}
\end{eqnarray}
where $Q$ 
represents a
supercharge and $P$ 
represents the
momentum (Hamiltonian in one dimension).  
Identifying (\ref{simp-susy-alg1}) as an operator relation, we can recognize $P$ as a 
translation generator of (space-)time. On the lattice,
 $P$ can be identified as a difference 
operator
$P=i\Delta_\pm$.
We define the forward and backward difference 
operators
as:
\begin{eqnarray}
\Delta_{\pm} 
\phi(x) 
&
\equiv
&
\pm\{\phi(x\pm a)-\phi(x)\},
\label{fb-diffop}
\end{eqnarray}
 where $\phi(x)$ 
denotes
a bosonic field and $a$ 
denotes the
lattice constant.
The forward
 difference operator 
applied on
 a product of two bosonic fields $\phi_1(x)$ 
 and $\phi_2(x)$ leads
 \begin{eqnarray}
\Delta_+
(\phi_1(x)\phi_2(x)) 
&=& \{
\Delta_+
\phi_1(x)\}\phi_2(x) + 
 \phi_1(x+a)
\Delta_+
 \phi_2(x),
 \label{Leibniz10}\\
 &=& \{
\Delta_+
 \phi_1(x)\}\phi_2(x+a) + 
 \phi_1(x)
\Delta_+
 \phi_2(x). 
 \label{Leibniz20}
\end{eqnarray}
Similar relations for the backward difference operator 
$\Delta_-$ hold.

As one can see in (\ref{Leibniz10}) and (\ref{Leibniz20}), 
the forward difference operator does not satisfy Leibniz rule due to the 
shift nature of the field coordinate. The same is true for the backward difference operator. 
As far as
the supercharge operator satisfies Leibniz rule, 
the lattice version of 
SUSY algebra (\ref{simp-susy-alg1}) causes breakdown of SUSY. 

At the very first lattice SUSY investigation\cite{Dondi-Nicolai}, this problem was recognized 
and considered
as the major difficulty 
in
constructing SUSY invariant 
lattice formulation. 
In fact,
it was claimed that Leibniz rule problem cannot be solved 
within the local field formulation\cite{Kato-Sakamoto-So}.

There are many works including the ones 
addressing a question: ``How much 
is the breaking of SUSY in the continuum limit?"
See, for example, \cite{old-SUSY-works1,old-SUSY-works2}. 

In order to cure this 
problem,
we proposed ``link approach formulation of lattice 
SUSY''\cite{DKKN1,DKKN2,DKKN3}. The essential idea is to introduce the following Ansatz: 
the supercharge generates a similar shift $a_Q$ when operating on the product of fields: 
\begin{eqnarray}
 Q(\phi_1(x)\phi_2(x)) = \{Q\phi_1(x)\}\phi_2(x) + 
 \phi_1(x+a_Q)Q\phi_2(x),
 \label{QLeibniz0}
 \end{eqnarray}
where $\phi_1(x)$ and $\phi_2(x)$ 
denote
bosonic fields for simplicity. 
Amongst two possible choices of  (\ref{Leibniz10}) and (\ref{Leibniz20}), we
have chosen (\ref{Leibniz10}) 
type of shift for $Q$ 
operation.
We cannot impose both types 
 (\ref{Leibniz10}) and (\ref{Leibniz20}) simultaneously.
By recognizing $\{Q\phi_1(x)\}$ as a fermionic field, 
a further $Q$ operation on (\ref{QLeibniz0}) 
leads to
\begin{eqnarray}
 Q^2(\phi_1(x)\phi_2(x)) = \{Q^2\phi_1(x)\}\phi_2(x) + 
 \phi_1(x+2a_Q)Q^2\phi_2(x).
 \label{Q2Leibniz}
 \end{eqnarray}
Thus, the
lattice version of SUSY algebra (\ref{simp-susy-alg1}) is recovered:
\begin{eqnarray}
Q^2=i
\Delta_+ ,
\label{simp-susy-alglat}
\end{eqnarray}
 if 
the condition
\begin{eqnarray}
2a_Q=a
\label{shift-rel1}
\end{eqnarray}
is satisfied.
From (\ref{simp-susy-alglat}) and (\ref{shift-rel1}) we can geometrically interpret that 
$i\Delta_+$ 
is a translation generator of single lattice constant $a$, 
while SUSY operator 
$Q$ can be interpreted as a half lattice constant $a_Q=a/2$ translation generator. 

An important observation here is
that the algebraic relation (\ref{QLeibniz0}) of fields includes information 
of lattice location of fields. 
Actually, the relation (\ref{QLeibniz0}) can be expressed as:
\begin{eqnarray}   
 \{Q(\phi_1\phi_2)\}_{x+a_Q,x} = \{Q\phi_1\}_{x+a_Q,x}\{\phi_2\}_{x,x} + 
 \{\phi_1\}_{x+a_Q,x+a_Q}\{Q\phi_2\}_{x+a_Q,x}, 
\label{QLeibnizMatrix}
\end{eqnarray}
where $a_Q$ is given in (\ref{shift-rel1}). 
As we can see, the
 bosonic fields are located on 
the original  lattice $x$
and dual lattice   $x+a_{Q}$
while fermionic fields $Q\phi_1$ and $Q\phi_2$ are 
located on the links connecting the original lattice 
and the dual lattice sites. 
The name of the formulation ``link approach'' is originated from this link nature.

If we keep only original lattice structure, the lattice SUSY algebra (\ref{simp-susy-alglat})
cannot be consistently defined. If we, however, introduce the dual lattice structure with 
link approach Ansatz,
 we can consistently define lattice SUSY algebra. 
When we state ``exact lattice SUSY'' in this  paper,
 we mean this algebraic consistency 
in the sense of link approach  with consistent semi-local extension of space-time points. 

Another 
obstacle to realizing SUSY
on the lattice is:

B) Chiral fermion doublers problem. 

If one defines massless chiral fermion on the lattice by naive lattice fermion formulation, 
copies of a original fermion (doublers) appear as physical particles. Since SUSY requires 
the same number of bosons and fermions, the increase of the number of fermions generates 
a source of SUSY breaking.

One may wonder that this problem would be solved by using different lattice fermion 
formulation. For example Wilson fermion formulation introduces species doublers mass 
at the cut-off level. It was, however, realized that in order to get exact cancellation of 
the SUSY vacuum energy at the quantum level, we need to introduce the same cut-off
mass structure for the corresponding bosons as fermions\cite{Kramer}. 

In principle, chiral fermion solution of lattice QCD can be used for the fermion formulation 
of lattice SUSY, but the bosonic counterpart of the formulation would be hopelessly 
complicated to cancel the SUSY vacuum energy exactly
(see, for example, \cite{Fujikawa}).
As proposed in the series of papers\cite{DKKN1,DKKN2,DKKN3},
we found elegant specific solutions.

Our proposal was to identify the physical degrees of freedom of chiral fermion doublers 
as super multiplets of extended SUSY. Dimension $D$ dependence of the number of 
chiral fermion doublers is $(2^D)$ which coincides with the extended SUSY degrees 
of freedom of $(2^{\frac{N}{2}}\cdot 2^{\frac{N}{2}})$ for $N=D$ in even dimensions:
\begin{eqnarray}
D=1, N=2 (2), \ \ \ D=2,N=2 (4),\ \ \ D=3,N=4 (8), \ \ \ D=4,N=4 (16). 
\nonumber
\end{eqnarray}
In particular 2-dimensional $N=2$ SUSY and 4-dimensional $N=4$ SUSY can be 
formulated on the lattice without chiral fermion problems. The odd dimensional 
counterparts can be treated similarly. 

In order to accommodate lattice chiral fermion doublers degrees of freedom 
into extended SUSY formulation,  we recognized that Dirac-K\"ahler 
twisting mechanism\cite{KT,KKU,KKM} does this job.
As for formulations of Dirac-K\"ahler fermions on a lattice, see also \cite{DK-fermion}.
In the topological field theory investigation,
 scalar BRST charge of quantization of topological 
field theory was found to be a part of supercharge of twisted extended 
SUSY\cite{W}. This notion was extended to include vector, tensor, $\cdots$, 
supercharges. In this way, the notion of Dirac-K\"ahler twisting 
procedure was proposed\cite{KT,KKU} .

As a simple example, $D=N=2$ lattice SUSY algebra is given in Euclidean 2-dimension as
\begin{eqnarray}
\{Q_{\alpha i},\overline{Q}_{j\beta}\}=2\delta_{ij}(\gamma)_{\alpha\beta}i
\Delta_{\pm\mu},
\label{N=2-SUSY-alg}
\end{eqnarray}
where $\overline{Q}=C^{-1}Q^TC$, and we can take the charge 
conjugation matrix $C=\bf{1}$.
Dirac-K\"ahler twisting is defined by the following $\gamma$-matrix expansion of 
supercharge:
\begin{eqnarray}
Q_{\alpha i}=({\bf{1}}Q+\gamma^\mu Q_\mu+\gamma^5\tilde{Q})_{\alpha i},
\end{eqnarray}
where $\gamma^5=i\gamma^1\gamma^2$. 
This leads to $D=N=2$ twisted SUSY algebra: 
\begin{eqnarray}
\{Q,Q_\mu\}=i
\Delta_{+\mu},
 \ \ \ \ \{\tilde{Q},Q_\mu\}=-i\epsilon_{\mu\nu}
\Delta_{-\nu},
\ \ \  \ \{\hbox{others}\}=0.
\label{twist-N=2-alg}
\end{eqnarray}

We can now impose similar Ansatz as (\ref{QLeibniz0}) for $Q_A=(Q,Q_\mu,\tilde{Q})$ 
with shift parameters $a_A=(a,a_\mu,\tilde{a})$.
Carrying out the similar manipulation as in (\ref{Q2Leibniz}) for two twisted supercharges 
of (\ref{twist-N=2-alg}), we obtain $D=N=2$ version of (\ref{shift-rel1}).  
See the details in \cite{DKKN1,DKKN2}. 

In order to formulate twisted SYM,
 we need to introduce the following 
gauge covariantization\cite{DKKN2}:  
\begin{eqnarray}
\Delta_{\pm\mu}
   \rightarrow \mp\mathcal{U}_{\pm\mu} 
   =\mp
(\mathcal{U}_{\pm\mu})
_{x\pm n_\mu,x } \ \ \ \
   Q_A \rightarrow \nabla_A = 
(\nabla_A)
_{x+a_A,x},
\label{cov-N=2alg}   
\end{eqnarray}
where the coordinate dependence is naturally introduced with a link nature of 
shift $n_\mu$ and $a_A$. 
The ordering of the coordinate is determined 
uniquely according to the choice of the definition (\ref{QLeibniz0}).
This property is the typical nature of this formulation.

The gauge 
transformations
of these link variables on a lattice are
given as
\begin{eqnarray}
(\mathcal{U}_{\pm\mu})_{x\pm n_\mu,x} \rightarrow 
G_{x\pm n_\mu}(\mathcal{U}_{\pm\mu})_{x\pm n_\mu,x} G^{-1}_x,  
\nonumber \\
(\nabla_A)_{x+ a_A,x} \rightarrow 
G_{x+ a_A}(\nabla_A)_{x+ a_A,x} G^{-1}_x,
\end{eqnarray}
where $G_x$ denotes a finite gauge transformation at the site $x$.
A gauge covariantized version of (\ref{twist-N=2-alg}) on a lattice is given by
\begin{eqnarray}
\{\nabla,\nabla_\mu\}_{x+a+a_\mu,x} &=&i(\mathcal{U}_{+\mu})_{x+n_\mu,x},
\nonumber \\
\{\tilde{\nabla},\nabla_\mu\}_{x+\tilde{a}+a_\mu,x} 
&=&i\epsilon_{\mu\nu}(\mathcal{U}_{-\nu})_{x-n_\nu,x}, 
\nonumber \\
\{\hbox{others}\}&=&0.
\label{cov-twist-N=2-alg}
\end{eqnarray}

Using (\ref{cov-twist-N=2-alg}) as starting constraints, we can examine graded 
Jacobi identities of $\nabla_A$'s. It turns out that the algebra closes up to  
four $\nabla_A$'s. 
Jacobi identities of three $\nabla_A$'s give defining equations of twisted 
fermions and SUSY transformation of link variables $\mathcal{U}_{\pm\mu}$. 
On the other hand,
 Jacobi identities of four $\nabla_A$'s give SUSY 
transformation of the twisted fermions. 

We define SUSY transformation  $s_A$ of a field $\varphi$ as,
\begin{eqnarray}
s_A(\varphi)_{x+a_\varphi,x}\equiv [\nabla_A,\varphi\}_{x+a_\varphi+a_A,x},
\label{susy-tr1}
\end{eqnarray}
where $a_\varphi$ is a shift of a field $\varphi$. We take commutator or 
anti-commutator depending on $\varphi$ being a boson or fermion.

The continuum $D=N=2$ twisted SYM action was found to have a 
complete exact form of all four twisted supercharges\cite{KKM}. 
The corresponding lattice SYM action can be derived similarly by applying 
the above lattice SUSY transformation just like a continuum construction. 
The final lattice action includes only closed loops in the sense of lattice 
coordinate chain as in (\ref{QLeibnizMatrix}) so that each term is gauge invariant.
Exact lattice SUSY invariance was considered to be obvious since the lattice 
action has all twisted supercharges exact form too and square of each super 
charge all vanishes. Here we claimed the exact lattice SUSY invariance.  
The details can be found in \cite{DKKN2,DKKN3}
\footnote{
There are  also other works which are based on twisted SUSY and respect only a scalar part of
the supercharges on a lattice, for example, \cite{other-twistedsusy}.   
}.

Here came critiques to our statement for exact lattice SUSY invariance 
of link approach formulation by Bruckmann, Kok and Catterall \cite{Bruckmann,BKC}, claiming:  \\ 
1) An ordering ambiguity of product of fields causes inconsistency\cite{Bruckmann}. \\
2) Gauge invariance is lost after SUSY operation in the SYM formulation\cite{BKC}. 

The simplest version 
of
1) can be seen in the following.  
SUSY transformation of a product of two bosonic Abelian fields $\phi_1(x)$ and 
$\phi_2(x)$, which do not carry a shift, 
should coincide irrespective of the order of the 
product: $Q(\phi_1(x)\phi_2(x))=Q(\phi_2(x)\phi_1(x))$. 
On the other hand, 
the link approach Ansatz require the following relations:
\begin{eqnarray}
Q(\phi_1(x)\phi_2(x))=(Q\phi_1)(x)\phi_2(x)+\phi_1(x+a_Q)(Q\phi_2)(x),
\nonumber \\
Q(\phi_2(x)\phi_1(x))=(Q\phi_2)(x)\phi_1(x)+\phi_2(x+a_Q)(Q\phi_1)(x).
\label{ord-amb}
\end{eqnarray}
Their cliam is that there
is an obvious inconsistency since 
the right-hand sides of (\ref{ord-amb}),
where the fields $\phi_{1}$ and $\phi_{2}$ reside in different sites, could not be equated.

The second problem 2) is closely connected with the definition of SUSY 
transformation (\ref{susy-tr1}). The lattice SYM action consists of  gauge invariant
closed loop terms. If we make SUSY transformation $s_A$ to the action, 
a link hole of the shift vector size $a_A$ will be created to each loop term following 
to (\ref{susy-tr1}). Thus, after SUSY transformation, remaining terms are gauge variant. 
Therefore, the transformed action loses gauge invariance and trivially vanishes
on the gauge invariant vacuum, regardless of the SUSY invariance of the action.  
This is their claim in \cite{BKC}.

There is, however, obvious exception for the above critiques 1) and 2). 
As far as 
the corresponding shift parameter $a_A=0$ 
is concerned,
the ordering ambiguity of fields 
disappears and a possible link hole cannot be generated so that gauge 
invariance is kept after SUSY transformation (\ref{susy-tr1}). 

When we proposed the link approach formulation of lattice SYM, there was 
already orbifold construction of lattice SYM by 
Cohen,
Kaplan, Katz and \"{U}nsal\cite{Kaplan}. 
See also related works\cite{Kaplan-related}.
We realized that the orbifold construction corresponds to the special case 
of the link formulation; the scalar shift parameter $a_A=0$, 
as mentioned in
\cite{DKKN2}. 
Accordingly the corresponding scalar supercharge is exactly conserved on the lattice.  
This point was explicitly revealed and stressed also by Damgaard and 
Matsuura\cite{D-M}, Catterall\cite{C}. 
See also related works\cite{relation-others}.

On the other hand,
in a case  where the
shift parameters are non-zero $a_A\ne 0$,
 the link nature of the 
corresponding supercharges are kept and thus the critiques 1) and 2)  
should be seriously addressed.
After a while,
we found out possible way out for the critiques 1) and 2) 
separately. 

First, as for the critique 1), 
if the following non-commutativity holds 
in 
(\ref{ord-amb}), 
\begin{eqnarray}
(Q\phi_1)(x)\phi_2(x)&=\phi_2(x+a_Q)(Q\phi_1)(x), \nonumber \\
(Q\phi_2)(x)\phi_1(x)&=\phi_1(x+a_Q)(Q\phi_2)(x),
\label{non-com-rel1}
\end{eqnarray}
then the inconsistency disappears: $Q(\phi_1(x)\phi_2(x))= Q(\phi_2(x)\phi_1(x))$. 
This 
can be interpreted
that a field carrying shift $a_Q$ generates noncommutative shift 
when changing the order of fields in a product. 
Mathematical consistency for this treatment
was fully investigated within the framework of Hopf Algebra\cite{DKS}. 
Although the ordering ambiguity problem can be solved in this way, the introduction 
of the noncommutativity is by hand at this stage and there remains a puzzle for  
the origin of the noncommutativity.
Generalization to the gauge theory was left uninvestigated.

The critique 2) is suggesting an 
introduction of link super parameters. 
The SUSY transformation $s_A\varphi$ given in (\ref{susy-tr1}) changes the 
shift nature of the original field $\varphi$ so that it cannot be identified 
as SUSY variation of the original field. 
Since the
shift nature should coincide  
with the original field after SUSY transformation, 
the link super parameter should 
be introduced to cancel the shift nature of $\nabla_A$. 
Furthermore the link super parameters should commute with all the fields. 
The link super parameters should take care of the gauge transformations of 
both ending sites of the link. 
Although these features of the super parameters were clearly indicated in \cite{DKKN3},
a confirmative understanding of these objects
has been a long standing 
difficult question that we have been struggling for long time.    

Since a possible solution of link approach to 1) and 2) may be highly nontrivial 
even if there is one, 
we started to consider different type of approaches to solve the starting obstacles A) 
and B). The above mentioned noncommutative geometry approach with 
Hopf algebra was a serious trial. Shift nature of Leibniz rule of the difference 
operator is consistently accommodated in the field theory of Hopf algebra\cite{DKS}. 
In this formulation, however, a straightforward 
non-Abelian gauge extension would share the same 
difficulty of 2).   

Matrix model investigations were tried to accommodate the obstacles A)\cite{ADKS,0806.0686}. In particular the derivation of the shift nature of coordinate   
was realized from particular 
noncommutativity\cite{ADKS}.
Besides, a formulation with a star product to accommodate the above non-commutativity
was proposed in 
\cite{Nagata-largeN}.   
 
We have also
tried completely different type of formulation from the link approach 
to accommodate the lattice Leibniz rule A) and chiral fermion doublers 
problem B). It is to identify momentum conservation as nonlocal product of 
distributive law of Leibniz rule in the coordinate space\cite{DKS2}. 
This type of idea was first recognized in the early stage of investigation\cite{Nojiri}. 
It turned out that we can formulate lattice field theory exactly conserved for all super 
charges in the case of infinitely extended lattice. So far the formulation was 
successfully formulated except for non-Abelian gauge theory. 
  
The main body of calculations of this paper was completed at the time of 
the completion of $N=4, D=3$ super Yang-Mills on the lattice in 2008\cite{DKKN3}, 
since it was a part of Ph.D thesis\cite{Nagata-PhD} of one of the authors of 
this paper, Kazuhiro Nagata. Since Bruckmann, Kok and Catterall critiques to 
the link approach came out around those time (2006-2007)\cite{Bruckmann,BKC}, 
we postponed to publish the link approach formulation of $N=D=4$ lattice SYM version 
until we may solve the critiques completely.      
 
As explained in the above we tried other formulations than the link approach 
to overcome the obstacles A) and B). We reached partial success to overcome the 
obstacles and to clear the critiques. 
In the meantime, our dearest collaborator, Alessandro D'Adda passed away, 
and we decided to come back again to the link approach formulation and try to 
find out complete understanding for the critiques 1) and 2). 

{\bf
We eventually reached to satisfactory understanding for the origin of the 
noncommutativity, 1), and found out a concrete and explicit expression of the 
link super parameter, 2). In this new understanding the solution for 1) and 2) are 
fundamentally related.  
}

This paper is organized as follows. 
In section 2, we first start from reviewing 
the Dirac-K\"ahler twisting procedure of $N=4$
in 4-dimensional continuum spacetime\cite{KKM}. 
After observing twisted SYM multiplet of $N=D=4$ in section 3,
we formulate $N=D=4$ Dirac-K\"ahler twisted SYM in continuum spacetime in section 4.
Then, we 
provide a
lattice Leibniz rule on 4-dimensional lattice,
proceed to formulate lattice counterpart of $N=D=4$ twisted SYM,
and explain a group and algebraic background of the link approach
in section 5.
In particular, in section 5, from the group and algebraic aspect, 
we explicitly show that how and why the two fundamental concepts of 
non-commutativity and link nature are tightly related each other.
A possible extension of five dimensional lift-up is 
also proposed 
in section 6,
and our reply to the criticism  \cite{Bruckmann,BKC} 
is explicitly mentioned in section 7.

\section{$N=D=4$ Dirac-K\"ahler Twisted SUSY Algebra}

\indent

In this section, we review the $N=D=4$ Dirac-K\"ahler twisted SUSY algebra
in continuum spacetime. Full analysis of the algebra is already given in \cite{KKM}.
See also \cite{Kato:2011yh} for related subjects.

We start from the following extended SUSY algebra in 4-dimensions,
\begin{eqnarray}
\{Q_{\alpha i},\overline{Q}_{j\beta}\} &=& 2\delta_{ij}(\gamma^{\mu})_{\alpha\beta}P_{\mu},
\label{N=D=4susy}
\end{eqnarray}
where $\overline{Q}_{j\beta}$ is defined by the following conjugation of $Q_{\alpha i}$,
\begin{eqnarray}
\overline{Q}_{i\alpha} &=& (C^{-1}Q^{T}C)_{i\alpha},
\end{eqnarray}
with charge conjugation matrix $C$ which satisfies
\begin{eqnarray}
\gamma_{\mu}^{T} &=& C\gamma_{\mu}C^{-1},\ \ \ \ \
C^{T} \ =\ -C. 
\end{eqnarray}
The remaining part of extended super Poincare algebra can be defined as
\begin{eqnarray}
[J_{\mu\nu},Q_{\alpha i}] &=& -\frac{i}{2}(\gamma_{\mu\nu})_{\alpha\beta}Q_{\beta i},\\[0pt]
[R_{\mu\nu},Q_{\alpha i}] &=& +\frac{i}{2}Q_{\alpha j}(\gamma_{\mu\nu})_{ji},\\[2pt]
[P_{\mu},Q_{\alpha i}] &=& 0,\\[2pt]
[J_{\mu\nu},P_{\rho}] &=& -i\delta_{\mu\nu\rho\sigma}P_{\sigma},\\[2pt]
[R_{\mu\nu},P_{\rho}] &=& 0,\\[2pt]
[P_{\mu},P_{\nu}] &=& 0,\\[2pt]
[J_{\mu\nu},J_{\rho\sigma}] &=& -i(\delta_{\mu\rho}J_{\nu\sigma}-\delta_{\nu\rho}J_{\mu\sigma}
-\delta_{\mu\sigma}J_{\nu\rho}+\delta_{\nu\sigma}J_{\mu\rho}),\\[2pt]
[R_{\mu\nu},R_{\rho\sigma}] &=& -i(\delta_{\mu\rho}R_{\nu\sigma}-\delta_{\nu\rho}R_{\mu\sigma}
-\delta_{\mu\sigma}R_{\nu\rho}+\delta_{\nu\sigma}R_{\mu\rho}),\\[2pt]
[J_{\mu\nu},R_{\rho\sigma}] &=& 0,
\end{eqnarray}
where $J_{\mu\nu}$ and $R_{\mu\nu}$ denote $SO(4)$ Lorentz and $SO(4) \subset SU(4)$ of 
$N=4$ internal rotation generators, respectively. Here in this paper we use the terminology 
of 4-dimensional Euclidean space-time as $SO(4)$ Lorentz space.

Dirac-K\"ahler twisting procedure can be done by
introducing ``twisted" Lorentz generator $J'_{\mu\nu}$ as
\begin{eqnarray}
J'_{\mu\nu} &=& J_{\mu\nu} + R_{\mu\nu}.
\end{eqnarray}
The corresponding twisted super Poincare algebra is most naturally
described through the following 
Dirac-K\"ahler expansion of supercharge $Q_{\alpha i}$,
\begin{eqnarray}
Q_{\alpha i} &=& \frac{1}{\sqrt{2}}(\mathbf{1}Q + \gamma^{\mu}Q_{\mu}
+\frac{1}{2}\gamma^{\mu\nu}Q_{\mu\nu} + \tilde{\gamma^{\mu}}\tilde{Q}_{\mu}
+ \gamma^{5}\tilde{Q})_{\alpha i}, \label{4DQ}
\end{eqnarray}
where $\gamma^{\mu\nu}\equiv\frac{1}{2}[\gamma^{\mu},\gamma^{\nu}]$,
$\tilde{\gamma}^{\mu}\equiv \gamma^{\mu}\gamma^{5}$, 
$\gamma^{5}\equiv \gamma^{1}\gamma^{2}\gamma^{3}\gamma^{4}$ 
with $\{\gamma^\mu,\gamma^\nu\}=2\delta^{\mu\nu}$.

SUSY algebra of (\ref{N=D=4susy}) can be written by using the above expansion as
\begin{eqnarray}
\{Q,Q_{\mu}\} &=& P_{\mu},\label{N=D=4tsusy1}\\[2pt]
\{Q_{\rho\sigma},Q_{\mu}\} &=& -\delta_{\rho\sigma\mu\nu}P_{\nu},
\label{N=D=4tsusy2}\\[2pt]
\{Q_{\rho\sigma},\tilde{Q}_{\mu}\} &=& +\epsilon_{\rho\sigma\mu\nu}P_{\nu},
\label{N=D=4tsusy3}\\[2pt]
\{\tilde{Q},\tilde{Q}_{\mu}\} &=& P_{\mu},\label{N=D=4tsusy4}\\[2pt]
\{others\} &=& 0, \label{N=D=4tsusy5}
\end{eqnarray}
and also for the remaining part of the algebra,
\begin{eqnarray}
[J_{\mu\nu},Q] &=& +\frac{i}{2}Q_{\mu\nu}, \label{Lorentz4D1}\\[2pt]
[J_{\mu\nu},Q_{\rho}]  &=&  -\frac{i}{2}
\delta_{\mu\nu\rho\sigma}Q_{\sigma}
-\frac{i}{2}\epsilon_{\mu\nu\rho\sigma}\tilde{Q}_{\sigma},\\[2pt]
[J_{\mu\nu},Q_{\rho\sigma}]&=&-\frac{i}{2}\delta_{\mu\nu\rho\sigma}Q
+\frac{i}{2}\epsilon_{\mu\nu\rho\sigma}\tilde{Q}
+\frac{i}{2}\delta_{\mu\nu\rho\lambda}Q_{\sigma\lambda}
-\frac{i}{2}\delta_{\mu\nu\sigma\lambda}Q_{\rho\lambda},\\[2pt]
[J_{\mu\nu},\tilde{Q}_{\rho}]&=&-\frac{i}{2}\epsilon_{\mu\nu\rho\sigma}
Q_{\sigma}-\frac{i}{2}\delta_{\mu\nu\rho\sigma}\tilde{Q}_{\sigma}\\[2pt]
[J_{\mu\nu},\tilde{Q}] &=& -\frac{i}{4}\epsilon_{\mu\nu\rho\sigma}
Q_{\rho\sigma}, \label{Lorentz4D5} \\[7pt]
[J_{\mu\nu},P_{\rho}]&=&-i\delta_{\mu\nu\rho\sigma}P_{\sigma},\\[10pt]
[J_{\mu\nu},J_{\rho\sigma}]&=&
-i(\delta_{\mu\rho}J_{\nu\sigma}-\delta_{\nu\rho}J_{\mu\sigma}
-\delta_{\mu\sigma}J_{\nu\rho}+\delta_{\nu\sigma}J_{\mu\rho}),
\end{eqnarray}
\begin{eqnarray}
[R_{\mu\nu},Q] &=& -\frac{i}{2}Q_{\mu\nu},\label{R4D1}\\[2pt]
[R_{\mu\nu},Q_{\rho}]  &=&  -\frac{i}{2}
\delta_{\mu\nu\rho\sigma}Q_{\sigma}
+\frac{i}{2}\epsilon_{\mu\nu\rho\sigma}\tilde{Q}_{\sigma},\\[2pt]
[R_{\mu\nu},Q_{\rho\sigma}]&=&+\frac{i}{2}\delta_{\mu\nu\rho\sigma}Q
-\frac{i}{2}\epsilon_{\mu\nu\rho\sigma}\tilde{Q}
+\frac{i}{2}\delta_{\mu\nu\rho\lambda}Q_{\sigma\lambda}
-\frac{i}{2}\delta_{\mu\nu\sigma\lambda}Q_{\rho\lambda},\\[2pt]
[R_{\mu\nu},\tilde{Q}_{\rho}]&=&+\frac{i}{2}\epsilon_{\mu\nu\rho\sigma}
Q_{\sigma}
-\frac{i}{2}\delta_{\mu\nu\rho\sigma}\tilde{Q}_{\sigma}\\[2pt]
[R_{\mu\nu},\tilde{Q}] &=& +\frac{i}{4}\epsilon_{\mu\nu\rho\sigma}
Q_{\rho\sigma},\label{R4D5}\\[7pt]
[R_{\mu\nu},P_{\rho}]&=&0,\\[5pt]
[R_{\mu\nu},R_{\rho\sigma}]&=&
-i(\delta_{\mu\rho}R_{\nu\sigma}-\delta_{\nu\rho}R_{\mu\sigma}
-\delta_{\mu\sigma}R_{\nu\rho}+\delta_{\nu\sigma}R_{\mu\rho}),\\[5pt]
[J_{\mu\nu},R_{\rho\sigma}]&=& [P_{\mu},P_{\nu}]\ =\ 0,
\end{eqnarray} 
where $\delta_{\mu\nu\rho\sigma}\equiv\delta_{\mu\rho}\delta_{\nu\sigma}
-\delta_{\mu\sigma}\delta_{\nu\rho}$. 
Notice that under twisted Lorentz generator $J'_{\mu}$,
each component of Dirac-K\"ahler twisted supercharge
transforms as
\begin{eqnarray}
[J'_{\mu\nu},Q] &=& 0,\\[2pt]
[J'_{\mu\nu},Q_{\rho}] &=& -i\delta_{\mu\nu\rho\sigma}Q_{\sigma},\\[2pt]
[J'_{\mu\nu},Q_{\rho\sigma}]&=& i\delta_{\mu\nu\rho\lambda}Q_{\sigma\lambda}
-i\delta_{\mu\nu\sigma\lambda}Q_{\rho\lambda},\\[2pt]
[J'_{\mu\nu},\tilde{Q}_{\rho}] &=& -i\delta_{\mu\nu\rho\sigma}\tilde{Q}_{\sigma},\\[2pt]
[J'_{\mu\nu},\tilde{Q}] &=& 0,
\end{eqnarray}
which means each component $(Q,Q_{\mu},Q_{\mu\nu},\tilde{Q}_{\mu},\tilde{Q})$ behaves
as (scalar, vector, 2nd rank tensor, (pseudo)vector, (pseudo)scalar), respectively, 
in the space where $J'_{\mu\nu}$ is rotation generators.

As pointed out in \cite{KKM}, the SUSY algebra (\ref{N=D=4tsusy1})-(\ref{N=D=4tsusy5})
can be decomposed into the following two independent set of $N=2\ D=4$ twisted SUSY algebra,
\begin{eqnarray}
\{Q^{+},Q^{+}_{\mu}\} &=& P_{\mu},\label{N=2D=4tsusy+1}\\[2pt]
\{Q^{+}_{\rho\sigma},Q^{+}_{\mu}\} &=& -\delta^{+}_{\rho\sigma\mu\nu}P_{\nu},
\label{N=2D=4tsusy+2}\\[5pt]
\{Q^{-},Q^{-}_{\mu}\} &=& P_{\mu},\label{N=2D=4tsusy-1}\\[2pt]
\{Q^{-}_{\rho\sigma},Q^{-}_{\mu}\} &=& -\delta^{-}_{\rho\sigma\mu\nu}P_{\nu},
\label{N=2D=4tsusy-2}\\[5pt]
\{others\} &=& 0,
\end{eqnarray}
where $\delta^{\pm}_{\mu\nu\rho\sigma}\equiv \delta_{\mu\nu\rho\sigma}\pm
\epsilon_{\mu\nu\rho\sigma}$ 
projects self-dual and anti self-dual part of 2nd rank tensors, 
and the supercharges $Q_{A}^{\pm}$ in l.h.s.
are defined as
\begin{eqnarray}
Q^{\pm}&\equiv&\frac{1}{\sqrt{2}}(Q\mp \tilde{Q}), \label{N=4decomp1}\\[2pt]
Q^{\pm}_{\rho}&\equiv&\frac{1}{\sqrt{2}}(Q_{\rho}\mp\tilde{Q}_{\rho}),\label{N=4decomp2}\\[2pt]
Q^{\pm}_{\mu\nu}&\equiv&\frac{1}{\sqrt{2}}(Q_{\mu\nu}\pm\frac{1}{2}\epsilon_{\mu\nu\rho\sigma}
Q_{\rho\sigma}).\label{N=4decomp3}
\end{eqnarray}

One can find superspace representations of twisted supercharges for the $N=D=4$  
$(Q,Q_{\mu},Q_{\rho\sigma},\tilde{Q}_{\mu},\tilde{Q})$ as well as for the supercovariant derivatives
$(D,D_{\mu},D_{\rho\sigma},\tilde{D}_{\mu},\tilde{D})$,
\begin{eqnarray}
Q &=& \frac{\partial}{\partial\theta}+\frac{i}{2}\theta^{\mu}\partial_{\mu},\\[2pt]
Q_{\mu} &=& \frac{\partial}{\partial\theta^{\mu}}
+\frac{i}{2}\theta \partial_{\mu}
-\frac{i}{4}\delta_{\mu\nu\rho\sigma}\theta^{\rho\sigma}\partial_{\mu},\\[2pt]
Q_{\rho\sigma} &=& \frac{\partial}{\partial\theta^{\rho\sigma}}
-\frac{i}{2}\delta_{\rho\sigma\alpha\beta}\theta^{\alpha}\partial_{\beta}
+\frac{i}{2}\epsilon_{\rho\sigma\alpha\beta}\tilde{\theta}^{\alpha}\partial_{\beta},\\[2pt]
\tilde{Q}_{\mu} &=& \frac{\partial}{\partial\tilde{\theta}^{\mu}}
+\frac{i}{2}\tilde{\theta}\partial_{\mu} 
+\frac{i}{4}\epsilon_{\mu\nu\rho\sigma}\theta^{\rho\sigma}\partial_{\nu},\\[2pt]
\tilde{Q} &=& \frac{\partial}{\partial\tilde{\theta}}  
+ \frac{i}{2}\tilde{\theta}^{\mu}\partial_{\mu}, 
\end{eqnarray}
\begin{eqnarray}
D &=& \frac{\partial}{\partial\theta}-\frac{i}{2}\theta^{\mu}\partial_{\mu},
\label{4DD1}\\[2pt]
D_{\mu} &=& \frac{\partial}{\partial\theta^{\mu}}
-\frac{i}{2}\theta \partial_{\mu}
+\frac{i}{4}\delta_{\mu\nu\rho\sigma}\theta^{\rho\sigma}\partial_{\mu},
\label{4DD2}\\[2pt]
D_{\rho\sigma} &=& \frac{\partial}{\partial\theta^{\rho\sigma}}
+\frac{i}{2}\delta_{\rho\sigma\alpha\beta}\theta^{\alpha}\partial_{\beta}
-\frac{i}{2}\epsilon_{\rho\sigma\alpha\beta}\tilde{\theta}^{\alpha}\partial_{\beta},
\label{4DD3}\\[2pt]
\tilde{D}_{\mu} &=& \frac{\partial}{\partial\tilde{\theta}^{\mu}}
-\frac{i}{2}\tilde{\theta}\partial_{\mu} 
-\frac{i}{4}\epsilon_{\mu\nu\rho\sigma}\theta^{\rho\sigma}\partial_{\nu},
\label{4DD4}\\[2pt]
\tilde{D} &=& \frac{\partial}{\partial\tilde{\theta}}  
- \frac{i}{2}\tilde{\theta}^{\mu}\partial_{\mu}, \label{4DD5}
\end{eqnarray}
which satisfy
\begin{eqnarray}
\{Q,Q_{\mu}\} &=& +i\partial_{\mu},\label{4DQ1}\\[2pt]
\{Q_{\rho\sigma},Q_{\mu}\} &=& -i\delta_{\rho\sigma\mu\nu}\partial_{\nu},\label{4DQ2}\\[2pt]
\{Q_{\rho\sigma},\tilde{Q}_{\mu}\} &=& +i\epsilon_{\rho\sigma\mu\nu}\partial_{\nu},
\label{4DQ3}\\[2pt]
\{\tilde{Q},\tilde{Q}_{\mu}\} &=& +i\partial_{\mu},\label{4DQ4}\\[2pt]
\{others\} &=& 0,\label{4DQ5}
\end{eqnarray}
\begin{eqnarray}
\{D,D_{\mu}\} &=& -i\partial_{\mu},\\[2pt]
\{D_{\rho\sigma},D_{\mu}\} &=& +i\delta_{\rho\sigma\mu\nu}\partial_{\nu},\\[2pt]
\{D_{\rho\sigma},\tilde{D}_{\mu}\} &=& -i\epsilon_{\rho\sigma\mu\nu}\partial_{\nu},\\[2pt]
\{\tilde{D},\tilde{D}_{\mu}\} &=& -i\partial_{\mu},\\[2pt]
\{others\} &=& 0,
\end{eqnarray}
where in the $D$-algebra the sign of $\partial_\mu$ is reversed. 

Supercharges $Q_{A}=(Q,Q_{\mu},Q_{\rho\sigma},\tilde{Q}_{\mu},\tilde{Q})$ and 
super derivatives $D_{B}=(D,D_{\mu},D_{\rho\sigma},\tilde{D}_{\mu},\tilde{D})$ 
anti-commute each other
\begin{eqnarray}
\{Q_{A},D_{B}\} &=& 0.
\end{eqnarray}

\section{Relation to the other types of $4D$ Twisted Algebra}

\indent

Topological twisting of extended supersymmetry was originally
proposed in \cite{W} for $N=2$ in 4-dimensional Euclidean spacetime
in the context of topological quantum field theory. 
Later on its extension to $N=4$ has been discussed in 
\cite{Yamron,Vafa-Witten,Marcus}.
Twisted superspace formulation of $N=2\ D=4$ SYM was given in \cite{Labastida}. 
There has been a large number of studies
especially relating to duality 
in the context of topological quantum field theory
\cite{Vafa-Witten,Dijkgraaf,Lozano,Blau-Thompson}.
Another aspect of twisted SUSY is 
related to BRST quantization procedure of gauge theories.
It was shown that instanton gauge fixing of topological Yang-Mills
action led to $N=2$ twisted SYM in 4-dimensions \cite{BS}.
Also, it was explicitly shown in 2-dimensions
that instanton gauge fixing
of generalized gauge theory \cite{KW} led to $N=D=2$
twisted SYM action, and resulting fermion structure 
turns out to be Dirac-K\"ahler fermion in 2-dimensions.
It was then recognized that there is a fundamental 
relation between twisting procedure and Dirac-K\"ahler fermion mechanism \cite{KT}.
Along with these understandings,
Dirac-K\"ahler twisted superspace formulations of SYM
have been proposed
in \cite{KKU,KKM}.

In this section, 
we summarize the twisting procedure of
$N=2$ and $N=4$ SYM multiplets in 4-dimensions 
proposed in \cite{W,Yamron,Marcus} 
and try to clarify the relations to Dirac-K\"ahler twisted SYM multiplets.

\subsection{$N=2\ D=4$ Twisting}

The twisting of $N=2\ D=4$ given in \cite{W} is defined as 
\begin{itemize}
\item to take $SU(2)'_L$ as the diagonal subgroup of
$SU(2)_L \otimes SU(2)_I$,
\end{itemize}     
where $SU(2)_L$ denotes the left-handed part of the Lorentz
representation $SU(2)_L\otimes SU(2)_R$ and $SU(2)_I$ the
internal symmetry of $N=2$ supersymmetric multiplet.
Denoting the indices by $(SU(2)_L,SU(2)_R,SU(2)_I)$
for the $N=2$ SYM multiplet and $(SU(2)'_L,SU(2)_R)$ for
the twisted SYM multiplet, one could see the manipulation of
twisting procedure as follows,

\vspace{10pt}
\hspace{48pt}\underline{$N=2$ 
SYM multiplet}
\hspace{25pt}\underline{$N=2$
 twisted SYM multiplet}
\begin{equation}
\begin{split}
\omega_{\mu}(\mathbf{2},\mathbf{2},\mathbf{1})
&\longrightarrow 
\omega_{\mu}(\mathbf{2},\mathbf{2})\\
\lambda^{i}_{\alpha}(\mathbf{2},\mathbf{1},\mathbf{2})
&\longrightarrow
\rho^{+}_{\mu\nu}(\mathbf{3},\mathbf{1})\oplus
\rho(\mathbf{1},\mathbf{1})\\
\lambda^{i\dot{\alpha}}(\mathbf{1},\mathbf{2},\mathbf{2})
&\longrightarrow
\lambda_{\mu}(\mathbf{2},\mathbf{2})\\
\phi(\mathbf{1},\mathbf{1},\mathbf{1})
&\longrightarrow
A(\mathbf{1},\mathbf{1})\\
\overline{\phi}(\mathbf{1},\mathbf{1},\mathbf{1})
&\longrightarrow
B(\mathbf{1},\mathbf{1})\\
D^{ij}(\mathbf{1},\mathbf{1},\mathbf{3})
&\longrightarrow
H^{+}_{\mu\nu}(\mathbf{3},\mathbf{1})
\end{split}
\label{N=2twisting}
\end{equation}
where the suffices $i,j=(1,2)$ denote the 
$N=2$ internal symmetry and
$\phi$, $\overline{\phi}$ are the scalar bosons
in $N=2$ ordinary SYM multiplet.
$D^{ij}$ is auxiliary field transforming as a triplet of $SU(2)_I$. 

Indices in the parentheses 
after the twisting denote
transformation properties under twisted $SO(4)$ Lorentz
transformation which is denoted as $J'_{\mu\nu}$
in the previous subsection.
 For instance,
$\rho^{+}_{\mu\nu}(\mathbf{3},\mathbf{1})$, 
which representing three components of twisted fermion,
should transform as a self-dual anti-symmetric 
2nd-rank tensor under twisted Lorentz transformation $J'_{\mu\nu}$.
In a similar way, $A(\mathbf{1},\mathbf{1})$ and $B(\mathbf{1},\mathbf{1})$
should transform as scalars under $J'_{\mu\nu}$ 

The above multiplet can be constructed in the 
Dirac-K\"ahler twisting procedure by
the $N=2$ decomposition of $N=D=4$ Dirac-K\"ahler twisted algebra.
One can actually show that the following
$N=2\ D=4$ twisted SYM constraints and Jacobi identity procedure
give the off-shell structure of the above multiplet \cite{KKM}. 
We first impose the following super connection ansatz: 
\begin{eqnarray}
\{\D^{+},\D^{+}\}&=& -iW \label{n2cc1}\\
\{\D_{\mu\nu}^{+},\Drs^{+}\}&=& -i\delta^{+}_{\mu\nu\rho\sigma}W \\
\{\Dm^{+},\Dn^{+}\} &=& -i \delta_{\mu\nu}F \\
\{\D^{+},\Dm^{+}\}&=& -i\D_{\underline{\mu}} \\
\{\Drs^{+},\Dm^{+}\}&=& +i\delta^{+}_{\rho\sigma\mu\nu}\D_{\underline{\nu}}, 
\label{n2cc2}\hspace{50pt}
\{others\} = 0,
\end{eqnarray}
or, if we take $SU(2)'_{R}$ as a diagonal subgroup of $SU(2)_{R}\otimes SU(2)_{I}$,
\begin{eqnarray}
\{\D^{-},\D^{-}\}&=& -iW \label{n2cc3}\\
\{\D_{\mu\nu}^{-},\Drs^{-}\}&=& -i\delta^{-}_{\mu\nu\rho\sigma}W \\
\{\Dm^{-},\Dn^{-}\} &=& -i \delta_{\mu\nu}F \\
\{\D^{-},\Dm^{-}\}&=& -i\D_{\underline{\mu}} \\
\{\Drs^{-},\Dm^{-}\}&=& +i\delta^{-}_{\rho\sigma\mu\nu}\D_{\underline{\nu}}, 
\label{n2cc4}\hspace{50pt}
\{others\} = 0,
\end{eqnarray}
where $\delta^{+}_{\mu\nu\rho\sigma}\equiv \delta_{\mu\nu\rho\sigma} \pm \epsilon_{\mu\nu\rho\sigma}$,
and 
\begin{eqnarray}
\D^{\pm}&\equiv&\frac{1}{\sqrt{2}}(\D\mp \tD)\\
\Dm^{\pm}&\equiv&\frac{1}{\sqrt{2}}(\Dm\mp \tDm)\\
\D^{\pm}_{\mu\nu}&\equiv&\frac{1}{\sqrt{2}}(\D_{\mu\nu}\pm \frac{1}{2}\epsilon_{\mu\nu\rho\sigma}\Drs)
\end{eqnarray}
are given by the Dirac-K\"{a}hler twisted super-covariant derivatives with super-connections,
\begin{eqnarray}
\D &\equiv& D -i\Gamma(x,\theta_{A}),
\label{4DDD1}\\[2pt]
\D_{\mu} &\equiv& D_{\mu} -i\Gamma_{\mu}(x,\theta_{A}),
\label{4DDD2}\\[2pt]
\D_{\rho\sigma} &\equiv& D_{\rho\sigma} -i\Gamma_{\rho\sigma}(x,\theta_{A}),
\label{4DDD3}\\[2pt]
\tilde{\D}_{\mu} &\equiv& \tilde{D}_{\mu} -i\tilde{\Gamma}_{\mu}(x,\theta_{A}),
\label{4DDD4}\\[2pt]
\tilde{\D} &\equiv& \tilde{D} -i\tilde{\Gamma}(x,\theta_{A}). 
\label{4DDD5}
\end{eqnarray}
 
The lowest components of the 
superfields in
the r.h.s.
of (\ref{n2cc1})-(\ref{n2cc4})  
coincide with the bosonic components of $N=2\ D=4$ SYM multiplet, $(\Dc_{\mu},A,B)$,
\begin{eqnarray}
\D_{\underline{\mu}}| &\equiv& \Dc_{\mu} \ =\ \partial_{\mu} -i\omega_{\mu},\\[2pt]
W| &=& A,\\[2pt]
F| &=& B,
\end{eqnarray}
where, the symbol $|$ denotes taking $\theta_{A}=0$ in the corresponding superfield.
It can be shown that starting from
the $N=2\ D=4$ selfdual constraints (\ref{n2cc1})-(\ref{n2cc2}),
one can formulate off-shell twisted $N=2\ D=4$ SYM multiplet 
whose bosonic components of the multiplet are given by
gauge field $\omega_{\mu}$, scalars $A$ and $B$ as well as
three d.o.f. of auxiliary field $H^{+}_{\mu\nu}$
while fermionic components are given by
$(\rho,\lambda_{\mu},\rho^{+}_{\mu\nu})$,
which makes complete agreement with the above
observation given in (\ref{N=2twisting}).
We do not go further into the detail of $N=2\ D=4$ SYM formulation since
a self-contained analysis in terms of Dirac-K\"ahler twisting 
was already given in \cite{KKM}
where the $N=2\ D=4$ twisted SYM multiplet,
off-shell closure of the algebra and
corresponding SUSY invariant action is formulated.

\subsection{$N=D=4$ Twisting}

$N=4$ on-shell gauge multiplet with $SU(4)$ internal symmetry consists of 
\begin{equation}
\begin{split}
&\bullet\ \   \omega_{\mu}
(\mathbf{2},\mathbf{2},\mathbf{1}) \\
&\bullet\ \  \lambda_{u}^{\alpha}
(\mathbf{2},\mathbf{1},\mathbf{4}) \\
&\bullet\ \  \bar{\lambda}^{u}_{\dot{\alpha}}
(\mathbf{1},\mathbf{2},\mathbf{\bar{4}}) \\
&\bullet\ \ \phi_{uv}
(\mathbf{1},\mathbf{1},\mathbf{6}) 
 \end{split}
\end{equation}
where the indices denote $(SU(2)_L,SU(2)_R,SU(4)_I)$
with the extended internal symmetry $SU(4)_I$
and $\phi_{uv}$ is the bosonic scalar field
represented as $\mathbf{6}$ of $SU(4)$,
\begin{equation}
\phi^{uv}=(\phi_{uv})^{\dagger}=
\phi_{vu}^{*}=-\frac{1}{2}
\epsilon^{uvwx}\phi_{wx},\ \ \ \
\epsilon^{1234}=1.
\end{equation} 

The twisting for $N=4$
comes from the observation that  
$SU(4)$ splits into its regular subgroups
such as 
\begin{equation}
SU(4)\supset SO(4) \simeq SU(2)_1\otimes SU(2)_2.
\end{equation}
Accordingly, the representation $\mathbf{4}$
of $SU(4)_I$ splits into the following
three types,
\begin{equation}
\begin{split}
 (i) \ \ & \mathbf{4}\rightarrow 
(\mathbf{2},\mathbf{1})\oplus
(\mathbf{2},\mathbf{1}) \\
 (i\hspace{-1pt}i) \ \ & \mathbf{4}\rightarrow 
(\mathbf{2},\mathbf{1})\oplus
(\mathbf{1},\mathbf{2})\\ 
 (i\hspace{-1pt}i\hspace{-1pt}i) 
\ \ & \mathbf{4}\rightarrow 
(\mathbf{2},\mathbf{1})\oplus
(\mathbf{1},\mathbf{1})\oplus
(\mathbf{1},\mathbf{1})
 \end{split}
\end{equation}
with the indices $(SU(2)_1,SU(2)_2)$.

The first case (\textit{i}) is called A-type which
 was first introduced by Yamron \cite{Yamron}
 and studied by Vafa and Witten in a strong coupling test of S-duality \cite{Vafa-Witten}.
The second case (\textit {ii}) 
is called B-type which
was first briefly introduced at the end of \cite{Yamron}.
We will see that the above derived Dirac-K\"ahler twisting
is identified as this B-type. Historically, the B-type twist was first extensively 
studied by Marcus in \cite{Marcus} as another type of TQFT
by means of complexifying the gauge fields. 
The third case (\textit{iii}) was also first introduced in \cite{Yamron}.
As we will see in the following,
the first two cases (\textit{i}) and  (\textit{ii}) 
lead
to two independent fermionic
scalar charges while
the third case (\textit{iii}) generates only one
scalar charge and so is called half-twisted theory \cite{Lozano}. 
As for the studies of $N=D=4$ twisting, see also 
\cite{Dijkgraaf}.
It is interesting to note that as pointed out in \cite{Kapustin:2006pk}, the B-type twisting has been playing an important role in the studies of geometric Langlands correspondence.

Twisting procedure for $N=4$ 
is to take $SU(2)'_L$ and $SU(2)'_R$
as diagonal subgroups of 
$SU(2)_L\otimes SU(2)_1$ and
$SU(2)_R\otimes SU(2)_2$ ,
respectively.

For case $(i)$,
\begin{equation}
\begin{split}
&
(\mathbf{2},\mathbf{1},\mathbf{4}) 
\rightarrow
2(\mathbf{1},\mathbf{1})
\oplus
2(\mathbf{3},\mathbf{1}) \\
&
(\mathbf{1},\mathbf{2},\mathbf{4}) 
\rightarrow
2(\mathbf{2},\mathbf{2})
\end{split}
\end{equation}

For case $(i\hspace{-1pt}i)$,
\begin{equation}
\begin{split}
&
(\mathbf{2},\mathbf{1},\mathbf{4}) 
\rightarrow
(\mathbf{1},\mathbf{1})
\oplus
(\mathbf{3},\mathbf{1})
\oplus
(\mathbf{2},\mathbf{2}) \\
&
(\mathbf{1},\mathbf{2},\mathbf{4}) 
\rightarrow
(\mathbf{2},\mathbf{2})
\oplus
(\mathbf{1},\mathbf{1})
\oplus
(\mathbf{1},\mathbf{3}).
\end{split}
\label{fermion}
\end{equation}

For case $(i\hspace{-1pt}i\hspace{-1pt}i)$,
\begin{equation}
\begin{split}
&
(\mathbf{2},\mathbf{1},\mathbf{4}) 
\rightarrow
(\mathbf{1},\mathbf{1})
\oplus
(\mathbf{3},\mathbf{1})
\oplus
2(\mathbf{2},\mathbf{1}) \\
&
(\mathbf{1},\mathbf{2},\mathbf{4}) 
\rightarrow
(\mathbf{2},\mathbf{2})
\oplus
2(\mathbf{1},\mathbf{2}).
\end{split}
\end{equation}
Let us turn to the $\mathbf{6}$ of scalars which is the
anti-symmetric part of $\mathbf{4}\otimes\mathbf{4}$.

For case $(i)$,
\begin{equation}
\begin{split}
\mathbf{4}\otimes\mathbf{4}
&\rightarrow
[(\mathbf{2},\mathbf{1})\oplus
(\mathbf{2},\mathbf{1})]
\otimes
[(\mathbf{2},\mathbf{1})\oplus
(\mathbf{2},\mathbf{1})] \\
& =
4(\mathbf{1},\mathbf{1})
\oplus
4(\mathbf{3},\mathbf{1}).
\end{split}
\end{equation}

For case $(i\hspace{-1pt}i)$,
\begin{equation}
\begin{split}
\mathbf{4}\otimes\mathbf{4}
&\rightarrow
[(\mathbf{2},\mathbf{1})\oplus
(\mathbf{1},\mathbf{2})]
\otimes
[(\mathbf{2},\mathbf{1})\oplus
(\mathbf{1},\mathbf{2})] \\
& =
2(\mathbf{1},\mathbf{1})
\oplus
(\mathbf{3},\mathbf{1})
\oplus
(\mathbf{1},\mathbf{3})
\oplus
2(\mathbf{2},\mathbf{2})
\end{split}
\end{equation}
Since twisted $\mathbf{6}$ should contain
two scalars $(\mathbf{1},\mathbf{1})$
of $N=2$ twisted vector-
multiplet, the rest of bosonic fields
is uniquely determined,  
\begin{equation}
\begin{split}
&\mathbf{6} \rightarrow
3(\mathbf{1},\mathbf{1})
\oplus
(\mathbf{3},\mathbf{1})
\hspace{20pt}\mathrm{for\ (\mathit{i})}\\
&\mathbf{6}\rightarrow
2(\mathbf{1},\mathbf{1})
\oplus
(\mathbf{2},\mathbf{2})
\hspace{20pt}\mathrm{for\ (\mathit{i}
\hspace{-1pt}\mathit{i})}.
\end{split}
\label{scalar}
\end{equation}
The on-shell twisted multiplets for A-type (\textit{i}) and B-type (\textit{ii})
are summarized in Table \ref{multi}. 
 
\renewcommand{\arraystretch}{2.0}
\renewcommand{\tabcolsep}{5pt}
\begin{table}
\begin{tabular}{c|l}
helicity & \ untwisted \hspace{100pt} A-type (\textit{i})
\hspace{110pt} B-type (\textit{ii})
 \\ \hline
+1 & $\omega_{\mu}(\mathbf{2},\mathbf{2},\mathbf{1})$ \hspace{100pt}
$\omega_{\mu}(\mathbf{2},\mathbf{2})$ \hspace{110pt}
$\omega_{\mu}(\mathbf{2},\mathbf{2})$ \\
$+\frac{1}{2}$ & $\lambda^{\alpha}_{u}(\mathbf{2},\mathbf{1},\mathbf{4})$ \hspace{80pt}
$\psi(\mathbf{1},\mathbf{1})$, $\psi'(\mathbf{1},\mathbf{1}),$ \hspace{60pt}
$\rho^{+}(\mathbf{1},\mathbf{1})$, $\rho^{+}_{\mu\nu}(\mathbf{3},\mathbf{1})$,
\\[-12pt] 
& \hspace{125pt}
$\psi^{+}_{\mu\nu}(\mathbf{3},\mathbf{1})$, $\psi'^{+}_{\mu\nu}(\mathbf{3},\mathbf{1})$ \hspace{80pt} 
$\lambda^{+}_{\mu}(\mathbf{2},\mathbf{2})$ \\
0 & $\phi_{uv}(\mathbf{1},\mathbf{1},\mathbf{6})$ \hspace{50pt}
$X(\mathbf{1},\mathbf{1})$, $Y(\mathbf{1},\mathbf{1})$, $Z(\mathbf{1},\mathbf{1})$, \hspace{50pt} 
$A(\mathbf{1},\mathbf{1})$, $B(\mathbf{1},\mathbf{1})$, \\[-12pt]
& \hspace{150pt} $B^{+}_{\mu\nu}(\mathbf{3},\mathbf{1})$ \hspace{110pt}
$V_{\mu}(\mathbf{2},\mathbf{2})$ \\
$-\frac{1}{2}$ & $\bar{\lambda}_{\dot{\alpha}}^{u}(\mathbf{1},\mathbf{2},\mathbf{\bar{4}})$ 
\hspace{75pt}
$\psi_{\mu}(\mathbf{2},\mathbf{2})$, $\psi'_{\mu}(\mathbf{2},\mathbf{2})$ \hspace{60pt}
$\rho^{-}(\mathbf{1},\mathbf{1})$, $\rho^{-}_{\mu\nu}(\mathbf{1},\mathbf{3})$,
\\[-12pt] 
& \hspace{315pt}
$\lambda^{-}_{\mu}(\mathbf{2},\mathbf{2})$ \\
$ -1
$ & $\omega_{\mu}(\mathbf{2},\mathbf{2},\mathbf{1})$ \hspace{100pt}
$\omega_{\mu}(\mathbf{2},\mathbf{2})$ \hspace{110pt}
$\omega_{\mu}(\mathbf{2},\mathbf{2})$ \\ \hline
& \hspace{260pt} Dirac-K\"{a}hler twisted multiplet

\end{tabular}
\caption{Twisted multiplets for A-type (\textit{i}) and B-type (\textit{ii})}
\label{multi}
\end{table}
On the other hand, we already know that the $N=4$ Dirac-K\"{a}hler twisted 
supercharges $(Q,Q_{\mu},Q_{\mu\nu},\tilde{Q}_{\mu},\tilde{Q})$
can be represented 
through the decompositions (\ref{N=4decomp1})-(\ref{N=4decomp3})
as
\begin{eqnarray}
&&Q^{+}(\mathbf{1},\mathbf{1}),\ \ Q^{+}_{\rho\sigma}(\mathbf{3},\mathbf{1}),
\ \ Q^{+}_{\mu}(\mathbf{2},\mathbf{2}), \\
&&Q^{-}(\mathbf{1},\mathbf{1}),\ \ Q^{-}_{\rho\sigma}(\mathbf{1},\mathbf{3}),
\ \ Q^{-}_{\mu}(\mathbf{2},\mathbf{2}), 
\end{eqnarray}
which, with Eq.(\ref{fermion}), implies that the Dirac-K\"{a}hler twisting procedure 
can be identified as B-type (\textit{ii}).
Accordingly, from the second line of Eq.(\ref{scalar}), our Dirac-K\"{a}hler twisted gauge multiplet 
is expected to contain twisted scalar fields $V_{\mu}(\mathbf{2},\mathbf{2})$
transforming as a 4-vector under twisted Lorentz transformation $J'_{\mu\nu}$
but as a scalar under untwisted Lorentz transformation $J_{\mu\nu}$,
as well as two scalar bosons $A(\mathbf{1},\mathbf{1})$ and $B(\mathbf{1},\mathbf{1})$.

This correspondence between Dirac-K\"ahler twisting and B-type twisting
is also pointed out in \cite{KKM2} in the formulation of
$N=2\ D=4$ twisted hypermultiplet in continuum spacetime. 
Also in the lattice context,
this correspondence has been pointed out \cite{Catterall,Catterall:2009it} in the formulation
of lattice SYM where
the scalar part of $N=4$ Dirac-K\"ahler twisted supercharge 
is exactly formulated on 4-dimensional lattice.

\section{$N=D=4$ Dirac-K\"ahler Twisted SYM in continuum spacetime}

\indent

Based on the observation in the last section, 
Dirac-K\"ahler twisted 
$N=D=4$ SYM super connection ansatz in the continuum spacetime can be found as, 
\begin{eqnarray}
\{\D,\D_{\mu}\} &=& -i\D_{+\mu},\label{4DSYMcc1}\\[2pt]
\{\D_{\rho\sigma},\D_{\mu}\} &=& +i\delta_{\rho\sigma\mu\nu}\D_{-\nu},\label{4DSYMcc2}\\[2pt]
\{\D_{\rho\sigma},\tilde{\D}_{\mu}\} &=& -i\epsilon_{\rho\sigma\mu\nu}
\D_{+\nu},\label{4DSYMcc3} \\[2pt]
\{\tilde{\D},\tilde{\D}_{\mu}\} &=& -i\D_{-\mu},\label{4DSYMcc4}
\end{eqnarray}
\begin{eqnarray}
\{\D,\tilde{\D}\} &=& -iW,\hspace{30pt}
\{\D_{\mu\nu},\D_{\rho\sigma}\} \ =\ +i\epsilon_{\mu\nu\rho\sigma} W,\label{4DSYMcc5}\\
\{\D_{\mu},\tilde{\D}_{\nu}\} &=& -i\delta_{\mu\nu}F, \hspace{30pt}
\{others\} \ =\ 0, \label{4DSYMcc6}
\end{eqnarray}
where $(\D,\D_{\mu},\D_{\rho\sigma},\tilde{\D}_{\mu},\tilde{\D})$
denote super covariant derivatives defined 
in (\ref{4DDD1})-(\ref{4DDD5}),
whose respective transformation properties under $J_{\mu\nu}$ and $R_{\mu\nu}$
are same as those of the supercharges $(Q,Q_{\mu},Q_{\rho\sigma},\tilde{Q}_{\mu},\tilde{Q})$,
defined in (\ref{Lorentz4D1})-(\ref{Lorentz4D5}) and (\ref{R4D1})-(\ref{R4D5}).
$\D_{\pm\mu}$ denotes the combination of gauge covariant derivative
superfield $\D_{\underline{\mu}}$ and Dirac-K\"ahler twisted 
scalar superfield $\mathcal{V}_{\mu}$,
\begin{eqnarray}
\D_{\pm\mu} &\equiv& \D_{\underline{\mu}}\pm \mathcal{V}_{\mu}, \label{nablapm1}
\end{eqnarray} 
whose lowest components are given by
\begin{eqnarray}
\D_{\underline{\mu}}| &=& \Dc_{\mu} \ \equiv\ \partial_{\mu} -i \omega_{\mu} \label{nablapm2},\\[2pt]
\mathcal{V}_{\mu}| &=& V_{\mu}. \label{nablapm3}
\end{eqnarray}
Note here that the gauge covariant superfield $\D_{\underline{\mu}}$
transforms as a vector under the untwisted Lorentz transformation $J_{\mu\nu}$ and 
transforms as a scalar under the internal rotation $R_{\mu\nu}$,
while the twisted scalar superfield $\mathcal{V}_{\mu}$
transforms as a scalar under the untwisted Lorentz transformation $J_{\mu\nu}$ and 
transforms as a vector under the internal rotation $R_{\mu\nu}$.
More explicitly, for example, taking commutators of $J_{\rho\sigma}$ with both hand sides of (\ref{4DSYMcc1}) and (\ref{4DSYMcc4}), and using the transformation properties of $\D$, 
$\tilde{\D}$, $\D_{\mu}$ and $\tilde{\D}_{\mu}$ under $J_{\rho\sigma}$, one obtains
\begin{eqnarray}
[J_{\rho\sigma},\D_{+\mu}] &=& -\frac{i}{2}\delta_{\rho\sigma\mu\nu}\D_{+\nu}
-\frac{i}{2}\delta_{\rho\sigma\mu\nu}\D_{-\nu}, \\[5pt]
[J_{\rho\sigma},\D_{-\mu}] &=& -\frac{i}{2}\delta_{\rho\sigma\mu\nu}\D_{+\nu}
-\frac{i}{2}\delta_{\rho\sigma\mu\nu}\D_{-\nu},
\end{eqnarray}
from which one has
\begin{eqnarray}
[J_{\rho\sigma},\D_{\underline{\mu}}] &=& -i\delta_{\rho\sigma\mu\nu}\D_{\underline{\nu}}, \\[5pt]
[J_{\rho\sigma},\mathcal{V}_{\mu}] &=& 0.
\end{eqnarray} 
In a similar manner, as for $R_{\mu\nu}$, one obtains
\begin{eqnarray}
[R_{\rho\sigma},\D_{+\mu}] &=& -\frac{i}{2}\delta_{\rho\sigma\mu\nu}\D_{+\nu}
+\frac{i}{2}\delta_{\rho\sigma\mu\nu}\D_{-\nu}, \\[5pt]
[R_{\rho\sigma},\D_{-\mu}] &=& +\frac{i}{2}\delta_{\rho\sigma\mu\nu}\D_{+\nu}
-\frac{i}{2}\delta_{\rho\sigma\mu\nu}\D_{-\nu},
\end{eqnarray}
from which one has
\begin{eqnarray}
[R_{\rho\sigma},\D_{\underline{\mu}}] &=& 0, \\[5pt]
[R_{\rho\sigma},\mathcal{V}_{\mu}] &=& -i\delta_{\rho\sigma\mu\nu}\mathcal{V}_{\nu}.
\end{eqnarray}
Thus, one can see that, under twisted Lorentz generator $J'_{\mu\nu} = J_{\mu\nu}+R_{\mu\nu}$,
both of $\D_{\underline{\mu}}$ and $\mathcal{V}_{\mu}$ transform as vectors,
\begin{eqnarray}
[J'_{\rho\sigma},\D_{\underline{\mu}}] &=& -i\delta_{\rho\sigma\mu\nu}\D_{\underline{\nu}}, \\[5pt]
[J'_{\rho\sigma},\mathcal{V}_{\mu}] &=& -i\delta_{\rho\sigma\mu\nu}\mathcal{V}_{\nu},
\end{eqnarray}
while, in a similar manner, one can see that  $W$ and $F$ transform as scalars under twisted Lorentz generator  $J'_{\mu\nu}$. 
These rotational properties are consistent with the observation for $N=D=4$ Dirac-K\"ahler twisted (in other words, B-type twisted)
SYM multiplet given in the last section.

In the following we denote the combination of the above lowest components as $\Dc_{\pm\mu}$,
\begin{eqnarray}
\Dc_{\pm\mu} &=& \Dc_{\mu}\pm V_{\mu}. \label{Dc}
\end{eqnarray}
$V_{\mu}$ denotes 4-component Dirac-K\"ahler twisted scalar fields 
in the bosonic part of the multiplet.
The lowest components of $W$ and $F$ denote two scalars,
\begin{eqnarray}
W| &=& A,\\[2pt]
F| &=& B.
\end{eqnarray}

Note that each of the relations in the ansatz (\ref{4DSYMcc1})-(\ref{4DSYMcc6}) 
transforms in a super covariant manner under the following super gauge transformation,
\begin{eqnarray}
(\D,\D_{\mu},\D_{\mu\nu},\tilde{\D}_{\mu},\tilde{\D}) & \rightarrow &
\mathcal{G}^{-1}\ (\D,\D_{\mu},\D_{\mu\nu},\tilde{\D}_{\mu},\tilde{\D})\ \mathcal{G}, \\
(\D_{\pm\mu}, W, F) & \rightarrow &
\mathcal{G}^{-1}\ (\D_{\pm\mu}, W, F)\ \mathcal{G},
\end{eqnarray}
where $\mathcal{G}$ denotes finite super gauge transformation which can be expressed 
in terms of arbitrary super field $\mathcal{K}$ as
\begin{eqnarray}
\mathcal{G}=e^{i\mathcal{K}}.
\end{eqnarray}
Due to the super gauge covariant ansatz (\ref{4DSYMcc1})-(\ref{4DSYMcc6}) as shown above, 
each step of the following derivations are manifestly (super) gauge covariant, which is one of the main advantages of the present formulation.

Jacobi identities for three fermionic $\D_{A}$ 
with the above constraints (\ref{4DSYMcc1})-(\ref{4DSYMcc6}) give
\begin{eqnarray}
[\D,W]\ =\ [\D_{\rho\sigma},W]\ =\ [\tilde{\D},W] &=& 0,\label{4DSYM1stJocabi1}\\[2pt]
[\D_{\mu},F]\ = \ [\tilde{\D}_{\mu},F] &=& 0,\\[2pt]
[\D,\D_{+\mu}]\ =\ [\tilde{\D},\D_{-\mu}] &=& 0,
\end{eqnarray}
\begin{eqnarray}
[\D_{\mu},\D_{+\nu}]+[\D_{\nu},\D_{+\mu}] &=& 0,\\[2pt]
[\tilde{\D}_{\mu},\D_{-\nu}]+[\tilde{\D}_{\nu},\D_{-\mu}] &=& 0,\\[2pt]
\delta_{\rho\sigma\nu\tau}[\D_{\mu},\D_{-\tau}]
+\delta_{\rho\sigma\mu\tau}[\D_{\nu},\D_{-\tau}] &=& 0,\\[2pt]
\epsilon_{\rho\sigma\nu\tau}[\tilde{\D}_{\mu},\D_{+\tau}]
+\epsilon_{\rho\sigma\mu\tau}[\tilde{\D}_{\nu},\D_{+\tau}] &=& 0,\\[2pt]
\delta_{\rho\sigma\lambda\tau}[\D_{\mu\nu},\D_{-\tau}]
+\delta_{\mu\nu\lambda\tau}[\D_{\rho},\D_{-\tau}]
+\epsilon_{\mu\nu\rho\sigma}[\D_{\lambda},W] &=& 0,\\[2pt]
\epsilon_{\rho\sigma\lambda\tau}[\D_{\mu\nu},\D_{+\tau}]
+\epsilon_{\mu\nu\lambda\tau}[\D_{\rho},\D_{+\tau}]
-\epsilon_{\mu\nu\rho\sigma}[\tilde{\D}_{\lambda},W] &=& 0,
\end{eqnarray}
\begin{eqnarray}
[\D_{\mu},W]+[\tilde{\D},\D_{+\mu}] &=& 0,\\[2pt]
[\tilde{\D}_{\mu},W]+[\D,\D_{-\mu}] &=& 0,\\[2pt]
\delta_{\mu\nu}[\D,F]+[\tilde{\D}_{\nu},\D_{+\mu}] &=& 0,\\[2pt]
\delta_{\mu\nu}[\tilde{\D},F]+[\D_{\mu},\D_{-\nu}] &=& 0,\\[2pt]
\delta_{\rho\sigma\mu\nu}[\D,\D_{-\nu}]-[\D_{\rho\sigma},\D_{+\mu}] &=& 0,\\[2pt]
\epsilon_{\rho\sigma\mu\nu}[\tilde{\D},\D_{+\nu}]+[\D_{\rho\sigma},\D_{-\mu}] &=& 0,\\[2pt]
\epsilon_{\rho\sigma\nu\tau}[\D_{\mu},\D_{+\tau}]
-\delta_{\rho\sigma\mu\tau}[\tilde{\D}_{\nu},\D_{-\tau}]
+\delta_{\mu\nu}[\D_{\rho\sigma},F] &=& 0, \label{4DSYM1stJocabi2}
\end{eqnarray}
from which one can define fermionic superfields 
$(\rho,\lambda_{\mu},\rho_{\mu\nu},\tilde{\lambda}_{\mu},\tilde{\rho})$
as non-vanishing elements,
\begin{eqnarray}
[\tilde{\D},\D_{+\mu}] &=& -\tilde{\lambda}_{\mu},\hspace{40pt}
[\D_{\mu},W] \ =\ +\tilde{\lambda}_{\mu}, \\[2pt]
[\D_{\rho\sigma},\D_{-\mu}] &=& +\epsilon_{\rho\sigma\mu\nu}\tilde{\lambda}_{\nu},\\[6pt]
[\D,\D_{-\mu}] &=& +\lambda_{\mu},\hspace{40pt}
[\tilde{\D}_{\mu},W]\ =\ -\lambda_{\mu},\\[2pt]
[\D_{\rho\sigma},\D_{+\mu}] &=& +\delta_{\rho\sigma\mu\nu}\lambda_{\nu},\\[6pt]
[\tilde{\D},F] &=& -\rho,\hspace{63pt}
[\D,F]\ =\ -\tilde{\rho},\\[2pt]
[\D_{\mu},\D_{-\nu}] &=& +\delta_{\mu\nu}\rho,\hspace{30pt}
[\tilde{\D}_{\mu},\D_{+\nu}]\ =\ +\delta_{\mu\nu}\tilde{\rho}, \\[6pt]
[\D_{\mu},\D_{+\nu}] &=& +\rho_{\mu\nu}, \hspace{40pt}
[\D_{\rho\sigma},F] \ =\ -\frac{1}{2}\epsilon_{\rho\sigma\alpha\beta}\rho_{\alpha\beta},\\[-2pt]
[\tilde{\D}_{\mu},\D_{-\nu}] &=& -\frac{1}{2}\epsilon_{\mu\nu\rho\sigma}\rho_{\rho\sigma}.
\end{eqnarray}  
The lowest components of 
$(\rho,\lambda_{\mu},\rho_{\mu\nu},\tilde{\lambda}_{\mu},\tilde{\rho})$
form $N=D=4$ Dirac-K\"ahler twisted fermions in the continuum spacetime.
In the following, for notational simplicity, we denote the same symbols to represent 
the corresponding lowest component of fermionic superfields. 

Jacobi identities for four fermionic $\D_{A}$ give
after some calculations,
\begin{eqnarray}
\{\D,\rho\} &=& -\frac{i}{2}([\D_{+\lambda},\D_{-\lambda}]-[W,F]),\\[2pt]
\{\D_{\rho},\rho\} &=& 0,\\[2pt]
\{\D_{\rho\sigma},\rho\} &=& -i[\D_{-\rho},\D_{-\sigma}],\\[2pt]
\{\tilde{\D}_{\rho},\rho\} &=& +i[\D_{-\rho},F],\\[2pt]
\{\tilde{\D},\rho\} &=& 0,
\end{eqnarray}
\begin{eqnarray}
\{\D,\lambda_{\mu}\} &=& 0,\\[2pt]
\{\D_{\rho},\lambda_{\mu}\} &=& -i[\D_{+\rho},\D_{-\mu}]
+\frac{i}{2}\delta_{\rho\mu}([\D_{+\lambda},\D_{-\lambda}]-[W,F]),\\[2pt]
\{\D_{\rho\sigma},\lambda_{\mu}\} &=& +i\epsilon_{\rho\sigma\mu\nu}[\D_{+\nu},W],\\[2pt]
\{\tilde{\D}_{\rho},\lambda_{\mu}\} &=& -\frac{i}{2}\epsilon_{\rho\mu\alpha\beta}
[\D_{+\alpha},\D_{+\beta}],\\[2pt]
\{\tilde{\D},\lambda_{\mu}\} &=& +i[\D_{-\mu},W],
\end{eqnarray}
\begin{eqnarray}
\{\D,\rho_{\mu\nu}\} &=& -i[\D_{+\mu},\D_{+\nu}],\\[2pt]
\{\D_{\rho},\rho_{\mu\nu}\} &=& -i\epsilon_{\rho\sigma\mu\nu}[\D_{-\sigma},F],\\[2pt]
\{\D_{\rho\sigma},\rho_{\mu\nu}\} &=& 
-i\delta_{\rho\sigma\mu\lambda}[\D_{+\nu},\D_{-\lambda}]
+i\delta_{\rho\sigma\nu\lambda}[\D_{+\mu},\D_{-\lambda}] \nonumber \\
&&+\frac{i}{2}\delta_{\rho\sigma\mu\nu}([\D_{+\lambda},\D_{-\lambda}]-[W,F]),\\[2pt]
\{\tilde{\D}_{\rho},\rho_{\mu\nu}\} &=& +i\delta_{\rho\sigma\mu\nu}[\D_{+\sigma},F],\\[2pt]
\{\tilde{\D},\rho_{\mu\nu}\} &=& +\frac{i}{2}\epsilon_{\mu\nu\alpha\beta}
[\D_{-\alpha},\D_{-\beta}]
\end{eqnarray}
\begin{eqnarray}
\{\D,\tilde{\lambda}_{\mu}\} &=& -i[\D_{+\mu},W],\\[2pt]
\{\D_{\rho},\tilde{\lambda}_{\mu}\} &=& +\frac{i}{2}\epsilon_{\rho\mu\alpha\beta}
[\D_{-\alpha},\D_{-\beta}],\\[2pt]
\{\D_{\rho\sigma},\tilde{\lambda}_{\mu}\} &=& +i\delta_{\rho\sigma\mu\nu}[\D_{-\nu},W],\\[2pt]
\{\tilde{\D}_{\rho},\tilde{\lambda}_{\mu}\} &=& -i[\D_{+\mu},\D_{-\rho}]
+\frac{i}{2}\delta_{\rho\mu}([\D_{+\lambda},\D_{-\lambda}]+[W,F]),\\[2pt]
\{\tilde{\D},\tilde{\lambda}_{\mu}\} &=& 0,
\end{eqnarray}
\begin{eqnarray}
\{\D,\tilde{\rho}\} &=& 0,\\[4pt]
\{\D_{\rho},\tilde{\rho}\} &=& +i[\D_{+\rho},F],\\[2pt]
\{\D_{\rho\sigma},\tilde{\rho}\} &=& +\frac{i}{2}\epsilon_{\rho\sigma\alpha\beta}
[\D_{+\alpha},\D_{+\beta}],\\[2pt]
\{\tilde{\D}_{\rho},\tilde{\rho}\} &=& 0,\\[2pt]
\{\tilde{\D},\tilde{\rho}\} &=& +\frac{i}{2}([\D_{+\lambda},\D_{-\lambda}]+[W,F]).
\end{eqnarray}

Twisted $N=D=4$ SUSY transformation laws for the component fields 
can be read off from,
\begin{eqnarray}
s_{A} \varphi &\equiv& [\D_{A},\varphi\}|_{\theta's=0}
\end{eqnarray}
where $\D_{A}$ represents any of fermionic supercovariant derivatives 
$(\D,\D_{\mu},\D_{\rho\sigma},\tilde{\D}_{\mu},\tilde{\D})$ 
and $s_{A}$ denotes the corresponding supercharge which operates
on the component fields.
$\varphi$ represents any of the component fields 
$(\Dc_{\pm\mu},A,B,\rho,\lambda_{\mu},\rho_{\rho\sigma},\tilde{\lambda}_{\mu},\tilde{\rho})$.
The results are summarized in Table \ref{N=D=4trans1} and \ref{N=D=4trans2}. 

\begin{table}
\renewcommand{\arraystretch}{1.6}
\renewcommand{\tabcolsep}{4pt}
\begin{center}
\begin{tabular}{|c||c|c|c|}
\hline
& $s$ & $s_{\rho\sigma}$ & $\tilde{s}$ \\ \hline
$\Dc_{+\mu}$ & $0$ & $+\delta_{\rho\sigma\mu\nu}\lambda_{\nu}$
& $-\tilde{\lambda}_{\mu}$ \\
$\Dc_{-\mu}$ & $+\lambda_{\mu}$  
& $+\epsilon_{\rho\sigma\mu\nu}\tilde{\lambda}_{\nu}$ 
& $0$ \\ 
$A$ & $0$ & $0$ & $0$ \\
$B$ & $-\tilde{\rho}$ & $-\frac{1}{2}\epsilon_{\rho\sigma\alpha\beta}\rho_{\alpha\beta}$
& $-\rho$ \\ \hline
$\rho$ & $-\frac{i}{2}([\Dc_{+\lambda},\Dc_{-\lambda}]-[A,B])$ 
& $-i[\Dc_{-\rho},\Dc_{-\sigma}]$ & $0$ \\
$\lambda_{\mu}$ & $0$ & $+i\epsilon_{\rho\sigma\mu\nu}[\Dc_{+\nu},A]$
& $+i[\Dc_{-\mu},A]$ \\
$\rho_{\mu\nu}$ & $-i[\Dc_{+\mu},\Dc_{+\nu}]$ &
$-i\delta_{\rho\sigma\mu\lambda}[\Dc_{+\nu},\Dc_{-\lambda}] 
+i\delta_{\rho\sigma\nu\lambda}[\Dc_{+\mu},\Dc_{-\lambda}]$
& $+\frac{i}{2}\epsilon_{\mu\nu\alpha\beta}[\Dc_{-\alpha},\Dc_{-\beta}]$ \\[-4pt]
&  & $+\frac{i}{2}\delta_{\rho\sigma\mu\nu}([\Dc_{+\lambda},\Dc_{-\lambda}]-[A,B])$ & \\
$\tilde{\lambda}_{\mu}$ & $-i[\Dc_{+\mu},A]$ 
& $+i\delta_{\rho\sigma\mu\nu}[\Dc_{-\nu},A]$
& $0$ \\
$\tilde{\rho}$ & $0$ 
& $+\frac{i}{2}\epsilon_{\rho\sigma\alpha\beta}[\Dc_{+\alpha},\Dc_{+\beta}]$
& $+\frac{i}{2}([\Dc_{+\lambda},\Dc_{-\lambda}]+[A,B])$ \\ \hline
\end{tabular}
\caption{Twisted $N=D=4$ SUSY transformation laws (1)}
\label{N=D=4trans1}
\vspace{30pt}

\begin{tabular}{|c||c|c|}
\hline
& $s_{\rho}$ & $\tilde{s}_{\rho}$  \\ \hline
$\Dc_{+\mu}$ &  $+\rho_{\rho\mu}$ 
& $+\delta_{\rho\mu}\tilde{\rho}$  \\
$\Dc_{-\mu}$ & $+\delta_{\rho\mu}\rho$  
& $-\frac{1}{2}\epsilon_{\rho\mu\alpha\beta}\rho_{\alpha\beta}$\\ 
$A$ & $+\tilde{\lambda}_{\rho}$ & $-\lambda_{\rho}$  \\
$B$ & $0$ & $0$ \\ \hline
$\rho$ & $0$
& $+i[\Dc_{-\rho},B]$  \\
$\lambda_{\mu}$ & $-i[\Dc_{+\rho},\Dc_{-\mu}]$ 
& $-\frac{i}{2}\epsilon_{\rho\mu\alpha\beta}[\Dc_{+\alpha},\Dc_{+\beta}]$ \\[-4pt]
& $+\frac{i}{2}\delta_{\rho\mu}([\Dc_{+\lambda},\Dc_{-\lambda}]-[A,B])$ & \\
$\rho_{\mu\nu}$ & $-i\epsilon_{\rho\sigma\mu\nu}[\Dc_{-\sigma},B]$
& $+i\delta_{\rho\sigma\mu\nu}[\Dc_{+\sigma},B]$ \\
$\tilde{\lambda}_{\mu}$ & $+\frac{i}{2}\epsilon_{\rho\mu\alpha\beta}[\Dc_{-\alpha},\Dc_{-\beta}]$
& $-i[\Dc_{+\mu},\Dc_{-\rho}]$\\[-4pt]
& & $+\frac{i}{2}\delta_{\rho\mu}([\Dc_{+\lambda},\Dc_{-\lambda}]+[A,B])$ \\
$\tilde{\rho}$ & $+i[\Dc_{+\rho},B]$ & $0$ \\ \hline
\end{tabular}
\caption{Twisted $N=D=4$ SUSY transformation laws (2)}
\label{N=D=4trans2}
\end{center}
\end{table}

An important notice here is that one cannot introduce any auxiliary field
to fulfill off-shell closure consistently with the above procedure,
which is actually the notorious aspect of $N=D=4$ supermultiplet.
One can show that, for bosonic contents of the multiplet 
$(\Dc_{\pm\mu},A,B)$, the resulting algebra closes
without any use of additional conditions,
\begin{eqnarray}
\{s,s_{\mu}\}\varphi_{B} &=& -i[\Dc_{+\mu},\varphi_{B}],\label{4DSYMalg1}\\[2pt]
\{s_{\rho\sigma},s_{\mu}\}\varphi_{B} 
&=& +i\delta_{\rho\sigma\mu\nu}[\Dc_{-\nu},\varphi_{B}],\label{4DSYMalg2}\\[2pt]
\{s_{\rho\sigma},\tilde{s}_{\mu}\}\varphi_{B} &=& -i\epsilon_{\rho\sigma\mu\nu}
[\Dc_{+\nu},\varphi_{B}],\label{4DSYMalg3} \\[2pt]
\{\tilde{s},\tilde{s}_{\mu}\}\varphi_{B} &=& -i[\Dc_{-\mu},\varphi_{B}],\label{4DSYMalg4}
\end{eqnarray}
\begin{eqnarray}
\{s,\tilde{s}\}\varphi_{B} &=& -i[A,\varphi_{B}],\hspace{43pt}
\{s_{\mu\nu},s_{\rho\sigma}\}\varphi_{B} \ =\ +i\epsilon_{\mu\nu\rho\sigma}[A,\varphi_{B}]
,\label{4DSYMalg5}\\
\{s_{\mu},\tilde{s}_{\nu}\}\varphi_{B} &=& -i\delta_{\mu\nu}[B,\varphi_{B}]
, \hspace{30pt}
\{others\}\varphi_{B} \ =\ 0, \label{4DSYMalg6}
\end{eqnarray}
where $\varphi_{B}$ represents any of the bosonic component of the multiplet,
$(\Dc_{\pm\mu},A,B)$.
Note here that closedness of the above resulting algebra is up to gauge transformation which is a natural consequence
of the super covariant formulation.

On the other hand, for fermionic components of the multiplet
$(\rho,\lambda_{\mu},\rho_{\rho\sigma},\tilde{\lambda}_{\mu},\tilde{\rho})$,
the resulting algebra closes on-shell
up to gauge transformation,
\begin{eqnarray}
\{s,s_{\mu}\}\varphi_{F} &\dot{=}& -i[\Dc_{+\mu},\varphi_{F}],\label{4DSYMalgf1}\\[2pt]
\{s_{\rho\sigma},s_{\mu}\}\varphi_{F} 
&\dot{=}& +i\delta_{\rho\sigma\mu\nu}[\Dc_{-\nu},\varphi_{F}],\label{4DSYMalgf2}\\[2pt]
\{s_{\rho\sigma},\tilde{s}_{\mu}\}\varphi_{F} &\dot{=}& -i\epsilon_{\rho\sigma\mu\nu}
[\Dc_{+\nu},\varphi_{F}],\label{4DSYMalgf3} \\[2pt]
\{\tilde{s},\tilde{s}_{\mu}\}\varphi_{F} &\dot{=}& -i[\Dc_{-\mu},\varphi_{F}],\label{4DSYMalgf4}
\end{eqnarray}
\begin{eqnarray}
\{s,\tilde{s}\}\varphi_{F} &\dot{=}& -i[A,\varphi_{F}],\hspace{43pt}
\{s_{\mu\nu},s_{\rho\sigma}\}\varphi_{F} \ \dot{=}\ +i\epsilon_{\mu\nu\rho\sigma}[A,\varphi_{F}]
,\label{4DSYMalgf5}\\
\{s_{\mu},\tilde{s}_{\nu}\}\varphi_{F} &\dot{=}& -i\delta_{\mu\nu}[B,\varphi_{F}]
, \hspace{30pt}
\{others\}\varphi_{F} \ \dot{=}\ 0, \label{4DSYMalgf6}
\end{eqnarray}
where $\varphi_{F}$ denotes any of fermionic component of the multiplet
$(\rho,\lambda_{\mu},\rho_{\rho\sigma},\tilde{\lambda}_{\mu},\tilde{\rho})$. 
The symbol $\dot{=}$ represents that the equality holds only up to the following 
equations  (\ref{eqm1})-(\ref{eqm5})
for fermionic components of the multiplet,
which will turn out to be equations of motion for the resulting action  (\ref{N=D=4SYMaction_cont}). 
\begin{eqnarray}
[\Dc_{+\mu},\lambda_{\mu}] + [A,\tilde{\rho}] &=& 0,\label{eqm1}\\[6pt]
[\Dc_{-\mu},\tilde{\lambda}_{\mu}] - [A,\rho] &=& 0,\label{eqm2}\\[6pt]
[\Dc_{+\mu},\rho] + [\Dc_{-\nu},\rho_{\mu\nu}] - [B,\tilde{\lambda}_{\mu}] &=& 0,
\label{eqm3}\\[2pt]
[\Dc_{-\mu},\tilde{\rho}]-\frac{1}{2}\epsilon_{\mu\nu\rho\sigma}
[\Dc_{+\nu},\rho_{\rho\sigma}] + [B,\lambda_{\mu}] &=& 0,\label{eqm4}\\[0pt]
\delta_{\mu\nu\rho\sigma}[\Dc_{-\rho},\lambda_{\sigma}]
+\frac{1}{2}\epsilon_{\mu\nu\rho\sigma}[A,\rho_{\rho\sigma}]
+\epsilon_{\mu\nu\rho\sigma}[\Dc_{+\rho},\tilde{\lambda}_{\sigma}] &=&0. \label{eqm5}
\end{eqnarray}

The above inevitable on-shell structure of $N=D=4$ gauge multiplet has been already well-known
and it was first pointed out in the context of superconnection method
for ordinary extended $N=4$ supersymmetry \cite{Sohnius}.
Now in the above calculations, we have explicitly seen the similar on-shell structure
for the Dirac-K\"ahler twisted $N=D=4$ SYM multiplet.
The resulting on-shell structure can be traced back to the absence of auxiliary
field in the multiplet which obeys from the Jacobi identities with the starting constraints
(\ref{4DSYMcc1})-(\ref{4DSYMcc6}).
There has been several argument for the possible off-shell structure of $N=D=4$ 
\cite{SSW}
and we comment here that the introduction of central charges may play
an important role in the context of superconnection formulation
to accommodate off-shell structure of $N=D=4$ twisted SYM multiplet \cite{Saito:2005pc}.  
It should also be noted that in \cite{Catterall:2013roa}, twisted SUSY transformations of the Dirac-K\"ahler twisted $N=4$ SYM multiplet in the continuum spacetime have been worked out in terms of 5-dimensional notation by paying a particular attention to R symmetries. 
However, here in the present formulation, the whole algebraic structure of Dirac-K\"ahler twisted SUSY of $N=D=4$ SYM
has clearly and entirely been revealed due to our manifestly covariant formulation.

$N=D=4$ Dirac-K\"ahler twisted SYM action can be found in such a way
that the equations (\ref{eqm1})-(\ref{eqm5}) come out as the equations of motion
of the action.
First, we can determine the fermion part of the action as
\begin{eqnarray}
S_{F}&=& \int d^{4}x\ \mathrm{tr}
\biggl[
-i\lambda_{\mu}[\Dc_{+\mu},\rho]
-i\tilde{\rho}[A,\rho]\nonumber \\[2pt]
&& -i\lambda_{\mu}[\Dc_{-\nu},\rho_{\mu\nu}]
+i\tilde{\rho}[\Dc_{-\mu},\tilde{\lambda}_{\mu}]
+i\lambda_{\mu}[B,\tilde{\lambda}_{\mu}]\nonumber \\[2pt]
&&-\frac{i}{2}\epsilon_{\mu\nu\rho\sigma}\tilde{\lambda}_{\mu}
[\Dc_{+\nu},\rho_{\rho\sigma}]
+\frac{i}{8}\epsilon_{\mu\nu\rho\sigma}\rho_{\mu\nu}[A,\rho_{\rho\sigma}]
\biggr]
\end{eqnarray}
from which the relations (\ref{eqm1})-(\ref{eqm5}) can be derived as the equations of motion.
Note that throughout this paper, the coupling constant $g$ is implicitly absorbed by rescaling the component fields, which doesn't affect the result of this paper.
The bosonic part of the action $S_{B}$ can be found in such a way that the sum of 
the fermionic and bosonic terms 
 vanishes under the operation of $s$, 
\begin{eqnarray}
S_{B} &=& \int d^{4}x\ \mathrm{tr}\ 
\biggl[
-\frac{1}{2}[\Dc_{+\mu},\Dc_{+\nu}][\Dc_{-\mu},\Dc_{-\nu}]
+\frac{1}{4}[\Dc_{+\mu},\Dc_{-\mu}][\Dc_{+\nu},\Dc_{-\nu}]\nonumber \\[2pt]
&&+\frac{1}{2}[\Dc_{+\mu},A][\Dc_{-\mu},B]
+\frac{1}{2}[\Dc_{-\mu},A][\Dc_{+\mu},B] +\frac{1}{4}[A,B][A,B]
\biggr].
\end{eqnarray}
Thus, 
the total action is given as the sum of $S_{F}$ and $S_{B}$ as,
\begin{eqnarray}
S^{N=D=4}_{TSYM} &=& S_{B} + S_{F} \nonumber \\[2pt]
&=& \int d^{4}x \ \mathrm{tr}\ \biggl[
-\frac{1}{2}[\Dc_{+\mu},\Dc_{+\nu}][\Dc_{-\mu},\Dc_{-\nu}]
+\frac{1}{4}[\Dc_{+\mu},\Dc_{-\mu}][\Dc_{+\nu},\Dc_{-\nu}]\nonumber \\[2pt]
&&+\frac{1}{2}[\Dc_{+\mu},A][\Dc_{-\mu},B]
+\frac{1}{2}[\Dc_{-\mu},A][\Dc_{+\mu},B] +\frac{1}{4}[A,B][A,B]\nonumber \\[6pt]
&&-i\lambda_{\mu}[\Dc_{+\mu},\rho]
-i\tilde{\rho}[A,\rho]\nonumber 
 -i\lambda_{\mu}[\Dc_{-\nu},\rho_{\mu\nu}]
+i\tilde{\rho}[\Dc_{-\mu},\tilde{\lambda}_{\mu}]
+i\lambda_{\mu}[B,\tilde{\lambda}_{\mu}] \nonumber \\[4pt]
&&-\frac{i}{2}\epsilon_{\mu\nu\rho\sigma}\tilde{\lambda}_{\mu}
[\Dc_{+\nu},\rho_{\rho\sigma}]
+\frac{i}{8}\epsilon_{\mu\nu\rho\sigma}\rho_{\mu\nu}[A,\rho_{\rho\sigma}]
\biggr]. \label{N=D=4SYMaction_cont}
\end{eqnarray}

One can show the invariance of the total action (\ref{N=D=4SYMaction_cont})
under each of the operations of $s$,  $s_{\mu}$, $s_{\mu\nu}$, 
$\tilde{s}_{\mu}$, and $\tilde{s}$,
with use of trace properties and Jacobi identities.
In other words, one can show that SUSY variations of 
the total action (\ref{N=D=4SYMaction_cont})
vanish w.r.t. all the supercharges 
$s_{A}=(s,  s_{\mu}, s_{\mu\nu}, \tilde{s}_{\mu}, \tilde{s}$),
\begin{eqnarray}
\delta_{A} S^{N=D=4}_{TSYM} (\varphi) &\equiv&
S^{N=D=4}_{TSYM} (\varphi+\delta_{A}\varphi) - S (\varphi) \\[5pt]
&=&0,
\end{eqnarray}
where $\varphi$ represents any of the component fields 
$(\Dc_{\pm\mu},A,B,\rho,\lambda_{\mu},\rho_{\rho\sigma},\tilde{\lambda}_{\mu},\tilde{\rho})$,
and we define the variation of the component fields as
\begin{eqnarray}
\delta_{A}\varphi &\equiv&
\xi_{A}
(s_{A}\varphi) \\
&=&
\xi_{A}
[\D_{A},\varphi\}.
\end{eqnarray}
In the above definition of the variation of the component fields, we introduced 
a set of Grassmann parameters $\xi_{A} = (\xi, \xi_{\mu}, \xi_{\mu\nu},\tilde{\xi}_{\mu},\tilde{\xi})$,
each of which anti-commutes with any of the fermionic covariant derivatives $\nabla_{B}=(\D,\D_{\mu},\D_{\mu\nu},\tilde{\D}_{\mu},\tilde{\D})$,
\begin{eqnarray}
\{ \xi_{A}, \nabla_{B} \}
 &=& 0.
\end{eqnarray}
Since all of the component fields are given as 
combinations of the fermionic covariant derivatives, 
the above assumption provides the following (anti)commutativity 
between $\xi_{A}$ and any of the component fields 
$\varphi = (\Dc_{\pm\mu},A,B,\rho,\lambda_{\mu},\rho_{\rho\sigma},\tilde{\lambda}_{\mu},\tilde{\rho})$,
\begin{eqnarray}
[ \xi_{A}, \varphi \}
 &=& 0.
\end{eqnarray}

It should be noted that the invariance of the action under the operations of twisted supercharges
can be shown without use of equations of motion, although the closure of the algebra holds up to equations of motion.

Remembering that $\Dc_{\pm\mu}$ denotes
usual covariant derivative plus/minus Dirac-K\"ahler twisted scalar $V_{\mu}$ (\ref{Dc}),
one sees that the action (\ref{N=D=4SYMaction_cont}) can be 
expressed in terms of $V_{\mu}$,
\begin{eqnarray}
S^{N=D=4}_{TSYM} 
&=& \int d^{4}x \ \mathrm{tr}\ \biggl[
\frac{1}{2}
F_{\mu\nu}F_{\mu\nu}
+[\Dc_{\mu},A][\Dc_{\mu},B]
+[\Dc_{\mu},V_{\nu}][\Dc_{\mu},V_{\nu}]\nonumber \\[2pt]
&& -\frac{1}{2}[V_{\mu},V_{\nu}][V_{\mu},V_{\nu}] - [V_{\mu},A][V_{\mu},B]
+\frac{1}{4}[A,B][A,B]\nonumber \\[8pt]
&&-i\lambda_{\mu}[\Dc_{\mu},\rho]
-i\lambda_{\mu}[V_{\mu},\rho]
-i\tilde{\rho}[A,\rho]\nonumber \\[8pt]
&&
 -i\lambda_{\mu}[\Dc_{\nu},\rho_{\mu\nu}]
+i\lambda_{\mu}[V_{\nu},\rho_{\mu\nu}]\nonumber \\[8pt]
&&+i\tilde{\rho}[\Dc_{\mu},\tilde{\lambda}_{\mu}]
-i\tilde{\rho}[V_{\mu},\tilde{\lambda}_{\mu}]
+i\lambda_{\mu}[B,\tilde{\lambda}_{\mu}] \nonumber \\[4pt]
&&-\frac{i}{2}\epsilon_{\mu\nu\rho\sigma}\tilde{\lambda}_{\mu}
[\Dc_{\nu},\rho_{\rho\sigma}]
-\frac{i}{2}\epsilon_{\mu\nu\rho\sigma}\tilde{\lambda}_{\mu}
[V_{\nu},\rho_{\rho\sigma}]
+\frac{i}{8}\epsilon_{\mu\nu\rho\sigma}\rho_{\mu\nu}[A,\rho_{\rho\sigma}]
\biggr], \label{N=D=4SYMaction_cont2}
\end{eqnarray}
where $F_{\mu\nu}\equiv i[\Dc_{\mu},\Dc_{\nu}]$ represents ordinary field strength.

In the context of topological quantum field theory \cite{Marcus,Lozano},
where only the scalar supercharge $s$ has a special importance,
it is usually introduced a few auxiliary fields 
in order to make $s^{2}$ part off-shell and make the action be
exact form w.r.t. the scalar supercharge $s$.
On the other hand, as mentioned above, here is no auxiliary field 
in the action as a direct consequence of $N=D=4$ twisted superconnection method.
Then, at least at present stage, 
we observe that
there is no way to 
express the action as an exact form w.r.t. any of the 
supercharges.

\section{Lattice Formulation of Dirac-K\"ahler Twisted $N=D=4$ SYM}

\subsection{Leibniz rule for Dirac-K\"ahler Twisted $N=D=4$ Algebra}

Let us begin the lattice observation
with the following $N=D=4$ Twisted SUSY algebra in the continuum spacetime,
(\ref{4DQ1})-(\ref{4DQ4}),
\begin{eqnarray}
\{Q,Q_{\mu}\}&=&+i\partial_{\mu}, \nonumber \\
\{Q_{\rho\sigma},Q_{\mu}\}&=&-i\delta_{\rho\sigma\mu\nu}\partial_{\nu}, \nonumber \\
\{Q_{\rho\sigma},\tilde{Q}_{\mu}\}&=&+i\epsilon_{\rho\sigma\mu\nu}\partial_{\nu},
\nonumber \\
\{\tilde{Q},\tilde{Q}_{\mu}\}&=&+i\partial_{\mu}, \nonumber
\end{eqnarray}
where $\delta_{\rho\sigma\mu\nu}\equiv 
\delta_{\rho\mu}\delta_{\sigma\nu}-\delta_{\rho\nu}\delta_{\sigma\mu}$.\\
The lattice counterpart of the above algebra is expected as
\begin{eqnarray}
\{Q,Q_{\mu}\}&=&+i\Delta_{\pm\mu}\\
\{Q_{\rho\sigma},Q_{\mu}\}&=&-i\delta_{\rho\sigma\mu\nu}\Delta_{\pm\nu}\\
\{Q_{\rho\sigma},\tilde{Q}_{\mu}\}&=&+i\epsilon_{\rho\sigma\mu\nu}\Delta_{\pm\nu}\\
\{\tilde{Q},\tilde{Q}_{\mu}\}&=&+i\Delta_{\pm\mu}
\end{eqnarray}
where the $\pm$ signs are to be fixed later on.

As it was discussed in details in \cite{DKKN1,DKKN2} that one should 
maintain the Leibniz rule to realize "exact" SUSY on a lattice. 
Let us remind some generic argument of the formulation.
Since we have only finite lattice spacings on a lattice, infinitesimal 
translations should be replaced by finite difference operators,
\begin{eqnarray}
P_{\mu}\ =\ i\partial_{\mu} \rightarrow i\Delta_{\pm\mu}
\end{eqnarray}
where $\Delta_{\pm\mu}$ denote forward and backward difference
operators, respectively.
The operation of $\Delta_{\pm\mu}$ on a function $\Phi(x)$
can be defined by the following type of ``shifted" commutators,
\begin{eqnarray}
(\Delta_{\pm\mu} \Phi(x)) &\equiv& \Delta_{\pm\mu} \Phi(x)
- \Phi(x\pm n_{\mu}) \Delta_{\pm\mu}, \label{D_Phi}
\end{eqnarray}
where $n_\mu$ is a unit vector, taking a value 1 for $\mu$-direction. 
The difference operators satisfy the following ``lattice" Leibniz rule,
\begin{eqnarray}
(\Delta_{\pm\mu} \Phi_1(x)\Phi_2(x)) &=& (\Delta_{\pm\mu}\Phi_1(x))\Phi_2(x)
+ \Phi_1(x\pm n_{\mu}) (\Delta_{\pm\mu}\Phi_2(x)), \label{D_Phi2}
\end{eqnarray}
where the $\Delta_{\pm\mu}$, locating on links from $x$ to $x\pm n_{\mu}$,
respectively, take unit values for generic $x$ as a matrix entry,
\begin{eqnarray}
\Delta_{\pm\mu}\ =\ (\Delta_{\pm\mu})_{x\pm n_{\mu},x} \ =\ \mp 1.
\end{eqnarray}
Since in the lattice formulation of SUSY we should embed 
the above properties of bosonic operators into the SUSY algebra, it is natural 
to assume that a lattice SUSY transformation can also be defined
as a ``shifted" (anti-)commutator of $Q_{A}$
located on a link from $x$ to $x+a_{A}$,
\begin{eqnarray}
(Q_A\Phi(x)) &\equiv& (Q_{A})_{x + a_{A},x}\Phi(x) 
- (-1)^{|\Phi|} \Phi(x + a_{A})(Q_{A})_{x+a_{A},x}, \label{ops}
\end{eqnarray}
where $|\Phi|$ represents $0$ or $1$ for bosonic or fermionic field $\Phi$,
respectively.
The operation of $Q_{A}$'s on a product of fields
accordingly gives,
\begin{eqnarray}
(Q_{A}\Phi_{1}(x)\Phi_{2}(x))
&=&
(Q_{A}\Phi_{1}(x))\Phi_{2}(x) 
+ (-1)^{|\Phi_{1}|} \Phi_{1}(x+a_{A})(Q_{A}\Phi_{2}(x)). \label{Q_Phi_Phi}
\end{eqnarray}
Since the supercharges $Q_{A}$'s are located on links,
it is then natural to define  
an anti-commutator of lattice supercharges 
as an successive connection of link operators,
\begin{eqnarray}
\{Q_{A},Q_{B}\}_{x+a_{A}+a_{B},x} =
(Q_{A})_{x+a_{A}+a_{B},x+a_{B}} (Q_{B})_{x+a_{B},x} 
+ (Q_{B})_{x+a_{A}+a_{B},x+a_{A}} (Q_{A})_{x+a_{A},x}. \label{link_com}
\nonumber \\
\end{eqnarray} 

By means of the above ingredients, 
lattice SUSY algebra could be expressed as
\begin{eqnarray}
\{Q_{A},Q_{B}\}_{x+a_{A}+a_{B},x} &=& 
(\Delta_{\pm\mu})_{x\pm n_{\mu},x}, \label{expected_alg} 
\end{eqnarray}
provided the following lattice Leibniz rule conditions hold
\begin{eqnarray}
a_{A}+ a_{B} &=& +n_{\mu} \hspace{20pt} for\ \ \ \Delta_{+\mu},  \label{llcond1}\\
a_{A}+ a_{B} &=& -n_{\mu} \hspace{20pt} for\ \ \ \Delta_{-\mu}, \label{llcond2}
\end{eqnarray}
which are the necessary conditions for the realization of lattice SUSY algebra
and eventually govern the structure of supersymmetric lattices.

By denoting the shift parameters associated with supercharges
$(Q,Q_{\mu},Q_{\rho\sigma},\tilde{Q}_{\mu},\tilde{Q})$ as 
$(a,a_{\mu},a_{\rho\sigma},\tilde{a}_{\mu},\tilde{a})$,
the $4D$ lattice Leibniz rule conditions can be 
written as
\begin{eqnarray}
a+a_{\mu} &=& \pm n_{\mu},\\[2pt]
a_{\rho\sigma} + a_{\mu} &=& \pm|\delta_{\rho\sigma\mu\nu}| n_{\nu},
\hspace{20pt} for\ \  \rho = \mu \ \ or\ \ \sigma = \mu\\[2pt]
a_{\rho\sigma} +\tilde{a}_{\mu} &=& \pm |\epsilon_{\rho\sigma\mu\nu}|n_{\nu}, \hspace{20pt}
for\ \  
\rho, \sigma, \mu \ {\it all\ different\ each\ other,}
\\[2pt]
\tilde{a} + \tilde{a}_{\mu} &=& \pm n_{\mu}.
\end{eqnarray} 
An important aspect of $N=D=4$ Dirac-K\"ahler twisted SUSY algebra
is that, as in $N=D=2$ and $N=4\ D=3$, there exist consistent solutions 
to satisfy the Leibniz rule requirement of 4-dimensional lattice, namely, 
\begin{eqnarray}
a &=& (arbitrary) \label{4DLeibnizsol1}\\ [5pt]
a_{\mu}&=&\eta^{(\mu)}n_{\mu}-a, \hspace{103pt} (no\ sum) \label{4DLeibnizsol2}\\[5pt]
a_{\mu\nu}&=&-\eta^{(\mu)}n_{\mu}-\eta^{(\nu)}n_{\nu}+a, \hspace{52pt} (no\ sum) 
\label{4DLeibnizsol3}\\[5pt]
\ta_{\mu}&=&\sum_{\lambda\neq\mu}\eta^{(\lambda)}n_{\lambda}-a, 
\label{4DLeibnizsol4} \\[-5pt]
\ta&=&-\sum_{\lambda=1}^{4}\eta^{(\lambda)}n_{\lambda}+a.\label{4DLeibnizsol5} 
\end{eqnarray}
which satisfy
\begin{eqnarray}
a+a_{\mu} &=& +\eta^{(\mu)}n_{\mu},\hspace{20pt}(\mu\ :\ no\ sum)
\label{4Dlatticecond1}\\[4pt]
a_{\rho\sigma} + a_{\mu} &=& -|\delta_{\rho\sigma\mu\nu}|\eta^{(\nu)}n_{\nu},
\hspace{20pt} for\ \  \rho = \mu \ \ or\ \ \sigma = \mu 
\label{4Dlatticecond2}\\[4pt]
a_{\rho\sigma} +\tilde{a}_{\mu} &=& + |\epsilon_{\rho\sigma\mu\nu}|\eta^{(\nu)}n_{\nu}, 
\hspace{20pt} for\ \  
\rho, \sigma, \mu \ {\it all\ different\ each\ other,}
\label{4Dlatticecond3}\\[4pt]
\tilde{a} + \tilde{a}_{\mu} &=& -\eta^{(\mu)} n_{\mu}, \hspace{20pt} (\mu\ :\ no\ sum)
\label{4Dlatticecond4}
\end{eqnarray}
where $\eta^{(\mu)}\ (\mu = 1\sim 4)$ denote sign parameter which takes the value of $\pm 1$.
Choice of $\eta^{(\mu)}$ defines an 
orientation of the coordinates.
If one takes $(\eta^{(1)},\eta^{(2)},\eta^{(3)},\eta^{(4)})=(+,+,+,+)$,
the corresponding $N=D=4$ Dirac-K\"ahler twisted lattice algebra can be expressed as
\begin{eqnarray}
\{Q,Q_{\mu}\}&=&+i\Delta_{+\mu}\\[2pt]
\{Q_{\rho\sigma},Q_{\mu}\}&=&-i\delta_{\rho\sigma\mu\nu}\Delta_{-\nu}\\[2pt]
\{Q_{\rho\sigma},\tilde{Q}_{\mu}\}&=&+i\epsilon_{\rho\sigma\mu\nu}\Delta_{+\nu}\\[2pt]
\{\tilde{Q},\tilde{Q}_{\mu}\}&=&+i\Delta_{-\mu}.
\end{eqnarray}
As in the case of $N=D=2$ and $N=4\ D=3$, there exists one parameter arbitrariness
for the solution of $a_{A}$.
One of the typical choice for $a_{A}$, symmetric choice, is obtained by setting 
$a=(+\frac{1}{2},+\frac{1}{2},+\frac{1}{2},+\frac{1}{2})$ in the case of sign parameters
$(\eta^{(1)},\eta^{(2)},\eta^{(3)},\eta^{(4)})=(+,+,+,+)$,

\begin{center}
\underline{Symmetric choice}
\end{center}
\begin{eqnarray}
a &=& (+\frac{1}{2},+\frac{1}{2},+\frac{1}{2},+\frac{1}{2}), \hspace{20pt}
a_{1}\ =\ (+\frac{1}{2},-\frac{1}{2},-\frac{1}{2},-\frac{1}{2}), \\[2pt]
a_{12} &=& (-\frac{1}{2},-\frac{1}{2},+\frac{1}{2},+\frac{1}{2}), \hspace{20pt}
a_{2}\ =\ (-\frac{1}{2},+\frac{1}{2},-\frac{1}{2},-\frac{1}{2}), \\[2pt] 
a_{13} &=& (-\frac{1}{2},+\frac{1}{2},-\frac{1}{2},+\frac{1}{2}), \hspace{20pt}
a_{3}\ =\ (-\frac{1}{2},-\frac{1}{2},+\frac{1}{2},-\frac{1}{2}), \\[2pt] 
a_{14} &=& (-\frac{1}{2},+\frac{1}{2},+\frac{1}{2},-\frac{1}{2}), \hspace{20pt}
a_{4}\ =\ (-\frac{1}{2},-\frac{1}{2},-\frac{1}{2},+\frac{1}{2}), \\[2pt] 
a_{23} &=& (+\frac{1}{2},-\frac{1}{2},-\frac{1}{2},+\frac{1}{2}), \hspace{20pt}
\tilde{a}_{4}\ =\ (+\frac{1}{2},+\frac{1}{2},+\frac{1}{2},-\frac{1}{2}), \\[2pt] 
a_{24} &=& (+\frac{1}{2},-\frac{1}{2},+\frac{1}{2},-\frac{1}{2}), \hspace{20pt}
\tilde{a}_{3}\ =\ (+\frac{1}{2},+\frac{1}{2},-\frac{1}{2},+\frac{1}{2}), \\[2pt] 
a_{34} &=& (+\frac{1}{2},+\frac{1}{2},-\frac{1}{2},-\frac{1}{2}), \hspace{20pt}
\tilde{a}_{2}\ =\ (+\frac{1}{2},-\frac{1}{2},+\frac{1}{2},+\frac{1}{2}), \\[2pt]
\tilde{a} &=& (-\frac{1}{2},-\frac{1}{2},-\frac{1}{2},-\frac{1}{2}), \hspace{20pt}
\tilde{a}_{1}\ =\ (-\frac{1}{2},+\frac{1}{2},+\frac{1}{2},+\frac{1}{2}), 
\end{eqnarray}
while the other typical choice, asymmetric choice, is given by setting
$a=(0,0,0,0)$ for the sign parameters 
$(\eta^{(1)},\eta^{(2)},\eta^{(3)},\eta^{(4)})=(+,+,+,+)$,
\begin{center}
\underline{Asymmetric choice}
\end{center}
\begin{eqnarray}
a &=& (0,0,0,0), \hspace{57pt}
a_{1}\ =\ (+1,0,0,0), \\[2pt]
a_{12} &=& (-1,-1,0,0), \hspace{38pt}
a_{2}\ =\ (0,+1,0,0), \\[2pt] 
a_{13} &=& (-1,0,-1,0), \hspace{38pt}
a_{3}\ =\ (0,0,+1,0), \\[2pt] 
a_{14} &=& (-1,0,0,-1), \hspace{38pt}
a_{4}\ =\ (0,0,0,+1), \\[2pt] 
a_{23} &=& (0,-1,-1,0), \hspace{38pt}
\tilde{a}_{4}\ =\ (+1,+1,+1,0), \\[2pt] 
a_{24} &=& (0,-1,0,-1), \hspace{38pt}
\tilde{a}_{3}\ =\ (+1,+1,0,+1), \\[2pt] 
a_{34} &=& (0,0,-1,-1), \hspace{38pt}
\tilde{a}_{2}\ =\ (+1,0,+1,+1), \\[2pt]
\tilde{a} &=& (-1,-1,-1,-1), \hspace{20pt}
\tilde{a}_{1}\ =\ (0,+1,+1,+1).
\end{eqnarray}

It is important to mention whether there is any possibility that twisted SUSY less than 16 supercharges could be
consistent with the 4-dimensional lattice Leibniz rule.
For example,
one may wonder if $N=2\ D=4$ twisted SUSY algebra with 8 supercharges
should also be consistent with  Leibniz rule requirement
on 4-dimensional lattice.
However, as noted in Ref.\cite{Nagata},
it is easy to show that
$N=2\ D=4$ twisted SUSY algebra cannot satisfy 
such a requirement on 4-dimensional lattice.
As far as we know, 8 supercharge algebra can satisfy
the requirement on at most 3-dimensional lattice,
which is $N=4\ D=3$ twisted SYM formulation given in Ref. 
\cite{DKKN3}. 
See also \cite{Nagata-Wu} for a $N=4\ D=3$ twisted SUSY formulation of lattice Chern-Simons.

\subsection{$N=D=4$ Dirac-K\"ahler Twisted SYM Constraints on a Lattice }
\label{lattice_action_section}

Although it is shown in the last subsection 
that $N=D=4$ Dirac-K\"ahler twisted SUSY algebra
itself can satisfy the Leibniz rule requirement of 4-dimensional lattice,
it is still non-trivial to ask how to formulate corresponding
lattice SYM.
In the following, we construct lattice SYM multiplet 
along the similar manner as in $N=D=2$ \cite{DKKN2} and $N=4\ D=3$ 
\cite{DKKN3,Nagata}
and then proceed to present lattice counterpart of $N=D=4$ Dirac-K\"ahler twisted
SYM action. 

First, as in the case of $N=D=2$ and $N=4\ D=3$, 
we introduce non-unitary link variables $\U_{\pm\mu}$ in 4-dimensions
as lattice counterparts of gauge covariant derivatives $\Dc_{\mu}$ plus/minus 
twisted scalar $V_{\mu}$,
\begin{eqnarray} 
\Dc_{\pm\mu}\ =\
\Dc_{\mu}\pm V_{\mu}\ \  \rightarrow \ \ 
\mp\ (\U_{\pm\mu})_{x\pm n_{\mu},x} 
&=& \mp\ (e^{\pm i(\omega_{\mu}\pm iV_{\mu})})_{x\pm n_{\mu},x},
\end{eqnarray}
which connect from $x$ to $x\pm n_{\mu}$ in a gauge covariant way
for generic value of $x$,
\begin{eqnarray}
(\U_{\pm\mu})_{x\pm n_{\mu},x} \rightarrow
G^{-1}(x\pm n_{\mu})(\U_{\pm\mu})_{x\pm n_{\mu},x}G(x).
\end{eqnarray}
Note that $\U_{\pm\mu}$ do not satisfy unitarity condition because
of the contribution of Dirac-K\"ahler twisted scalar $V_{\mu}$,
\begin{eqnarray}
\U_{+\mu}\U_{-\mu} \neq 1. \label{4Dnon_unitary}
\end{eqnarray}
We also introduce the fermionic gauge link variables as lattice counterparts
of fermionic covariant derivatives, 
$\D_{A}=(\D,\D_{\mu},\D_{\mu\nu},\tilde{\D}_{\mu},\tilde{\D})$, 
\begin{eqnarray}
\D_{A} \rightarrow (\D_{A})_{x+a_{A},x} 
\end{eqnarray}
which connect 
from $x$ to $x+a_{A}$ in a gauge covariant way\begin{eqnarray}
(\D_{A})_{x+a_{A},x} \rightarrow
G^{-1}(x+a_{A})(\D_{A})_{x+a_{A},x}G(x).
\end{eqnarray}
Then we impose the following $N=D=4$ Dirac-K\"ahler twisted SYM 
super connection ansatz on a lattice,
\begin{eqnarray}
\{\D,\D_{\mu}\}_{x+a+a_{\mu},x} &=& +i(\U_{+\mu})_{x+n_{\mu},x},\label{4DSYMlatc1}\\[2pt]
\{\D_{\rho\sigma},\D_{\mu}\}_{x+a_{\rho\sigma}+a_{\mu},x} 
&=& +i\delta_{\rho\sigma\mu\nu}(\U_{-\nu})_{x-n_{\nu},x},\label{4DSYMlatc2}\\[2pt]
\{\D_{\rho\sigma},\tilde{\D}_{\mu}\}_{x+a_{\rho\sigma}+\tilde{a}_{\mu},x}
 &=& +i\epsilon_{\rho\sigma\mu\nu}
(\U_{+\nu})_{x+n_{\nu},x},\label{4DSYMlatc3} \\[2pt]
\{\tilde{\D},\tilde{\D}_{\mu}\}_{x+\tilde{a}+\tilde{a}_{\mu},x} 
&=& -i(\U_{-\mu})_{x-n_{\mu},x},\label{4DSYMlatc4}
\end{eqnarray}
\begin{eqnarray}
\{\D,\tilde{\D}\}_{x+a+\tilde{a},x} &=& -i(W)_{x+a+\tilde{a},x},\label{4DSYMlatc5}\\[2pt]
\{\D_{\mu\nu},\D_{\rho\sigma}\}_{x+a_{\mu\nu}+a_{\rho\sigma},x}
 &=& +i\epsilon_{\mu\nu\rho\sigma} (W)_{x+a_{\mu\nu}+a_{\rho\sigma},x},\label{4DSYMlatc6}\\[2pt]
\{\D_{\mu},\tilde{\D}_{\nu}\}_{x+a_{\mu}+\tilde{a}_{\nu},x}
 &=& -i\delta_{\mu\nu}(F)_{x+a_{\mu}+\tilde{a}_{\nu},x}, \label{4DSYMlatc7}\\[2pt]
\{others\} &=& 0, \label{4DSYMlatc8}
\end{eqnarray}
where $W$ and $F$ represent lattice counterparts of scalar fields in the continuum spacetime,
$A$ and $B$.
All the anti-commutators in l.h.s. should be understood as link anti-commutators
as in $N=D=2$ 
\cite{DKKN2}
and $N=4\ D=3$ 
\cite{DKKN3,Nagata}.

Although the expressions of the constraints (\ref{4DSYMlatc1})-(\ref{4DSYMlatc8})
might seem just to replace
the continuum covariant derivatives by
the corresponding lattice gauge covariant link variables
in the continuum expressions 
(\ref{4DSYMcc1})-(\ref{4DSYMcc6}),
we should pay careful attentions especially for gauge covariance aspect
of the constraints on the lattice.  

First, in order for the constraints (\ref{4DSYMlatc1})-(\ref{4DSYMlatc4}) to maintain
the gauge covariance on the lattice, the ending sites should coincide each other for 
each constraint. This requirement gives
\begin{eqnarray}
a+a_{\mu} &=& +n_{\mu},\label{4DSYMlatgauge1}\\[2pt]
a_{\rho\sigma} + a_{\mu} &=& -|\delta_{\rho\sigma\mu\nu}|n_{\nu},
\hspace{30pt} for \ \ \rho = \mu \ \ or \ \ \sigma = \mu\\[2pt]
a_{\rho\sigma} +\tilde{a}_{\mu} &=& + |\epsilon_{\rho\sigma\mu\nu}|n_{\nu}, 
\hspace{30pt} for \ \ 
\rho, \sigma, \mu\ {\it all\ different\ each\ other}
\\[2pt]
\tilde{a} + \tilde{a}_{\mu} &=& -n_{\mu}, 
\end{eqnarray}
which are nothing but the Leibniz rule relations (\ref{4Dlatticecond1})-(\ref{4Dlatticecond4})
for the sign parameter choice  
$(\eta^{(1)},\eta^{(2)},\eta^{(3)},\eta^{(4)})=(+,+,+,+)$.
Therefore the gauge covariance of (\ref{4DSYMlatc1})-(\ref{4DSYMlatc4})
is maintained as far as the shift parameters $a_{A}$ are given by 
\begin{eqnarray}
a &=& (arbitrary) \label{4DSYMLeibnizsol1}\\ [5pt]
a_{\mu}&=& +n_{\mu}-a,  \label{4DSYMLeibnizsol2}\\[5pt]
a_{\mu\nu}&=&-n_{\mu}-n_{\nu}+a,  \label{4DSYMLeibnizsol3}\\[5pt]
\ta_{\mu}&=& +\sum_{\lambda\neq\mu}n_{\lambda}-a, \label{4DSYMLeibnizsol4} \\[-5pt]
\ta&=&-\sum_{\lambda=1}^{4}n_{\lambda}+a.\label{4DSYMLeibnizsol5} 
\end{eqnarray}

As for the constraints (\ref{4DSYMlatc5})-(\ref{4DSYMlatc6})
associated with $W$,
one can see from the above expressions for the shift parameter 
(\ref{4DSYMLeibnizsol1})-(\ref{4DSYMLeibnizsol5}) that 
\begin{eqnarray}
a+\tilde{a} &=& -\sum_{\lambda=1}^{4}n_{\lambda}+2a,\\[2pt]
a_{\mu\nu}+a_{\rho\sigma} &=& -\sum_{\lambda=1}^{4}n_{\lambda}+2a,\hspace{10pt}
for\ \ 
\mu, \nu, \rho, \sigma \ {\it all\ different\ each\ other,}
\end{eqnarray}
namely, $W$ is identically located on a link from $x$ to $x-\sum n_{\lambda}+2a$
or equivalently from $x$ to $x+a+\tilde{a}$.
Also for the constraints (\ref{4DSYMlatc7}) associated with $F$,
one can see from the above expressions of $a_{A}$ that
\begin{eqnarray}
a_{\mu} + \tilde{a}_{\nu} &=& +\sum_{\lambda=1}^{4}n_{\lambda} - 2a
\hspace{30pt} for\ \ \mu = \nu
\end{eqnarray} 
namely $F$ is identically located on a link from $x$ to $x+\sum n_{\lambda}-2a$ or
equivalently from $x$ to $x-a-\tilde{a}$.
Notice that $W$ and $F$ are located on 
oppositely oriented links each other
regardless of any particular choice of $a$, while they reduce to a site,
namely from $x$ to $x$,
if one takes generic symmetric choice which satisfies
\begin{eqnarray}
a+\tilde{a} &=& 0.
\end{eqnarray}

Once we impose the lattice constraints of $N=D=4$ Dirac-K\"ahler twisted SYM
(\ref{4DSYMlatc1})-(\ref{4DSYMlatc8}),
all the information of corresponding lattice multiplet
can be given through the resulting Jacobi identities.
Jacobi identities for three fermionic link variables
with the constraints
(\ref{4DSYMlatc1})-(\ref{4DSYMlatc8})
are given by (\ref{4DSYM1stJocabi1})-(\ref{4DSYM1stJocabi2})
but with the replacement,
\begin{eqnarray}
\D_{\pm\mu} &\rightarrow& \mp\ \U_{\pm\mu}
\end{eqnarray}
and fermionic covariant derivatives $\D_{A}$ by the 
corresponding fermionic link variables
as well as usual (anti-)commutators by the corresponding link (anti-)commutators.

Just along the similar manner as in the continuum observations,
non-vanishing fermionic components turn out to be
\begin{eqnarray}
[\tilde{\D},\U_{+\mu}] &=& +\tilde{\lambda}_{\mu},\hspace{40pt}
[\D_{\mu},W] \ =\ +\tilde{\lambda}_{\mu}, \\[2pt]
[\D_{\rho\sigma},\U_{-\mu}] &=& +\epsilon_{\rho\sigma\mu\nu}\tilde{\lambda}_{\nu},\\[6pt]
[\D,\U_{-\mu}] &=& +\lambda_{\mu},\hspace{40pt}
[\tilde{\D}_{\mu},W]\ =\ -\lambda_{\mu},\\[2pt]
[\D_{\rho\sigma},\U_{+\mu}] &=& -\delta_{\rho\sigma\mu\nu}\lambda_{\nu},\\[6pt]
[\tilde{\D},F] &=& -\rho,\hspace{63pt}
[\D,F]\ =\ -\tilde{\rho},\\[2pt]
[\D_{\mu},\U_{-\nu}] &=& +\delta_{\mu\nu}\rho,\hspace{30pt}
[\tilde{\D}_{\mu},\U_{+\nu}]\ =\ -\delta_{\mu\nu}\tilde{\rho}, \\[6pt]
[\D_{\mu},\U_{+\nu}] &=& -\rho_{\mu\nu}, \hspace{40pt}
[\D_{\rho\sigma},F] \ =\ -\frac{1}{2}\epsilon_{\rho\sigma\alpha\beta}\rho_{\alpha\beta},\\[-2pt]
[\tilde{\D}_{\mu},\U_{-\nu}] &=& -\frac{1}{2}\epsilon_{\mu\nu\rho\sigma}\rho_{\rho\sigma},
\end{eqnarray}  
where we just omit link indices for simplicity and all the commutators should
be understood as link commutators.
One can show that
link natures for the above non-vanishing fermionic components 
$(\rho,\lambda_{\mu},\rho_{\mu\nu},\tilde{\lambda}_{\mu},\tilde{\rho})$
are just opposite to those of 
$(\D,\D_{\mu},\D_{\mu\nu},\tilde{\D}_{\mu},\tilde{\D})$,
which are summarized in Table \ref{N=4shift_fermions}.

\begin{table}
\begin{center}
\renewcommand{\arraystretch}{1.4}
\renewcommand{\tabcolsep}{6pt}
\begin{tabular}{c|ccc|ccccc|ccccc}
& $\U_{\pm\mu}$ & $W$ & $F$ &$\rho$ & $\lambda_{\mu}$ & $\rho_{\mu\nu}$ 
& $\tilde{\lambda}_{\mu}$ & $\tilde{\rho}$ 
& $\D$ & $\D_{\mu}$ & $\D_{\mu\nu}$ & $\tilde{\D}_{\mu}$ & $\tilde{\D}$ \\ \hline
shift & $\pm n_{\mu}$ & $a+\tilde{a}$ & $-a-\tilde{a}$ &
$-a$ & $-a_{\mu}$ & $-a_{\mu\nu}$ & $-\tilde{a}_{\mu}$ & $-\tilde{a}$
& $+a$ & $+a_{\mu}$ & $+a_{\mu\nu}$ & $+\tilde{a}_{\mu}$ & $+\tilde{a}$
\end{tabular}
\caption{Shift nature for  Dirac-K\"ahler twisted $N=D=4$ lattice SYM multiplet}
\label{N=4shift_fermions}
\end{center}
\end{table}

Jacobi identities for four fermionic link variables 
can be carried out in a similar way as in the continuum case 
but with link (anti-)commutators.
Dirac-K\"ahler twisted $N=D=4$ SUSY transformation laws on the lattice
can also be read off as in $N=D=2$ or $N=4\ D=3$ from
\begin{eqnarray}
(s_{A}\varphi)_{x+a_{\varphi}+a_A,x} = 
s_{A}(\varphi)_{x+a_{\varphi},x}\equiv
[\D_{A},\varphi\}_{x+a_{\varphi}+a_A,x},
\end{eqnarray}
where $(\varphi)_{x+a_{\varphi},x}$ denotes any components of the lattice multiplet.
The results are summarized in Table \ref{N=D=4translat1} and \ref{N=D=4translat2}.
\begin{table}
\renewcommand{\arraystretch}{1.6}
\renewcommand{\tabcolsep}{4pt}
\begin{center}
\begin{tabular}{|c||c|c|c|}
\hline
& $s$ & $s_{\rho\sigma}$ & $\tilde{s}$ \\ \hline
$\U_{+\mu}$ & $0$ & $-\delta_{\rho\sigma\mu\nu}\lambda_{\nu}$
& $+\tilde{\lambda}_{\mu}$ \\
$\U_{-\mu}$ & $+\lambda_{\mu}$  
& $+\epsilon_{\rho\sigma\mu\nu}\tilde{\lambda}_{\nu}$ 
& $0$ \\ 
$W$ & $0$ & $0$ & $0$ \\
$F$ & $-\tilde{\rho}$ & $-\frac{1}{2}\epsilon_{\rho\sigma\alpha\beta}\rho_{\alpha\beta}$
& $-\rho$ \\ \hline
$\rho$ & $+\frac{i}{2}([\U_{+\lambda},\U_{-\lambda}]+[W,F])$ 
& $-i[\U_{-\rho},\U_{-\sigma}]$ & $0$ \\
$\lambda_{\mu}$ & $0$ & $-i\epsilon_{\rho\sigma\mu\nu}[\U_{+\nu},W]$
& $+i[\U_{-\mu},W]$ \\
$\rho_{\mu\nu}$ & $-i[\U_{+\mu},\U_{+\nu}]$ &
$+i\delta_{\rho\sigma\mu\lambda}[\U_{+\nu},\U_{-\lambda}] 
-i\delta_{\rho\sigma\nu\lambda}[\U_{+\mu},\U_{-\lambda}]$
& $+\frac{i}{2}\epsilon_{\mu\nu\alpha\beta}[\U_{-\alpha},\U_{-\beta}]$ \\[-4pt]
&  & $-\frac{i}{2}\delta_{\rho\sigma\mu\nu}([\U_{+\lambda},\U_{-\lambda}]+[W,F])$ & \\
$\tilde{\lambda}_{\mu}$ & $+i[\U_{+\mu},W]$ 
& $+i\delta_{\rho\sigma\mu\nu}[\U_{-\nu},W]$
& $0$ \\
$\tilde{\rho}$ & $0$ 
& $+\frac{i}{2}\epsilon_{\rho\sigma\alpha\beta}[\U_{+\alpha},\U_{+\beta}]$
& $-\frac{i}{2}([\U_{+\lambda},\U_{-\lambda}]-[W,F])$ \\ \hline
\end{tabular}
\caption{Twisted $N=D=4$ SUSY transformation laws on the lattice (1)}
\label{N=D=4translat1}
\vspace{30pt}
\begin{tabular}{|c||c|c|}
\hline
& $s_{\rho}$ & $\tilde{s}_{\rho}$  \\ \hline
$\U_{+\mu}$ &  $-\rho_{\rho\mu}$ 
& $-\delta_{\rho\mu}\tilde{\rho}$  \\
$\U_{-\mu}$ & $+\delta_{\rho\mu}\rho$  
& $-\frac{1}{2}\epsilon_{\rho\mu\alpha\beta}\rho_{\alpha\beta}$\\ 
$W$ & $+\tilde{\lambda}_{\rho}$ & $-\lambda_{\rho}$  \\
$F$ & $0$ & $0$ \\ \hline
$\rho$ & $0$
& $+i[\U_{-\rho},F]$  \\
$\lambda_{\mu}$ & $+i[\U_{+\rho},\U_{-\mu}]$ 
& $-\frac{i}{2}\epsilon_{\rho\mu\alpha\beta}[\U_{+\alpha},\U_{+\beta}]$ \\[-4pt]
& $-\frac{i}{2}\delta_{\rho\mu}([\U_{+\lambda},\U_{-\lambda}]+[W,F])$ & \\
$\rho_{\mu\nu}$ & $-i\epsilon_{\rho\sigma\mu\nu}[\U_{-\sigma},F]$
& $-i\delta_{\rho\sigma\mu\nu}[\U_{+\sigma},F]$ \\
$\tilde{\lambda}_{\mu}$ & $+\frac{i}{2}\epsilon_{\rho\mu\alpha\beta}[\U_{-\alpha},\U_{-\beta}]$
& $+i[\U_{+\mu},\U_{-\rho}]$\\[-4pt]
& & $-\frac{i}{2}\delta_{\rho\mu}([\U_{+\lambda},\U_{-\lambda}]-[W,F])$ \\
$\tilde{\rho}$ & $-i[\U_{+\rho},F]$ & $0$ \\ \hline
\end{tabular}
\caption{Twisted $N=D=4$ SUSY transformation laws on the lattice (2)}
\label{N=D=4translat2}
\end{center}
\end{table}

Since there does not appear any auxiliary fields in the lattice SYM multiplet 
just as in the continuum multiplet,
the resulting $N=D=4$ Dirac-K\"ahler twisted algebra for component fields
closes only on-shell. 
\begin{eqnarray}
\{s,s_{\mu}\}(\varphi)_{x+a_{\varphi},x} 
&\dot{=}& +i[\U_{+\mu},\varphi]_{x+n_{\mu}+a_{\varphi},x},
\label{4DSYMlatalgf1}\\[2pt]
\{s_{\rho\sigma},s_{\mu}\}(\varphi)_{x+a_{\varphi},x} 
&\dot{=}& +i\delta_{\rho\sigma\mu\nu}[\U_{-\nu},\varphi]_{x-n_{\nu}+a_{\varphi},x},
\label{4DSYMlatalgf2}\\[2pt]
\{s_{\rho\sigma},\tilde{s}_{\mu}\}(\varphi)_{x+a_{\varphi},x} 
&\dot{=}& +i\epsilon_{\rho\sigma\mu\nu}
[\U_{+\nu},\varphi]_{x+n_{\nu}+a_{\varphi},x},\label{4DSYMlatalgf3} \\[2pt]
\{\tilde{s},\tilde{s}_{\mu}\}(\varphi)_{x+a_{\varphi},x} 
&\dot{=}& -i[\U_{-\mu},\varphi]_{x-n_{\mu}+a_{\varphi},x},
\label{4DSYMlatalgf4}
\end{eqnarray}
\begin{eqnarray}
\{s,\tilde{s}\}(\varphi)_{x+a_{\varphi},x} &\dot{=}& -i[W,\varphi]_{x+a+\tilde{a}+a_{\varphi},x},
\label{4DSYMlatalgf5}\\[2pt]
\{s_{\mu\nu},s_{\rho\sigma}\}(\varphi)_{x+a_{\varphi},x} 
&\dot{=}& +i\epsilon_{\mu\nu\rho\sigma}[W,\varphi]_{x+a+\tilde{a}+a_{\varphi},x},
\label{4DSYMlatalgf6}\\
\{s_{\mu},\tilde{s}_{\nu}\}(\varphi)_{x+a_{\varphi},x}
&\dot{=}& -i\delta_{\mu\nu}[F,\varphi]_{x-a-\tilde{a}+a_{\varphi},x},
\label{4DSYMlatalgf7}\\[2pt]
\{others\}(\varphi)_{x+a_{\varphi},x} &\dot{=}& 0, \label{4DSYMlatalgf8}
\end{eqnarray}
where $(\varphi)_{x+a_{\varphi},x}$ denotes any of the component 
of the lattice SYM multiplet
while the symbol $\dot{=}$ represents that the equality holds only up to
the following lattice counterparts 
of equations of motion. If $\varphi$ is fermionic component
$(\rho,\lambda_{\mu},\rho_{\rho\sigma},\tilde{\lambda}_{\mu},\tilde{\rho})$,
\begin{eqnarray}
[\U_{+\mu},\lambda_{\mu}] - [W,\tilde{\rho}] &=& 0,\label{eqmlat1}\\[6pt]
[\U_{-\mu},\tilde{\lambda}_{\mu}] - [W,\rho] &=& 0,\label{eqmlat2}\\[6pt]
[\U_{+\mu},\rho] - [\U_{-\nu},\rho_{\mu\nu}] + [F,\tilde{\lambda}_{\mu}] &=& 0,
\label{eqmlat3}\\[2pt]
[\U_{-\mu},\tilde{\rho}]+\frac{1}{2}\epsilon_{\mu\nu\rho\sigma}
[\U_{+\nu},\rho_{\rho\sigma}] + [F,\lambda_{\mu}] &=& 0,\label{eqmlat4}\\[0pt]
\delta_{\mu\nu\rho\sigma}[\U_{-\rho},\lambda_{-\sigma}]
+\frac{1}{2}\epsilon_{\mu\nu\rho\sigma}[W,\rho_{\rho\sigma}]
-\epsilon_{\mu\nu\rho\sigma}[\U_{+\rho},\tilde{\lambda}_{\sigma}] &=&0, \label{eqmlat5}
\end{eqnarray}
where all the link indices are omitted for simplicity.

Throughout the above calculations, one can see that the gauge covariance is 
rigidly kept. For example, the lattice counterpart of equation of motion (\ref{eqmlat1})
actually represents,
\begin{eqnarray}
[\U_{+\mu},\lambda_{\mu}]_{x+n_{\mu}-a_{\mu},x} 
- [W,\tilde{\rho}]_{x+a+\tilde{a}-\tilde{a},x} &=& 0,
\end{eqnarray}
by remembering that the shift of $\lambda_{\mu}$, $W$ and $\tilde{\rho}$,
is given by $-a_{\mu}$, $a+\tilde{a}$ and $-\tilde{a}$, respectively.
The ending sites for both terms coincide each other,
which ensures the gauge covariance on the lattice, 
if one reminds $a$ and $a_{\mu}$ are subject to the condition (\ref{4DSYMlatgauge1}).

Now we proceed to discuss about constructing 
lattice action of $N=D=4$ Dirac-K\"ahler twisted SYM.
In the above formulation, the lattice multiplet 
$(\U_{\pm\mu},W,F,\rho,\lambda_{\mu},\rho_{\mu\nu},\tilde{\lambda}_{\mu},\tilde{\rho})$
is obtained only on-shell, which means
there is no manifest way to construct the corresponding lattice SYM action.  
However, the expression of continuum $N=D=4$ Dirac-K\"ahler twisted SYM action 
(\ref{N=D=4SYMaction_cont}) suggests that the lattice counterpart
can be expressed as
\begin{eqnarray}
S^{N=D=4}_{lat.\ TSYM} 
&=& \sum_{x} \ \mathrm{tr}\ \biggl[
\frac{1}{2}[\U_{+\mu},\U_{+\nu}]_{x,x-n_{\mu}-n_{\nu}}[\U_{-\mu},\U_{-\nu}]_{x-n_{\mu}-n_{\nu},x} 
\nonumber\\[0pt]
&&-\frac{1}{4}[\U_{+\mu},\U_{-\mu}]_{x,x}[\U_{+\nu},\U_{-\nu}]_{x,x}
-\frac{1}{4}[W,F]_{x,x}[W,F]_{x,x} \nonumber \\[2pt]
&&+\frac{1}{2}[\U_{+\mu},W]_{x,x-n_{\mu}-a-\tilde{a}}[\U_{-\mu},F]_{x-n_{\mu}-a-\tilde{a},x} 
\nonumber \\[2pt]
&&+\frac{1}{2}[\U_{-\mu},W]_{x,x+n_{\mu}-a-\tilde{a}}[\U_{+\mu},F]_{x+n_{\mu}-a-\tilde{a},x}
\nonumber \\[6pt]
&&-i(\lambda_{\mu})_{x,x+a_{\mu}}[\U_{+\mu},\rho]_{x+a_{\mu},x} 
+i\tilde{\rho}_{x,x+\tilde{a}}[W,\rho]_{x+\tilde{a},x}
-i(\lambda_{\mu})_{x,x+a_{\mu}}[F,\tilde{\lambda}_{\mu}]_{x+a_{\mu},x} 
\nonumber \\[8pt]
&&+i(\lambda_{\mu})_{x,x+a_{\mu}}[\U_{-\nu},\rho_{\mu\nu}]_{x+a_{\mu},x}
-i(\tilde{\rho})_{x,x+\tilde{a}}[\U_{-\mu},\tilde{\lambda}_{\mu}]_{x+\tilde{a},x}
\nonumber \\[2pt]
&&-\frac{i}{2}\epsilon_{\mu\nu\rho\sigma}(\tilde{\lambda}_{\mu})_{x,x+\tilde{a}_{\mu}}
[\U_{+\nu},\rho_{\rho\sigma}]_{x+\tilde{a}_{\mu},x}
-\frac{i}{8}\epsilon_{\mu\nu\rho\sigma}(\rho_{\mu\nu})_{x,x+a_{\mu\nu}}
[W,\rho_{\rho\sigma}]_{x+a_{\mu\nu},x}
\biggr],\nonumber \\ \label{N=D=4SYMaction_lat}
\end{eqnarray}
which is given by replacing 
the continuum multiplet by the corresponding lattice multiplet
and the ordinary (anti-)commutators by link (anti-)commutators
in the action (\ref{N=D=4SYMaction_cont}).

\begin{figure}
\begin{center}
\includegraphics[width=40mm]{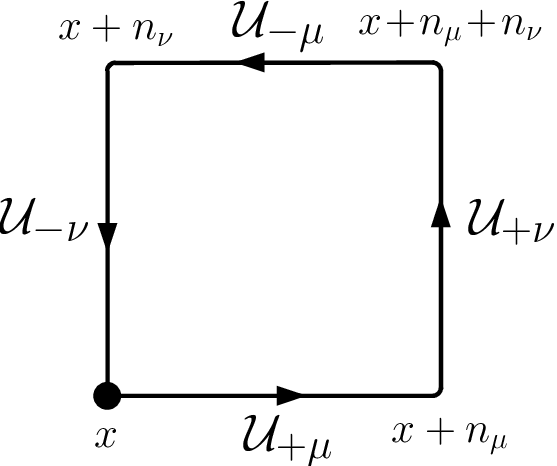}
\includegraphics[width=40mm]{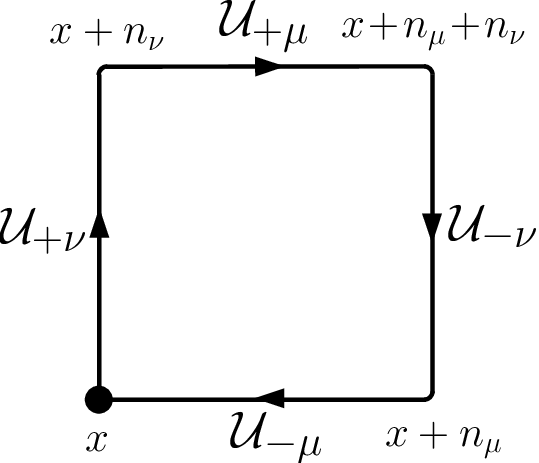}
\caption{Plaquettes}
\label{4DSYMPlaquettes}
\end{center}
\end{figure}

\begin{figure}
\begin{center}
\includegraphics[width=35mm]{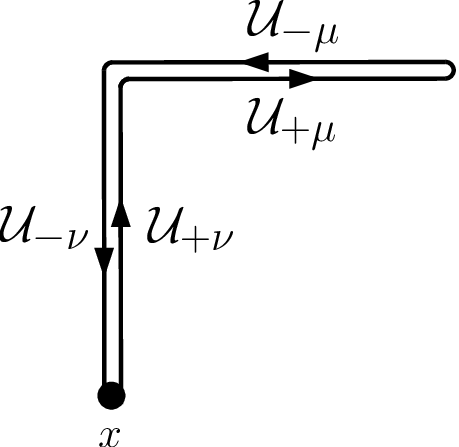}
\includegraphics[width=35mm]{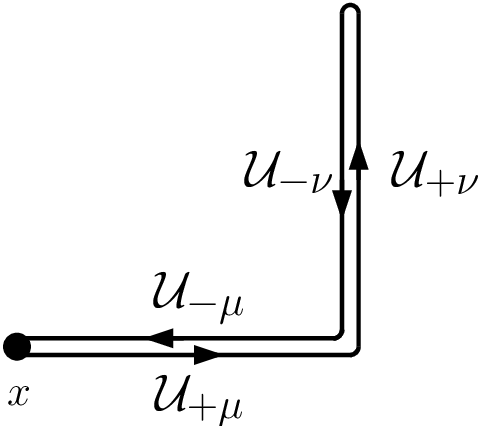}
\includegraphics[width=35mm]{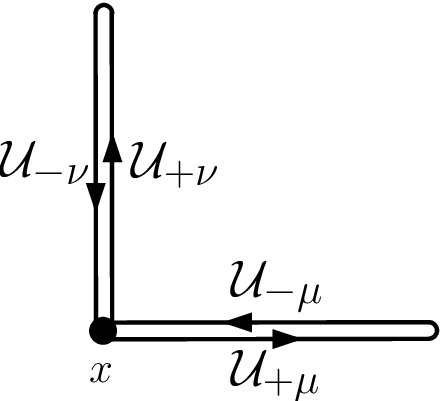}
\caption{Zero area loops}
\label{4DSYMzero_area}
\end{center}
\end{figure}

An important feature of the above lattice action (\ref{N=D=4SYMaction_lat}) 
is that each term forms closed loop from $x$ to $x$, which ensures manifest
gauge invariance. 
One can also see that the boson part of the action contains
ordinary plaquette terms in 4-dimensions (Fig.\ref{4DSYMPlaquettes})
as well as zero-area loops (Fig.\ref{4DSYMzero_area}) which is originated
from the non-unitary nature of bosonic gauge link variables,
$\U_{+\mu}\U_{-\mu}\neq 1$.
 

As in the continuum action (\ref{N=D=4SYMaction_cont})
discussed in the previous subsection,
on-shell structure of the $N=D=4$ Dirac-K\"ahler lattice SYM multiplet
$(\U_{\pm\mu},W,F,\rho,\lambda_{\mu},\rho_{\mu\nu},\tilde{\lambda}_{\mu},\tilde{\rho})$
prohibits us to express the lattice action (\ref{N=D=4SYMaction_lat})
as any of the lattice supercharges exact form.

Correspondingly, SUSY invariance of the action (\ref{N=D=4SYMaction_lat})
is not manifestly ensured on the lattice and
should rather be examined through explicit component-wise transformations.
First, we define a SUSY variation of the action $S(\varphi)$
as
\begin{eqnarray}
\delta_{A} S (\varphi) &\equiv&
S (\varphi+\delta_{A}\varphi) - S (\varphi), \label{SUSY_val_def}
\end{eqnarray}
where $\varphi$ represents any of the component fields 
$(\U_{\pm\mu},W,F,\rho,\lambda_{\mu},\rho_{\rho\sigma},\tilde{\lambda}_{\mu},\tilde{\rho})$,
and we define the variation of the component fields as
\begin{eqnarray}
(\delta_{A}\varphi)_{x+a_{\varphi},x} &\equiv&
(\xi_{A})_{x+a_{\varphi},x+a_{\varphi}+a_A}
(s_{A}\varphi)_{x+a_{\varphi}+a_A,x} \\
&=& 
(\xi_{A})_{x+a_{\varphi},x+a_{\varphi}+a_A}
s_{A}(\varphi)_{x+a_{\varphi},x} \\
&=&
(\xi_{A})_{x+a_{\varphi},x+a_{\varphi}+a_A}
[\D_{A},\varphi\}_{x+a_{\varphi}+a_A,x},
\end{eqnarray}
where $(\varphi)_{x+a_{\varphi},x}$ again denotes any components of the lattice multiplet
with explicit link indices.
In the above definition of the variation of the component fields, we introduced 
a set of Grassmann parameters $\xi_{A} = (\xi, \xi_{\mu}, \xi_{\mu\nu},\tilde{\xi}_{\mu},\tilde{\xi})$
with link nature from $x$ to $x-a_{A}$ 
which is an opposite link nature of the corresponding fermionic covariant derivative 
$(\nabla_{A})_{x+a_{A},x}$.
We also assume that the Grassmann parameter $(\xi_{A})_{x-a_{A},x}$ 
anti-commutes with any of the fermionic covariant derivatives $\nabla_{B}=(\D,\D_{\mu},\D_{\mu\nu},\tilde{\D}_{\mu},\tilde{\D})$
in the sense of link (anti)commutators,
\begin{eqnarray}
\{ \xi_{A}, \nabla_{B} \}_{x+a_{B}-a_{A},x}
 &=&
(\xi_{A})_{x+a_{B}-a_{A},x+a_{B}} (\nabla_{B})_{x+a_{B},x} \nonumber \\[5pt] 
&& +
 (\nabla_{B})_{x+a_{B}-a_{A},x-a_{A}} (\xi_{A})_{x-a_{A},x} \nonumber \label{NC_xi_nabla} \\[5pt]
 &=& 0.
\end{eqnarray}
Remembering that all of the component fields are given as 
combinations of the fermionic covariant derivatives, 
one can see the above assumption automatically provides the following (anti)commutativity 
between $\xi_{A}$ and any of the component fields 
$\varphi = (\U_{\pm\mu},W,F,\rho,\lambda_{\mu},\rho_{\rho\sigma},\tilde{\lambda}_{\mu},\tilde{\rho})$,
\begin{eqnarray}
[ \xi_{A}, \varphi \}_{x+a_{\varphi}-a_{A},x}
 &=&
(\xi_{A})_{x+a_{\varphi}-a_{A},x+a_{\varphi}} (\varphi)_{x+a_{\varphi},x} \nonumber \\[5pt] 
&& \pm
 (\varphi)_{x+a_{\varphi}-a_{A},x-a_{A}} (\xi_{A})_{x-a_{A},x} \\[5pt]
 &=& 0. \label{non-comm_xi1}
\end{eqnarray}

Based on the above setup, let us explicitly see the SUSY invariance of
the lattice action (\ref{N=D=4SYMaction_lat}). 
In the following explanations, for notational simplicity, we often omit the symbols of trace and summation over $x.$
First, we number each of the 
terms appeared in the lattice action  (\ref{N=D=4SYMaction_lat}) as follows,
\begin{eqnarray}
S^{N=D=4}_{lat.\ TSYM} 
&=& \sum_{x} \ \mathrm{tr}\ \biggl[
\underbrace{
\frac{1}{2}[\U_{+\mu},\U_{+\nu}]_{x,x-n_{\mu}-n_{\nu}}[\U_{-\mu},\U_{-\nu}]_{x-n_{\mu}-n_{\nu},x}
}_{\textcircled{\scriptsize{1}}}
\nonumber\\[0pt]
&&
\underbrace{
-\frac{1}{4}[\U_{+\mu},\U_{-\mu}]_{x,x}[\U_{+\nu},\U_{-\nu}]_{x,x}
}_{\textcircled{\scriptsize{2}}}
\
\underbrace{
-\frac{1}{4}[W,F]_{x,x}[W,F]_{x,x}
}_{\textcircled{\scriptsize{3}}}
 \nonumber \\[2pt]
&&
\underbrace{
+\frac{1}{2}[\U_{+\mu},W]_{x,x-n_{\mu}-a-\tilde{a}}[\U_{-\mu},F]_{x-n_{\mu}-a-\tilde{a},x} 
}_{\textcircled{\scriptsize{4}}}
\nonumber \\[2pt]
&&
\underbrace{
+\frac{1}{2}[\U_{-\mu},W]_{x,x+n_{\mu}-a-\tilde{a}}[\U_{+\mu},F]_{x+n_{\mu}-a-\tilde{a},x}
}_{\textcircled{\scriptsize{5}}}
\nonumber \\[6pt]
&&
\underbrace{
-i(\lambda_{\mu})_{x,x+a_{\mu}}[\U_{+\mu},\rho]_{x+a_{\mu},x} 
}_{\textcircled{\scriptsize{6}}}
\
\underbrace{
+i\tilde{\rho}_{x,x+\tilde{a}}[W,\rho]_{x+\tilde{a},x}
}_{\textcircled{\scriptsize{7}}}
\
\underbrace{
-i(\lambda_{\mu})_{x,x+a_{\mu}}[F,\tilde{\lambda}_{\mu}]_{x+a_{\mu},x} 
}_{\textcircled{\scriptsize{8}}}
\nonumber \\[8pt]
&&
\underbrace{
+i(\lambda_{\mu})_{x,x+a_{\mu}}[\U_{-\nu},\rho_{\mu\nu}]_{x+a_{\mu},x}
}_{\textcircled{\scriptsize{9}}}
\
\underbrace{
-i(\tilde{\rho})_{x,x+\tilde{a}}[\U_{-\mu},\tilde{\lambda}_{\mu}]_{x+\tilde{a},x}
}_{\textcircled{\scriptsize{10}}}
\nonumber \\[2pt]
&&
\underbrace{
-\frac{i}{2}\epsilon_{\mu\nu\rho\sigma}(\tilde{\lambda}_{\mu})_{x,x+\tilde{a}_{\mu}}
[\U_{+\nu},\rho_{\rho\sigma}]_{x+\tilde{a}_{\mu},x}
}_{\textcircled{\scriptsize{11}}}
\
\underbrace{
-\frac{i}{8}\epsilon_{\mu\nu\rho\sigma}(\rho_{\mu\nu})_{x,x+a_{\mu\nu}}
[W,\rho_{\rho\sigma}]_{x+a_{\mu\nu},x}
}_{\textcircled{\scriptsize{12}}}
\biggr].\nonumber \\ \label{N=D=4SYMaction_lat_num}
\end{eqnarray}
Then, for example, the $s$ variation of \textcircled{\scriptsize{1}}, denoted as 
$\delta\,\textcircled{\scriptsize{1}}$, is given by
\begin{eqnarray}
\delta\, \textcircled{\scriptsize{1}} &=&
\underbrace{
\frac{1}{2} [\delta\,\U_{+\mu},\U_{+\nu}]_{x,x-n_{\mu}-n_{\nu}} [\U_{-\mu},\U_{-\nu}]_{x-n_{\mu}-n_{\nu},x}
}_{\textcircled{\scriptsize{1}}_1}  \\
&&
\underbrace{
+ \frac{1}{2} [\U_{+\mu},\delta\,\U_{+\nu}]_{x,x-n_{\mu}-n_{\nu}} [\U_{-\mu},\U_{-\nu}]_{x-n_{\mu}-n_{\nu},x}
}_{\textcircled{\scriptsize{1}}_2} \\
&&
\underbrace{
+ \frac{1}{2} [\U_{+\mu},\U_{+\nu}]_{x,x-n_{\mu}-n_{\nu}} [\delta\,\U_{-\mu},\U_{-\nu}]_{x-n_{\mu}-n_{\nu},x}
}_{\textcircled{\scriptsize{1}}_3} \\
&&
\underbrace{
+ \frac{1}{2} [\U_{+\mu},\U_{+\nu}]_{x,x-n_{\mu}-n_{\nu}} [\U_{-\mu},\delta\,\U_{-\nu}]_{x-n_{\mu}-n_{\nu},x}
}_{\textcircled{\scriptsize{1}}_4},
\end{eqnarray}
where $\textcircled{\scriptsize{1}}_1 = \textcircled{\scriptsize{1}}_2= 0$, because of 
the vanishing $s$ variation of $\U_{+\mu}$,  
\begin{eqnarray}
\delta\,\U_{+\mu} &=& 
(\delta\,\U_{+\mu})_{x+n_{\mu},x} 
\ =\
(\xi) _{x+n_{\mu},x+n_{\mu}+a} (s\,\U_{+\mu})_{x+n_{\mu}+a,x} \\[5pt]
&=&  0,
\end{eqnarray}
while $\textcircled{\scriptsize{1}}_3$ and $\textcircled{\scriptsize{1}}_4$ together provide
\begin{eqnarray}
\textcircled{\scriptsize{1}}_3 + \textcircled{\scriptsize{1}}_4 
&=&
(\xi)_{x, x+a} [\U_{+\mu},\U_{+\nu}]_{x+a, x-n_{\mu}-n_{\nu}+a}
[\lambda_{\mu},\U_{-\nu}]_{x-n_{\mu}-n_{\nu}+a,x},
\end{eqnarray}
where we used the $s$ variation of $\U_{-\mu}$,
\begin{eqnarray}
\delta\, \U_{-\mu} &=&
(\delta\, \U_{-\mu})_{x-n_{\mu}, x} \ =\
(\xi)_{x-n_{\mu}, x-n_{\mu}+a}
(s\,\U_{-\mu})_{x-n_{\mu}+a,x} \\[5pt]
&=& (\xi)_{x-n_{\mu}, x-n_{\mu}+a}
(\lambda_{\mu})_{x-n_{\mu}+a,x},
\end{eqnarray}
and the link commutativities between $\xi$ and $\U_{\pm\mu}$,
\begin{eqnarray}
[\xi, \U_{\pm \mu}]_{x\pm n_{\mu} +a, x}  
&=& (\xi)_{x\pm n_{\mu} +a, x\pm n_{\mu}}(\U_{\pm \mu})_{x\pm n_{\mu}, x}  
-(\U_{\pm \mu})_{x\pm n_{\mu}+a, x+a}  (\xi)_{x+a,x} \nonumber \\[5pt]
&=& 0. \label{link_comm_U_xi}
\end{eqnarray}
Next, as for $\textcircled{\scriptsize{2}}$, the $s$ variation of $\textcircled{\scriptsize{2}}$, denoted as 
$\delta\,\textcircled{\scriptsize{2}}$, is given by
\begin{eqnarray}
\delta\, \textcircled{\scriptsize{2}} &=&
\underbrace{
- \frac{1}{4} [\delta\,\U_{+\mu},\U_{-\mu}]_{x,x} [\U_{+\nu},\U_{-\nu}]_{x,x}
}_{\textcircled{\scriptsize{2}}_1} 
\
\underbrace{
- \frac{1}{4} [\U_{+\mu},\delta\,\U_{-\mu}]_{x,x} [\U_{+\nu},\U_{-\nu}]_{x,x}
}_{\textcircled{\scriptsize{2}}_2} \ \ \ \\
&&
\underbrace{
- \frac{1}{4} [\U_{+\mu},\U_{-\mu}]_{x,x} [\delta\,\U_{+\nu},\U_{-\nu}]_{x,x}
}_{\textcircled{\scriptsize{2}}_3} 
\
\underbrace{
- \frac{1}{4} [\U_{+\mu},\U_{-\mu}]_{x,x} [\U_{+\nu},\delta\,\U_{-\nu}]_{x,x}
}_{\textcircled{\scriptsize{2}}_4},
\end{eqnarray}
where $\textcircled{\scriptsize{2}}_1 = \textcircled{\scriptsize{2}}_3=0$,
again because of 
the vanishing $s$ variation of $\U_{+\mu}$, 
\begin{eqnarray}
\delta\,\U_{+\mu} &=& 
(\delta\,\U_{+\mu})_{x+n_{\mu},x} 
\ =\
(\xi) _{x+n_{\mu},x+n_{\mu}+a} (s\,\U_{+\mu})_{x+n_{\mu}+a,x} \\[5pt]
&=&  0,
\end{eqnarray}
while $\textcircled{\scriptsize{2}}_2$ and $\textcircled{\scriptsize{2}}_4$ together provide
\begin{eqnarray}
\textcircled{\scriptsize{2}}_2 + \textcircled{\scriptsize{2}}_4 
&=&
-\frac{1}{2}
(\xi)_{x, x+a} [\U_{+\mu},\lambda_{\mu}]_{x+a, x}
[\U_{+\nu},\U_{-\nu}]_{x,x},
\end{eqnarray}
where we again used the $s$ variation of $\U_{-\mu}$,
\begin{eqnarray}
\delta\, \U_{-\mu} &=&
(\delta\, \U_{-\mu})_{x-n_{\mu}, x} \ =\
(\xi)_{x-n_{\mu}, x-n_{\mu}+a}
(s\,\U_{-\mu})_{x-n_{\mu}+a,x} \\[5pt]
&=& (\xi)_{x-n_{\mu}, x-n_{\mu}+a}
(\lambda_{\mu})_{x-n_{\mu}+a,x},
\end{eqnarray}
and the link commutativities between $\xi$ and $\U_{\pm\mu}$,
\begin{eqnarray}
[\xi, \U_{\pm \mu}]_{x\pm n_{\mu} +a, x}  
&=& (\xi)_{x\pm n_{\mu} +a, x\pm n_{\mu}}(\U_{\pm \mu})_{x\pm n_{\mu}, x}  
-(\U_{\pm \mu})_{x\pm n_{\mu}+a, x+a}  (\xi)_{x+a,x} \nonumber \\[5pt]
&=& 0, 
\end{eqnarray}
as well as cyclic permutation properties under the trace,
\begin{eqnarray}
(\xi)_{x, x+a} [\U_{+\mu},\lambda_{\mu}]_{x+a, x}
[\U_{+\nu},\U_{-\nu}]_{x,x}
&=&
[\U_{+\nu},\U_{-\nu}]_{x,x}
(\xi)_{x, x+a} [\U_{+\mu},\lambda_{\mu}]_{x+a, x}. \ \ \ \ \
\end{eqnarray}

In a similar manner, one can proceed to calculate the $s$ variations of all the terms in
 (\ref{N=D=4SYMaction_lat}), and obtains,
\begin{eqnarray}
&&\ \hspace{-30pt}\delta\, S^{N=D=4}_{lat.\ TSYM}  \nonumber \\[5pt]
&=& \sum_{x} \ \mathrm{tr}\ \biggl[
\underbrace{
\xi\, [\U_{+\mu},\U_{+\nu}][\lambda_{\mu},\U_{-\nu}]
}_{\textcircled{\scriptsize{1}}'}
\
\underbrace{
- \frac{1}{2}\xi\, [\U_{+\mu},\lambda_{\mu}][\U_{+\nu},\U_{-\nu}]
}_{\textcircled{\scriptsize{2}}'} \nonumber \\[5pt]
&&
\underbrace{
+ \frac{1}{2} \xi\, [W, \tilde{\rho}][W,F]
}_{\textcircled{\scriptsize{3}}'} 
\
\underbrace{
+ \frac{1}{2}\xi\, [\U_{+\mu},W][\lambda_{\mu},F]
}_{\textcircled{\scriptsize{4}}'_{1}} 
\
\underbrace{
-\frac{1}{2}\xi\,[\U_{+\mu},W][\U_{-\mu}, \tilde{\rho}]
}_{\textcircled{\scriptsize{4}}'_{2}} 
\nonumber \\[5pt]
&&
\underbrace{
+\frac{1}{2} \xi\, [\lambda_{\mu},W][\U_{+\mu},F]
}_{\textcircled{\scriptsize{5}}'_{1}}
\
\underbrace{
 -\frac{1}{2}\xi\, [\U_{-\mu},W][\U_{+\mu},\tilde{\rho}]
}_{\textcircled{\scriptsize{5}}'_{2}} \nonumber \\[5pt]
&&
\underbrace{
-\frac{1}{2}\xi\, \lambda_{\mu}[\U_{+\mu},[\U_{+\nu},\U_{-\nu}]]
}_{\textcircled{\scriptsize{6}}'_{1}} 
\
\underbrace{
-\frac{1}{2}\xi\, \lambda_{\mu}[\U_{+\mu},[W,F]]
}_{\textcircled{\scriptsize{6}}'_{2}} \nonumber \\[5pt]
&&
\underbrace{
+\frac{1}{2}\xi\, \tilde{\rho}\,[W, [\U_{+\mu},\U_{-\mu}]]
}_{\textcircled{\scriptsize{7}}'_{1}}
\
\underbrace{
+\frac{1}{2}\xi\, \tilde{\rho}\,[W, [W,F]]
}_{\textcircled{\scriptsize{7}}'_{2}} \nonumber \\[5pt]
&&
\underbrace{
- i \xi\, \lambda_{\mu} \{\tilde{\rho},\tilde{\lambda}_{\mu}\}
}_{\textcircled{\scriptsize{8}}'_{1}}
\
\underbrace{
-\xi\, \lambda_{\mu} [F, [\U_{+\mu},W]]
}_{\textcircled{\scriptsize{8}}'_{2}} \nonumber \\[5pt]
&&
\underbrace{
-i\xi\, \lambda_{\mu}\{\lambda_{\nu},\rho_{\mu\nu}\}
}_{\textcircled{\scriptsize{9}}'_{1}}
\
\underbrace{
-\xi\, \lambda_{\mu}[\U_{-\nu},[\U_{+\mu},\U_{+\nu}]]
}_{\textcircled{\scriptsize{9}}'_{2}} \nonumber \\
&&
\underbrace{
+i \xi\, \tilde{\rho}\, \{\lambda_{\mu},\tilde{\lambda}_{\mu}\}
}_{\textcircled{\scriptsize{10}}'_{1}}
\
\underbrace{
-\xi\, \tilde{\rho}\ [\U_{-\mu},[\U_{+\mu},W]]
}_{\textcircled{\scriptsize{10}}'_{2}} \nonumber \\
&&
\underbrace{
+\frac{1}{2}\epsilon_{\mu\nu\rho\sigma}\,\xi\, [\U_{+\mu},W][\U_{+\nu},\rho_{\rho\sigma}]
}_{\textcircled{\scriptsize{11}}'_{1}}
\
\underbrace{
+\frac{1}{2}\epsilon_{\mu\nu\rho\sigma}\,\xi\, \tilde{\lambda}_{\mu}[\U_{+\nu},[\U_{+\rho},\U_{+\sigma}]]
}_{\textcircled{\scriptsize{11}}'_{2}} \nonumber \\
&&
\underbrace{
-\frac{1}{8}\epsilon_{\mu\nu\rho\sigma}\,\xi\,[\U_{+\mu},\U_{+\nu}][W,\rho_{\rho\sigma}]
}_{\textcircled{\scriptsize{12}}'_{1}}
\
\underbrace{
+\frac{1}{8}\epsilon_{\mu\nu\rho\sigma}\,\xi\, \rho_{\mu\nu}[W,[\U_{+\rho},\U_{+\sigma}]]
}_{\textcircled{\scriptsize{12}}'_{2}}
\biggr],
\end{eqnarray}
where we omitted the link indices for notational simplicity.
Utilizing the link (anti)commutativities between $\xi$ and the component fields,
as well as the cyclic permutation properties under the trace and summation over $x$
as explained above, one obtains,
\begin{eqnarray}
\textcircled{\scriptsize{1}}'
+ \textcircled{\scriptsize{9}}'_{2}
&=& 0, \\[5pt]
\textcircled{\scriptsize{2}}'
+ \textcircled{\scriptsize{6}}'_{1}
&=& 0, \\[5pt]
\textcircled{\scriptsize{3}}'
+ \textcircled{\scriptsize{7}}'_{2} 
&=& 0, \\[5pt]
\textcircled{\scriptsize{4}}'_{1}
+ \textcircled{\scriptsize{5}}'_{1} 
+ \textcircled{\scriptsize{6}}'_{2} 
+ \textcircled{\scriptsize{8}}'_{2} 
&=& 0, \\[5pt]
\textcircled{\scriptsize{8}}'_{1}
+ \textcircled{\scriptsize{10}}'_{1} 
&=& 0, \\[5pt] 
\textcircled{\scriptsize{9}}'_{1}
&=& 0,  \\[5pt]
\textcircled{\scriptsize{11}}'_{2}
&=& 0. 
\end{eqnarray}
For example, we have, for $\textcircled{\scriptsize{9}}'_{2}$ and 
$\textcircled{\scriptsize{1}}'$,
with the explicit link indices,
\begin{eqnarray}
\textcircled{\scriptsize{9}}'_{2} &=& 
-(\xi)_{x,x+a} (\lambda_{\mu})_{x+a, x+n_{\mu}}[\U_{-\nu},[\U_{+\mu},\U_{+\nu}]]_{x+n_{\mu},x} \\[5pt]
&=& - (\xi)_{x,x+a} (\lambda_{\mu})_{x+a, x+n_{\mu}} (\U_{-\nu})_{x+n_{\mu}, x+n_{\mu}+n_{\nu}}
[\U_{+\mu},\U_{+\nu}]_{x+n_{\mu}+n_{\nu},x} \nonumber  \\[5pt]
&& + (\xi)_{x,x+a} (\lambda_{\mu})_{x+a, x+n_{\mu}}
 [\U_{+\mu},\U_{+\nu}]_{x+n_{\mu},x-n_{\nu}}(\U_{-\nu})_{x-n_{\nu},x} \label{2nd_line} \\[5pt]
 &=&  - (\xi)_{x,x+a} (\lambda_{\mu})_{x+a, x+n_{\mu}} (\U_{-\nu})_{x+n_{\mu}, x+n_{\mu}+n_{\nu}}
[\U_{+\mu},\U_{+\nu}]_{x+n_{\mu}+n_{\nu},x} \nonumber \\[5pt]
&& + (\xi)_{x,x+a} (\U_{-\nu})_{x+a,x+a+n_{\nu}}
 (\lambda_{\mu})_{x+a+n_{\nu}, x+n_{\mu}+n_{\nu}}
 [\U_{+\mu},\U_{+\nu}]_{x+n_{\mu}+n_{\nu},x} \ \ \ \ \label{3rd_line} \\[5pt]
 &=& -(\xi)_{x,x+a} [\U_{-\nu},\lambda_{\mu}]_{x+a,x+n_{\mu}+n_{\nu}}
  [\U_{+\mu},\U_{+\nu}]_{x+n_{\mu}+n_{\nu},x} \\[5pt]
 &=& -(\xi)_{x,x+a}
  [\U_{+\mu},\U_{+\nu}]_{x+a,x+a-n_{\mu}-n_{\nu}}
  [\U_{-\nu},\lambda_{\mu}]_{x+a-n_{\mu}-n_{\nu},x} \\[5pt]
 &=& - 
 \textcircled{\scriptsize{1}}',
\end{eqnarray}
where, from the second to the fifth line, we utilized the link (anti)commutativities between $\xi$ and the component fields,
as well as the cyclic permutation properties under the trace and summation over $x$.
More explicitly, the second term of the third line (\ref{3rd_line})
can be obtained, from the second term of the  second line (\ref{2nd_line}), by 
\begin{eqnarray}
&& (\xi)_{x,x+a} (\lambda_{\mu})_{x+a, x+n_{\mu}}
 [\U_{+\mu},\U_{+\nu}]_{x+n_{\mu},x-n_{\nu}}(\U_{-\nu})_{x-n_{\nu},x} \nonumber \\[5pt]
&&\underset{under\ tr}{=} 
(\U_{-\nu})_{x-n_{\nu},x} (\xi)_{x,x+a} (\lambda_{\mu})_{x+a, x+n_{\mu}}
 [\U_{+\mu},\U_{+\nu}]_{x+n_{\mu},x-n_{\nu}} \nonumber  \\[5pt]
&&\underset{under\ sum.\ over\ x}{=}
(\U_{-\nu})_{x,x+n_{\nu}} (\xi)_{x+n_{\nu},x+n_{\nu}+a} (\lambda_{\mu})_{x+n_{\nu}+a, x+n_{\nu}+n_{\mu}}
 [\U_{+\mu},\U_{+\nu}]_{x+n_{\nu}+n_{\mu},x} \nonumber   \\[5pt]
&& \ \ \ = 
(\xi)_{x,x+a} (\U_{-\nu})_{x+a,x+a+n_{\nu}}
 (\lambda_{\mu})_{x+a+n_{\nu}, x+n_{\mu}+n_{\nu}}
 [\U_{+\mu},\U_{+\nu}]_{x+n_{\mu}+n_{\nu},x}, 
\end{eqnarray}
where in the last line, 
we utilized the link commutativity between $\xi$ and $\U_{-\nu}$, as shown in (\ref{link_comm_U_xi}).

As for $\textcircled{\scriptsize{4}}'_{2}$,  $\textcircled{\scriptsize{5}}'_{2}$, 
$\textcircled{\scriptsize{7}}'_{1}$,  and  $\textcircled{\scriptsize{10}}'_{2}$, 
by further utilizing the Jacobi identity,
\begin{eqnarray}
[\U_{-\mu},[\U_{+\mu},W]] + [\U_{+\mu},[W, \U_{-\mu}]] +[W,[\U_{-\mu},\U_{+\mu}]] &=& 0,
\end{eqnarray}
one obtains
\begin{eqnarray}
\textcircled{\scriptsize{4}}'_{2}
+ \textcircled{\scriptsize{5}}'_{2} 
+ \textcircled{\scriptsize{7}}'_{1} 
+ \textcircled{\scriptsize{10}}'_{2} 
&=& 0.
\end{eqnarray}
Similarly, as for $\textcircled{\scriptsize{11}}'_{1}$, 
$\textcircled{\scriptsize{12}}'_{1}$, and $\textcircled{\scriptsize{12}}'_{2}$,
utilizing the Jacobi identity,
\begin{eqnarray}
[W,[\U_{+\rho},\U_{+\sigma}]]+[\U_{+\rho},[\U_{+\rho},W]] + [\U_{+\sigma},[W,\U_{+\rho}]]
&=& 0,
\end{eqnarray}
leads us to
\begin{eqnarray}
\textcircled{\scriptsize{11}}'_{1}
+ \textcircled{\scriptsize{12}}'_{1} 
+ \textcircled{\scriptsize{12}}'_{2} 
&=& 0, 
\end{eqnarray}
from which we conclude that 
\begin{eqnarray}
\delta\, S^{N=D=4}_{lat.\ TSYM} &=& 0.
\end{eqnarray}

By tedious but similar and straightforward calculations utilizing: \\ 
$\cdot$ link (anti)commutativities between $\xi_{\mu}, \xi_{\mu\nu},\tilde{\xi}_{\mu},\tilde{\xi}$ and the component fields;\\
$\cdot$ cyclic permutation properties under the trace and summation over $x$; and \\
$\cdot$ Jacobi identities,\\ 
one can show that the exact SUSY invariance of the lattice action
(\ref{N=D=4SYMaction_lat}) w.r.t. the rest of  all the supercharges $(s,s_{\mu},s_{\mu\nu},\tilde{s}_{\mu},\tilde{s})$
, namely,
\begin{eqnarray}
\delta_{\mu}\, S^{N=D=4}_{lat.\ TSYM} & = & 
\delta_{\mu\nu}\, S^{N=D=4}_{lat.\ TSYM} \ =\ 
\tilde{\delta}_{\mu}\, S^{N=D=4}_{lat.\ TSYM} \ =\
\tilde{\delta}\, S^{N=D=4}_{lat.\ TSYM} \ =\
0. \nonumber \\
\end{eqnarray}

In order to support the above calculations, 
the summation over $x$ needs to appropriately cover the corresponding lattice  structures.
Namely, if we take the symmetric choice of $a_{A}$,
the summation should cover both of 
the original lattice
(the integer sites)
and 
the dual lattice
(half-integer sites), 
\begin{eqnarray}
\sum_{x} &=& \sum_{(m_1,m_2,m_3,m_4)}
+\sum_{(m_1+\frac{1}{2},m_2+\frac{1}{2},m_3+\frac{1}{2},m_4+\frac{1}{2})}, 
\end{eqnarray}
while, if we take the asymmetric choice of $a_{A}$,
it suffices that the summation covers only the integer sites, 
since the original and dual lattices are overlapping with each other,
\begin{eqnarray}
\sum_{x} &=& \sum_{(m_1,m_2,m_3,m_4)}, 
\end{eqnarray}
where each of $m_1,m_2,m_3,m_4$ denotes any integer.

Notice here that, in constructing and proving the SUSY invariant action,
we do NOT rely on the interchanging property among the component fields,
while utilizing the 
cyclic permutation properties under the trace and summation over $x$.
Therefore, the critique posed in  \cite{Bruckmann} (mentioned as claim 1) and (\ref{ord-amb}) in the Introduction)  is not applied in the present framework.

We would also
 like to mention here 
the formulation of Dirac-K\"ahler twisted
SYM on the lattice given in \cite{Catterall,Catterall:2013roa,Catterall:2009it,Catterall:2012yq,Catterall:2014vka},
where only the scalar part of the supercharge $s$
is exactly considered on the lattice.
Their formulation basically corresponds to 
a case that one takes the asymmetric choice of $a_{A}$, namely, $a=0$, and
respects only a SUSY variation under the scalar supercharge $s$.  
Although the present formulation has SUSY invariance with respect to all the supercharges
by employing the link Grassmann parameters as shown above, even if we concentrate on 
a SUSY variation with a Grassmann parameter without link nature,
we still have an option to determine which supercharge to be selected. 
More particularly, even without employing the link nature of the Grassmann parameter,
the present formulation accommodates the following SUSY invariant nature of the resulting action, 
due to the one-parameter arbitrariness
of shift parameter $a_{A}$ associated with lattice supercharges $s_{A}$,
\begin{eqnarray}
\delta\ S^{N=D=4}_{lat.\ TSYM} &=& 0, \hspace{30pt} with \ \ \ a\ =\ 0, \label{inv_s}\\[2pt]
\delta_{\mu}\ S^{N=D=4}_{lat.\ TSYM} &=& 0, \hspace{30pt} with \ \ \ a_{\mu}\ =\ 0,\\[2pt]
\delta_{\mu\nu}\ S^{N=D=4}_{lat.\ TSYM} &=& 0, \hspace{30pt} with \ \ \ a_{\mu\nu}\ =\ 0,\\[2pt]
\tilde{\delta}_{\mu}\ S^{N=D=4}_{lat.\ TSYM} &=& 0, \hspace{30pt} with \ \ \ \tilde{a}_{\mu}\ =\ 0,\\[2pt]
\tilde{\delta}\ S^{N=D=4}_{lat.\ TSYM} &=& 0, \hspace{30pt} with \ \ \ \tilde{a}\ =\ 0.
\label{inv_tildes}
\end{eqnarray}

It is also important to comment about a gauge covariant (invariant)  aspect of the action. 
As mentioned above, although the lattice action (\ref{N=D=4SYMaction_lat}) itself has a manifest gauge invariance, a gauge covariant (invariant) nature of the SUSY variation of the action (\ref{SUSY_val_def}), $\delta_{A}S(\varphi)$, 
needs to be carefully examined.
From the explicit calculation of $\delta\ S^{N=D=4}_{lat.\ TSYM}$ as shown above,
one can see that any term included in the SUSY variation of the action
form closed loops, for example, from $x$ to $x$, 
due to the link nature of the Grassmann parameter $(\xi_{A})_{x-a_{A},x}$.
Here, the link Grassmann parameter $(\xi_{A})_{x-a_{A},x}$ is expected to transform
under gauge transformation as,
\begin{eqnarray}
(\xi_{A})_{x-a_{A},x} &\rightarrow& G^{-1}(x-a_{A})(\xi_{A})_{x-a_{A},x}G(x),
\end{eqnarray}
while, at the same time, the non-commutative nature of $\xi_{A}$ (\ref{non-comm_xi1})
applied to the gauge transformation functions $G(x)$ leads to the following 
gauge invariant nature of the link Grassmann variable,
\begin{eqnarray}
(\xi_{A})_{x-a_{A},x} &\rightarrow& G^{-1}(x-a_{A})(\xi_{A})_{x-a_{A},x}G(x) \\[5pt]
&=& (\xi_{A})_{x-a_{A},x} G^{-1}(x)G(x)\\[5pt] 
&=&  (\xi_{A})_{x-a_{A},x}. \label{gauge_inv_xi1}
\end{eqnarray}
We will provide an exemplary representation of $\xi_{A}$ at the end of next subsection.

\subsection{Group and algebraic interpretation of link formulation} 
\label{GAI}

\indent

Throughout this article and our lower dimensional formulations given in \cite{DKKN2,DKKN3}, we have constructed the lattice super Yang-Mills action based on a link approach
where the gauge covariance is kept manifest at every stage of the calculation.
As we have already mentioned, the link approach may be understood as another aspect of the non-commutative supercharge formulation given in \cite{DKKN1} where non-commutative supercharges have been introduced to be consistent with the lattice Leibniz rule.
Although, as we have shown in the previous subsection, the link approach already provides
a sufficient framework to derive a super Yang-Mills multiplet, SUSY transformation laws, and 
SUSY invariance of the resulting action on the lattice, it is worthwhile to make a step forward and 
investigate more detailed aspects of the formulation.

In this subsection, as an example of such investigations, we show that the gauge covariant link (anti)commutators may naturally arise from a promotion of the bosonic gauge covariant derivative to a gauge covariant group element,
which may provide a group and algebraic theoretical interpretation of the link formulation.

Let us begin with the following superconnection relation in the continuum spacetime,
which corresponds to a generic expression of (\ref{4DSYMcc1})-(\ref{4DSYMcc4}),
\begin{eqnarray}
\{\D_{A},\D_{B} \} &=& f_{AB}^{\mu}\D_{\pm\mu} \label{Nabla1}
\end{eqnarray}
where $\nabla_{A}$ and $\nabla_{B}$ denote any of the fermionic super-covariant derivatives 
$(\nabla,\nabla_{\mu},\nabla_{\rho\sigma},\tilde{\nabla}_{\mu},\tilde{\nabla})$
given by (\ref{4DDD1})-(\ref{4DDD5}), 
and $\nabla_{\pm\mu}$ denotes corresponding bosonic covariant derivatives 
given by (\ref{nablapm1})-(\ref{nablapm3}), 
while $f_{AB}^{\mu}$ denotes a corresponding numerical coefficient.
In the following, 
we express $\nabla_{\pm\mu}$ in terms of bosonic superconnections $\Gamma^{\pm}_{\mu}$,
\begin{eqnarray}
\nabla_{\pm\mu} &=& \nabla_{\underline{\mu}}\pm \mathcal{V}_{\mu}
\ \equiv\
\partial_{\mu}-i\Gamma^{\pm}_{\mu}
\end{eqnarray}
where, according to  (\ref{nablapm1})-(\ref{nablapm3}), the lowest components are given by,
\begin{eqnarray}
\nabla_{\pm\mu} |
 &=& \partial_{\mu}-i\Gamma^{\pm}_{\mu}| 
\ =\ \nabla_{\underline{\mu}} | \pm \mathcal{V}_{\mu} |\\[5pt]
&=& \partial_{\mu}-i (\omega_{\mu}\pm iV_{\mu}).
\end{eqnarray}
Recall that, at this stage, the subscript  $\pm$ of  $\nabla_{\pm\mu}$ merely indicates
the sign in front of $V_{\mu}$,
and the relation (\ref{Nabla1}) is given in the continuum spacetime.
Super gauge covariance of  the elements appeared in the relation  (\ref{Nabla1}) is given  by
\begin{eqnarray}
(\D_{A}, \D_{B}, \D_{\pm\mu}) \rightarrow
\mathcal{G}^{-1}(x)(\D_{A}, \D_{B}, \D_{\pm\mu})\, \mathcal{G}(x).
\end{eqnarray}
where we explicitly write the argument $x$ to stress that the covariance is given locally.

Since the bosonic gauge fields $\omega_{\mu}$ are embedded in the bosonic covariant derivative $\nabla_{\pm\mu}$, we may expect that a lattice counterpart of the superconnection relation (\ref{Nabla1})
is derived by starting from 
a promotion of algebraic object in the r.h.s. to a group element, namely, 
exponentiating
the $\nabla_{\pm\mu}$,
\begin{eqnarray}
\{\D_{A},\D_{B} \} &=& -f^{\mu}_{AB}\, e^{-\D_{+\mu}},  \ \ or \label{Nabla2-1} \\[5pt]
\{\D_{A},\D_{B} \} &=& +f^{\mu}_{AB}\, e^{+\D_{-\mu}},  \label{Nabla2-2}
\end{eqnarray}
where we take the 
coefficients in front of  $\nabla_{\pm\mu}$, 
which will be the
lattice constant, as unit length for notational simplicity, and
the signs 
of the coefficients
are intentionally chosen to be consistent with the 
derivations given in the previous subsection.
Note that the following descriptions  provide 
explanations that bridge continuum theory and lattice theory
which is
given by (\ref{Link_comm1}) and (\ref{Link_comm2}).
Once the relations  (\ref{Link_comm1}) and (\ref{Link_comm2})
are obtained, the algebraic relations can be entirely described by lattice coordinates.
Meanwhile, in the following derivations, and
with the above understanding, 
we make use of 
the notation $e^{\pm\partial_{\mu}}$ which 
provide finite translation operations
acting on the continuum spacetime.

Although
the right hand sides 
of the relations (\ref{Nabla2-1}) and (\ref{Nabla2-2})
belong to group elements while the left hand sides belong to algebraic elements.
as we will see in the following, these sorts of ``algebraic decompositions of group elements'' 
actually make sense when the lattice Leibniz rule conditions are satisfied.
First, 
notice that 
any k-th order of the bosonic gauge covariant derivative, $(\D_{\pm\mu})^{k}$, is still a locally  covariant object,
\begin{eqnarray}
(\D_{\pm\mu})^{k} &\rightarrow&
\mathcal{G}^{-1}(x) \D_{\pm\mu}
\mathcal{G}(x) \mathcal{G}^{-1}(x)\D_{\pm\mu}\mathcal{G}(x)
\cdots
\mathcal{G}^{-1}(x)\D_{\pm\mu}\mathcal{G}(x) 
\ =\ \mathcal{G}^{-1}(x) (\D_{\pm\mu})^{k}
\mathcal{G}^{-1}(x). \nonumber \\
\end{eqnarray}
Thus, one sees that
the gauge covariance of (\ref{Nabla2-1}) and (\ref{Nabla2-2}) still remains local, 
\begin{eqnarray}
e^{\mp\D_{\pm\mu}} \rightarrow
\mathcal{G}^{-1}(x)\, e^{\mp\D_{\pm\mu}}\, \mathcal{G}(x). \label{local_cov}
\end{eqnarray}
This may have the following interpretation.
 Namely, in the exponentiaion
of the bosonic gauge covariant derivatives, $e^{\mp \D_{\pm\mu}}$, 
the bosonic superconnections $\pm i\Gamma^{\pm}_{\mu}$ 
have contributions of connecting a point $x$ to its neighboring points
$x\pm n_{\mu}$ in a gauge covariant manner, 
while the derivative operators $\mp \partial_{\mu}$
have contributions of pulling back the neighboring points $x\pm n_{\mu}$
to the original point $x$. Both of these contributions ensure the eventual local gauge covariance of $e^{\mp \D_{\pm\mu}}$, or equivalently, of 
(\ref{Nabla2-1}) and (\ref{Nabla2-2}). 

In order to look at the above features more precisely, one may extract the 
leading order contributions of the derivative operators $\mp \partial_{\mu}$
from  $e^{\mp \D_{\pm\mu}}$,
\begin{eqnarray}
e^{\mp \D_{\pm\mu}} &=&
(\U_{\pm \mu})_{x, x\mp n_{\mu}} e^{\mp \partial_{\mu}}
\end{eqnarray}
where we notationally introduced
bosonic gauge link variables $(\U_{\pm\mu})_{x,x\mp n_{\mu}}$ 
defined as
\begin{eqnarray}
(\U_{+\mu})_{x, x-n_{\mu}} &\equiv& e^{-\D_{+\mu}}\, e^{+\partial_{\mu}}
\ =\ e^{-\partial_{\mu}+i\Gamma^{+}_{\mu}}\, e^{+\partial_{\mu}} \label{U1}\\[5pt]
&=& e^{+i\Gamma^{+}_{\mu}- \frac{i}{2}[\partial_{\mu},\Gamma^{+}_{\mu}] + \cdots}, 
\ \ \ (\mu: {\rm no\ sum}),  \label{U2}  \\[5pt]
(\U_{-\mu})_{x, x+n_{\mu}} &=& e^{+\D_{-\mu}}\, e^{-\partial_{\mu}} 
\ = \ e^{+\partial_{\mu}-i\Gamma^{-}_{\mu}}\, e^{-\partial_{\mu}} \label{U3}\\[5pt]
&=& e^{-i\Gamma^{-}_{\mu}- \frac{i}{2}[\partial_{\mu},\Gamma^{-}_{\mu}] + \cdots}, 
\ \ \ (\mu: {\rm no\ sum}).
\label{U4} 
\end{eqnarray}
where, as mentioned above,  $\Gamma^{\pm}_{\mu}= \Gamma^{\pm}_{\mu}(x,\theta_{A},\theta_{B})$ represent the bosonic superconnections
whose lowest components are given by $\Gamma^{\pm}_{\mu}|_{\theta=0}=\omega_{\mu}\pm i V_{\mu}$, and  the dots `$\cdots$' represents higher order terms including  $\partial_{\mu}$ and $\Gamma^{\pm}_{\mu}$ which can be obtained by Baker-Campbell-Hausdorff formula.
Due to the local gauge covariant nature of $e^{\mp \D_{\pm\mu}}$ (\ref{local_cov}), the above introduced bosonic gauge link variables
should have the following link covariant properties,
\begin{eqnarray}
(\U_{\pm\mu})_{x,x\mp n_{\mu}} &\rightarrow& \mathcal{G}^{-1}(x)\, (\U_{\pm\mu})_{x,x\mp n_{\mu}}\,  \mathcal{G}(x\mp n_{\mu}).
\label{gauge_cov_1}
\end{eqnarray} 
The above link covariant properties are compatible with those of the 
ordinary gauge link variables used in lattice gauge theory. 
In a similar manner, the inverse of the bosonic gauge link variables can be defined as
\begin{eqnarray}
(\U^{-1}_{+\mu})_{x-n_{\mu},x} &\equiv& e^{-\partial_{\mu}}\, e^{+\D_{+\mu}}\ 
\ =\ e^{-\partial_{\mu}}\, e^{+\partial_{\mu}-i\Gamma^{+}_{\mu}}  \label{U1-1}\\[5pt]
&=& e^{-i\Gamma^{+}_{\mu}+ \frac{i}{2}[\partial_{\mu},\Gamma^{+}_{\mu}] + \cdots}, 
\ \ \ (\mu: {\rm no\ sum}),  \label{U2-1}  \\[5pt]
(\U^{-1}_{-\mu})_{x+n_{\mu},x} &=& e^{+\partial_{\mu}}\, e^{-\D_{-\mu}} 
\ = \  e^{+\partial_{\mu}}\, e^{-\partial_{\mu}+i\Gamma^{-}_{\mu}} \label{U3-1}\\[5pt]
&=& e^{+i\Gamma^{-}_{\mu}+ \frac{i}{2}[\partial_{\mu},\Gamma^{-}_{\mu}] + \cdots}. 
\ \ \ (\mu: {\rm no\ sum})
\label{U4-1} 
\end{eqnarray}
whose covariant nature is given by
\begin{eqnarray}
(\U_{\pm\mu}^{-1})_{x\mp n_{\mu},x} &\rightarrow& \mathcal{G}^{-1}(x\mp n_{\mu})\, 
(\U_{\pm\mu}^{-1})_{x\mp n_{\mu},x}\,  \mathcal{G}(x). \label{gauge_cov_2}
\end{eqnarray} 

Viewing the above explicit expressions of the bosonic gauge link variables 
$(\U_{\pm\mu})_{x,x\mp n_{\mu}}$ given in  (\ref{U1}) - (\ref{U4-1}),
one can 
clearly
see 
that 
the bosonic gauge link variables $(\U_{\pm\mu})_{x,x\mp n_{\mu}}$
are obtained by removing the leading-order contributions of  
derivative operators $\mp \partial_{\mu}$
from $e^{\mp \D_{\pm\mu}}$.
In other words,
the 
leading order
contributions of shift operations embedded in the exponentialized bosonic
gauge covariant derivatives $e^{\mp\D_{\pm\mu}}$ are canceled by the shift elements $e^{\pm\partial_{\mu}}$, 
while at the same time,
due to
the local gauge covariant properties of $e^{\mp\D_{\pm\mu}}$ in (\ref{local_cov}), 
the explicit forms (\ref{U1}) - (\ref{U4-1})
provide the 
link nature
of $(\U_{\pm\mu})_{x,x\mp n_{\mu}}$ and $(\U_{\pm\mu}^{-1})_{x\mp n_{\mu},x}$
as in (\ref{gauge_cov_1}) and  (\ref{gauge_cov_2}).

In terms of the above notations, 
taking a commutator between $e^{\mp \D_{\pm\mu}}$
and an arbitrary function $\Phi(x)$
naturally provides 
the following 
gauge covariant differentiations,
\begin{eqnarray}
[e^{-\D_{+\mu}}, \Phi(x)] &=& 
e^{-\D_{+\mu}}\Phi(x) -\Phi(x)e^{-\D_{+\mu}} \label{eq1}  \\[5pt]
&=& 
[e^{-\D_{+\mu}} e^{+\partial_{\mu}} \Phi(x-n_{\mu})e^{-\partial_{\mu}} e^{+\D_{+\mu}}
- \Phi(x) ]e^{-\D_{+\mu}}
\\[5pt]
&=&
[(\U_{+\mu})_{x, x-n_{\mu}}\Phi(x-n_{\mu})(\U_{+\mu}^{-1})_{x-n_{\mu},x} 
- \Phi(x) ]e^{-\D_{+\mu}},
\label{eq22} \ \ \ \ \ \ \ \ \ \ \\[10pt]
[e^{+\D_{-\mu}}, \Phi(x)] &=& 
e^{+\D_{-\mu}}\Phi(x) -\Phi(x)e^{+\D_{-\mu}} \label{eq1-2}  \\[5pt]
&=& 
[e^{+\D_{-\mu}} e^{-\partial_{\mu}} \Phi(x+n_{\mu})e^{+\partial_{\mu}} e^{-\D_{-\mu}}
- \Phi(x) ]e^{-\D_{+\mu}}
\\[5pt]
&=&
[(\U_{-\mu})_{x, x+n_{\mu}}\Phi(x+n_{\mu})(\U_{-\mu}^{-1})_{x+n_{\mu},x} - \Phi(x) ]e^{+\D_{-\mu}}.
\label{eq22-2} \ \ \ \ \ \ \ \ \ \ 
\end{eqnarray}
Here, the adjoint transformation property of $\Phi(x)$ under gauge transformation 
is naturally obtained.
Noticing that all of the lattice SYM multiplet
derived in the present formulation transform adjointly under gauge transformation,
one may recognize that the above derivation is 
compatible with
all of the lattice SYM multiplet
in the present formulation.

In terms of the bosonic gauge link variables $(\U_{\pm\mu})_{x,x\mp n_{\mu}}$, now the algebra  (\ref{Nabla2-1}) and (\ref{Nabla2-2}) can be expressed as
\begin{eqnarray}
\{\D_{A}(x),\D_{B}(x) \} &=& -f^{\mu}_{AB}\, (\U_{+\mu})_{x,x-n_{\mu}} e^{-\partial_{\mu}}, \\[5pt]
 &=& -f^{\mu}_{AB}\,  e^{-\partial_{\mu}}\, (\U_{+\mu})_{x+n_{\mu},x},
\ \ \ or \\[5pt]
\{\D_{A}(x),\D_{B}(x) \} &=& +f^{\mu}_{AB}\, (\U_{-\mu})_{x,x+n_{\mu}} e^{+\partial_{\mu}}, \\[5pt]
 &=& +f^{\mu}_{AB}\,  e^{+\partial_{\mu}}\, (\U_{-\mu})_{x-n_{\mu},x}
\end{eqnarray}
where we explicitly write the argument $x$ to keep track of the link nature of the elements.
The shift elements in the r.h.s can be consistently divided and rearranged to each of $\D_{A}$ and $\D_{B}$ as
\begin{eqnarray}
&& e^{+a_{A}\cdot \partial}\, \D_{A}(x+a_{B})\, e^{+a_{B}\cdot \partial}\, \D_{B}(x)  \nonumber \\[5pt]
&&\ \ \  +  e^{+a_{B}\cdot \partial}\, \D_{B}(x+a_{A})\, e^{+a_{A}\cdot \partial}\,  \D_{A}(x)
\ = \ -f^{\mu}_{AB}\, (\U_{+\mu})_{x+n_{\mu},x}, \ \ \ \ \  \label{link_ex1}
\end{eqnarray}
in a case where $a_{A}+a_{B} = + n_{\mu}$ is satisfied, or
\begin{eqnarray}
&& e^{+a_{A}\cdot \partial}\, \D_{A}(x+a_{B})\, e^{+a_{B}\cdot \partial}\, \D_{B}(x)  \nonumber \\[5pt]
&&\ \ \  +  e^{+a_{B}\cdot \partial}\, \D_{B}(x+a_{A})\, e^{+a_{A}\cdot \partial}\,  \D_{A}(x)
\ = \ + f^{\mu}_{AB}\, (\U_{-\mu})_{x-n_{\mu},x}, \ \ \ \ \ \label{link_ex2}
\end{eqnarray}
in a case where $a_{A}+a_{B} = - n_{\mu}$ is satisfied.
In the relations (\ref{link_ex1}) and (\ref{link_ex2}), we used the notation $a_{A}\cdot \partial = (a_{A})_{\nu}\, \partial_{\nu}$.
Then, if we express $e^{+a_{A}\cdot \partial}\, \D_{A}(x)$ as $(\D_{A})_{x+a_{A},x}$, and
$e^{+a_{B}\cdot \partial}\, \D_{B}(x)$ as $(\D_{B})_{x+a_{B}, x}$, 
the relations (\ref{link_ex1}) and  (\ref{link_ex2}), can be respectively expressed,
in terms of link anticommutators, as  
\begin{eqnarray}
\{ \D_{A}, \D_{B}\}_{x+a_{A}+a_{B},x}
&=& -f^{\mu}_{AB}\, (\U_{+\mu})_{x+n_{\mu},x} , \ \ {\rm for}\ \  a_{A}+a_{B} = + n_{\mu}, \ \ \ \label{Link_comm1}\\[5pt]
\{ \D_{A}, \D_{B}\}_{x+a_{A}+a_{B},x}
&=& +f^{\mu}_{AB}\,  (\U_{-\mu})_{x-n_{\mu},x} ,  \ \ {\rm for}\ \  a_{A}+a_{B} = - n_{\mu}, \ \ \  \label{Link_comm2}
\end{eqnarray}
which are nothing but the generic expressions of the lattice SYM constraints (\ref{4DSYMlatc1})
-(\ref{4DSYMlatc4}).
In the subsection \ref{lattice_action_section},
the link nature of the lattice SYM constraints (\ref{4DSYMlatc1})-(\ref{4DSYMlatc4}) is introduced as an ansatz, Here, the link nature was systematically derived starting from the exponentiation of the bosonic covariant derivative $\nabla_{\pm\mu}$, (\ref{Nabla2-1}) and (\ref{Nabla2-2}).
Note also that although the above derivation starts from the continuum spacetime, the resulting structure of the link anticommutators (\ref{Link_comm1}) and  (\ref{Link_comm2}) 
does not rely on the continuum spacetime and can be described purely in terms of the discrete lattice coordinates.

It is worthwhile to mention here the rest of the $N=D=4$ lattice SYM constraints (\ref{4DSYMlatc1}) - (\ref{4DSYMlatc7}). Since in  (\ref{4DSYMlatc5}) - (\ref{4DSYMlatc7}), the gauge link objects are absent in the r.h.s., but the scalars $W$ and $F$ are present, one may wonder that above interpretation may not be applied to these scalar sectors. 
However, as we will see in the next section,  the $N=D=4$ lattice SYM constraints (\ref{4DSYMlatc1}) - (\ref{4DSYMlatc7}) can naturally be understood as dimensional reduced constraints from $N=4\ D=5$ where all the constraint relations are given with gauge link objects.
In this sense, the above interpretation may be applicable to the lattice twisted SYM constraints of $N=D=4$ and $N=4\ D=5$. Also, since the lattice twisted SYM constraints of $N=D=2$ and $N=4\ D=3$ are given with gauge link objects as shown in  \cite{DKKN2,DKKN3}, the above interpretation may be applied to these lower dimensional models.

It is also important to mention here a non-commutative aspect of the algebra 
(\ref{Nabla2-1}) and (\ref{Nabla2-2}). Before promoting the bosonic gauge covariant derivatives 
$\D_{\pm\mu}$ to their exponentiations, the algebra (\ref{Nabla1}) is invariant under 
a scale transformation ${\bf{D}}$,
\begin{eqnarray}
[{\bf{D}}, \D_{A}] &=& \frac{1}{2}\D_{A}, \\[2pt]
[{\bf{D}}, \D_{B}] &=& \frac{1}{2}\D_{B}, \\[5pt]
[{\bf{D}}, \D_{\pm\mu}] &=& \D_{\pm\mu}, 
\end{eqnarray}
where we assign the scaling weights $\frac{1}{2}, \frac{1}{2}, 1$ to $\D_{A}, \D_{B}, \D_{\pm\mu}$, respectively.
Once we promote the bosonic gauge covariant derivatives 
$\D_{\pm\mu}$ to their exponentiations,
the algebra (\ref{Nabla2-1}) and (\ref{Nabla2-2}) are not invariant under the scale transformation, since we introduced a particular length which is the lattice constant.
However, if we consider the following transformation,
\begin{eqnarray}
[{\bf{D}}, \D_{A}] &=& d_{A} \D_{A}, \label{Dnabla_cond1}\\[5pt]
[{\bf{D}}, \D_{B}] &=& d_{B} \D_{B}, \label{Dnabla_cond2}
\end{eqnarray}
and, for example, operate finite transformation elements of ${\bf{D}}$ on both sides of (\ref{Nabla2-1}) and (\ref{Nabla2-2}), 
we have
\begin{eqnarray}
e^{d_{A}+d_{B} } \{\D_{A},\D_{B}\} &=& 
- f^{\mu}_{AB}\,
e^{{\bf{D}}} 
e^{-\D_{+\mu}} 
e^{-{\bf{D}}} ,\\[5pt]
e^{d_{A}+d_{B} } \{\D_{A},\D_{B}\} &=& 
+ f^{\mu}_{AB}\,
e^{{\bf{D}}}\, e^{+ \D_{-\mu}}\, e^{-{\bf{D}}},
\end{eqnarray}
Using (\ref{Nabla2-1}) and (\ref{Nabla2-2}) again on the left hand sides, 
we obtain,
\begin{eqnarray}
e^{d_{A}+d_{B} }\, e^{- \D_{+\mu}} &=& 
e^{{\bf{D}}}\, e^{- \D_{+\mu}}\, e^{-{\bf{D}}}, \\[5pt]
e^{d_{A}+d_{B} }\, e^{+ \D_{-\mu}} &=& 
e^{{\bf{D}}}\, e^{+ \D_{-\mu}}\, e^{-{\bf{D}}},
\end{eqnarray}
or equivalently,
\begin{eqnarray} 
e^{{\bf{D}}}\, e^{- \D_{+\mu}} &=& 
e^{d_{A}+d_{B} }\, e^{- \D_{+\mu}}\,
e^{{\bf{D}}}, \\[5pt]
e^{{\bf{D}}}\, e^{+ \D_{-\mu}} &=& 
e^{d_{A}+d_{B} }\, e^{+ \D_{-\mu}}\,
e^{{\bf{D}}}, 
\end{eqnarray}
which can be regarded as Weyl - `tHooft algebra or group commutator relations 
between $\D_{\pm\mu}$ and ${\bf{D}}$,
whose degree of non-commutativity is expressed as $d_{A}+d_{B}$.
The non-commutative nature between ${\bf{D}}$ and $\D_{\pm\mu}$
can further be revealed if we switch off the gauge fields,
\begin{eqnarray} 
e^{{\bf{D}}}\, e^{- \partial_{\mu}}&=& 
e^{d_{A}+d_{B} }\, e^{-\partial_{\mu}}\,
e^{{\bf{D}}},  \ \ for \ \   e^{-\D_{+\mu}},
\\[5pt]
e^{{\bf{D}}}\, e^{+ \partial_{\mu}}&=& 
e^{d_{A}+d_{B} }\, e^{+\partial_{\mu}}\,
e^{{\bf{D}}}, \ \  for  \ \  e^{+\D_{-\mu}},
\end{eqnarray}
which has special solutions 
\begin{eqnarray}
[\partial_{\mu}, {\bf{D}}] &=& d_{A}+d_{B},  \ \ \ \ \ \ for \ \   e^{-\D_{+\mu}}, \label{NCC1}\\[5pt]
[\partial_{\mu}, {\bf{D}}] &=& -(d_{A}+d_{B}),   \ \ for \ \   e^{+\D_{-\mu}}, \label{NCC2}
\end{eqnarray}
which obviously imply that the operator ${\bf{D}}$ can serve as a position operator,
if $d_{A}$ and $d_{B}$ satisfy some particular conditions.
Actually if we express  ${\bf{D}}$ as a vector ${\bf{D}}_{\nu}$, and
require 
\begin{eqnarray}
(d_{A}+d_{B})_{\nu} = +(n_{\mu})_{\nu},   \ \ for \ \   e^{-\D_{+\mu}},  \label{NCC_sol1}\\[5pt]
(d_{A}+d_{B})_{\nu} = -(n_{\mu})_{\nu},   \ \ for \ \   e^{+\D_{-\mu}},  \label{NCC_sol2}
\end{eqnarray}
respectively, then the relations (\ref{NCC1}) and (\ref{NCC2}) can be expressed 
in a unified manner as
\begin{eqnarray}
[\partial_{\mu}, {\bf{D}}_{\nu}] &=& (n_{\nu})_{\mu} \ =\ \delta_{\nu\mu}.
\label{FQC}
\end{eqnarray} 
Noticing that the requirements (\ref{NCC_sol1}) and (\ref{NCC_sol2}) 
are nothing but the lattice Leibniz rule conditions,
 which can be satisfied by the following solutions,
\begin{eqnarray}
d_{A} &=& a_{A}, \ \ \ \ d_{B}\ =\ a_{B},
\end{eqnarray}
one may recognize 
that
the relations (\ref{Dnabla_cond1}) and (\ref{Dnabla_cond2})  expressed as a non-commutative
property between $x$ and $\nabla_{A}$, 
$\nabla_{B}$,
\begin{eqnarray}
[x_{\nu},\D_{A}] &=& (a_{A})_{\nu}\D_{A},  \label{NC_a_nabla}\\[5pt]
[x_{\nu},\D_{B}] &=& (a_{B})_{\nu}\D_{B},  \label{NC_a_nabla2}
\end{eqnarray} 
can be regarded
as a symmetry of the algebra (\ref{Nabla2-1}) and (\ref{Nabla2-2}),
provided
\begin{eqnarray}
e^{x_{\nu}}\, e^{- \D_{+\mu}} &=& 
e^{\delta_{\nu\mu}}\, e^{- \D_{+\mu}}\,  e^{x_{\nu}}, \label{WtHAlg1}  \\[5pt]
e^{x_{\nu}}\, e^{+ \D_{-\mu}} &=& 
e^{-\delta_{\nu\mu}}\, e^{+ \D_{-\mu}}\,  e^{x_{\nu}},\label{WtHAlg2}
 \end{eqnarray}
are satisfied.
We observe that
investigating the  above Weyl - `tHooft algebra (\ref{WtHAlg1}) (\ref{WtHAlg2}) under the presence of the gauge fields may need careful considerations from  superspace aspects and/or 
 matrix model aspects, which is beyond the initial scope of this paper, and will be presented elsewhere.

It is also important to recognize that the above derivation may imply that, after the exponentiation of $\nabla_{\pm\mu}$, the scale transformation generator $\bf{D}$ splits into a corresponding set of position operators each of which is associated with the respective axis of the spacetime coordinates through the first quantization condition (\ref{FQC}).
In other words, it may be expressed that the conformal (scale) direction has been 
turned
into 
a set of coordinate directions. 
Interpreting the lattice coordinates in relation to the scale invariance and the quantization condition in this way may provide new insights into the nature of SUSY-compatible discrete spacetime.
Investigating these aspects may require further  consideration and will be presented elsewhere.

We conclude our consideration of the group and algebraic interpretation of the link formulation
with a couple of immediate observations.
One may observe that in terms of the position operators $x$  satisfying (\ref{NC_a_nabla}), 
the above-mentioned fermionic link covariant derivative
\begin{eqnarray}
(\D_{A})_{x+a_{A},x} &=& e^{+a_{A}\cdot \partial}\, \D_{A}(x), 
\end{eqnarray}
may be regarded as a ``shiftless'' operator in the sense that 
\begin{eqnarray}
[x_{\nu}, e^{+a_{A}\cdot \partial}\, \D_{A}(x)] &=&0.
\end{eqnarray}
Another observation is that in the absence of the gauge fields, the relation (\ref{NC_a_nabla}) may imply  
\begin{eqnarray}
\bigl[x_{\nu}, \, \frac{\partial}{\partial \theta_{A}} - \frac{1}{2}f^{\mu}_{AB}\, \theta_{B}\, e^{-\partial_{\mu}} \bigr]
&=& (a_{A})_{\nu}\, \bigl(\, \frac{\partial}{\partial \theta_{A}} - \frac{1}{2}f^{\mu}_{AB}\, \theta_{B}\, e^{-\partial_{\mu}}\bigr),
\end{eqnarray}
in a case where $a_{A}+a_{B} = + n_{\mu}$ is satisfied, or
\begin{eqnarray}
\bigl[x_{\nu},\,  \frac{\partial}{\partial \theta_{A}} + \frac{1}{2}f^{\mu}_{AB}\, \theta_{B}\, e^{+\partial_{\mu}}\bigr]
&=& (a_{A})_{\nu}\, \bigl(\, \frac{\partial}{\partial \theta_{A}} + \frac{1}{2}f^{\mu}_{AB}\, \theta_{B}\, e^{+\partial_{\mu}}\bigr),
\end{eqnarray}
in a case where $a_{A}+a_{B} = - n_{\mu}$ is satisfied.
In either case, we obtain,
\begin{eqnarray}
[x_{\nu}, \frac{\partial}{\partial \theta_{A}} ] &=&  (a_{A})_{\nu}\frac{\partial}{\partial \theta_{A}}, 
\ \ \ \ 
[x_{\nu}, \theta_{A} ] \ =\ -(a_{A})_{\nu}\, \theta_{A}. 
\end{eqnarray}
Note here that the above non-commutativities between $x_{\nu}$ and $\frac{\partial}{\partial \theta_{A}}$, as well as $x_{\nu}$ and $\theta_{A}$ stem from the presence of  $e^{\pm\D_{\pm\mu}}$ in  the right hand sides of 
the starting algebra (\ref{Nabla2-1}) and (\ref{Nabla2-2}).
In this regard, if we start from a simple anti-commutativity between Grassmann parameters 
$\xi_{A}$ and $\xi_{B}$, 
\begin{eqnarray}
\{ \xi_{A}, \xi_{B}\} &=& 0,
\end{eqnarray}
it is natural to obtain a corresponding Grassmann link parameter, 
\begin{eqnarray}
(\xi_{A})_{x-a_{A},x} &\equiv& e^{-a_{A}\cdot \partial} \xi_{A}, \ \ \ \  
{\rm with}  \ \ \ 
[x_{\nu}, \xi_{A} ] \ =\  0, \label{NC_xi_def}
\end{eqnarray}
which has a global property in the sense of 
\begin{eqnarray}
(\xi_{A})_{x-a_{A},x} &=& (\xi_{A})_{x-a_{A}+a_{B},x+a_{B}},
\end{eqnarray}
and has a non-commutative property in the sense of
\begin{eqnarray}
(\xi_{A})_{x+a_{B}-a_{A},x+a_{B}} (\nabla_{B})_{x+a_{B},x} 
 +
 (\nabla_{B})_{x+a_{B}-a_{A},x-a_{A}} (\xi_{A})_{x-a_{A},x}  
 &=& 0,  
\end{eqnarray}
and has a gauge invariant nature in the sense of 
\begin{eqnarray}
(\xi_{A})_{x-a_{A},x} &\rightarrow& G^{-1}(x-a_{A})(\xi_{A})_{x-a_{A},x}G(x) \\[5pt]
&=& (\xi_{A})_{x-a_{A},x} G^{-1}(x)G(x)\\[5pt] 
&=&  (\xi_{A})_{x-a_{A},x}, \label{gauge_inv_xi2}
\end{eqnarray}
which means that the $(\xi_{A})_{x-a_{A},x}$ defined in (\ref{NC_xi_def})
may serve as the Grassmann link parameter introduced in (\ref{NC_xi_nabla}).

At last, we would like to emphasize that, although the above group and algebraic interpretation may provide a
deep insight into the essence of lattice supersymmetry, 
it does not necessarily mean the 
the continuum spacetime
is required to construct the link formulation of  lattice supersymmetry.
As we have shown in the previous subsection, the link approach, which does not rely on 
the existence of continuum spacetime, already provides 
a sufficient framework to derive a super Yang-Mills multiplet, SUSY transformation laws, 
and SUSY invariance of the resulting action on the lattice.

\section{Five dimensional lift-up of $N=D=4$ SYM}

A general aspect of 16 supercharge system is that
the theory possesses 10 d.o.f. of bosonic fields
including
gauge d.o.f..
For example, $D=10\ N=1$ SYM has
10 components of gauge field
$A_{\mu}\ (\mu =1 \sim 10)$ 
while $D=N=4$ SYM has 
4 components of gauge field
$A_{\mu}\ (\mu = 1\sim 4)$ as well as 6 d.o.f. of scalars.
From the experience of $N=D=2$ and $N=4\ D=3$ SYM on a lattice, it is natural to expect, 
also for the 16 supercharge system, that
each bosonic link variable carries one gauge and one scalar component
of the multiplet. 

This observation leads us to the 
plausible assumption that the 16 supercharge lattice gauge theory
may be most appropriately described by the 'non-unitary' gauge link variable 
of five dimensions,
\begin{eqnarray}
\U_{\pm\mu}=e^{\pm i (A_{\mu}\pm i\phi^{(\mu)})}, \hspace{30pt} (\mu = 1\sim 5),
\end{eqnarray}
which has five components of gauge 
fields $A_{\mu}$
and 
five components of scalar fields $\phi^{(\mu)}$,
respectively.
In the following, the corresponding 
gauge covariant formulations
 of $N=4\ D=5$ SYM with 16 supercharges
in the continuum spacetime and on a lattice are
given.

\subsection{$N=4\ D=5$ SYM Constraints
in the continuum spacetime}
Let us begin with the following $N=4\ D=5$ twisted SYM constraints with 16 fermionic $\D_{A}$'s
in the continuum spacetime,
\begin{eqnarray}
\{\D,\D_{\mu}\} &=& -i\D_{+\mu} \label{c1}\\[2pt]
\{\D_{\rho\sigma},\D_{\mu}\} &=& +i\delta_{\rho\sigma\mu\nu}\D_{-\nu}\label{c2} \\[2pt]
\{\D_{\rho\sigma},\D_{\mu\nu}\} &=& -i\epsilon_{\rho\sigma\mu\nu\lambda}\D_{+\lambda}
\label{c3}
\end{eqnarray}
where $\D_{\pm\mu}\equiv \D_{\underline{\mu}}\pm
\phi^{(\mu)}$,
$\delta_{\mu\nu\rho\sigma}\equiv\delta_{\mu\rho}\delta_{\nu\sigma}
-\delta_{\mu\sigma}\delta_{\nu\rho}$
and $\epsilon_{\mu\nu\rho\sigma\lambda}$ is a five dimensional totally anti-symmetric
tensor with $\epsilon_{12345}\equiv +1$.
\begin{center}
\renewcommand{\arraystretch}{1.4}
\renewcommand{\tabcolsep}{10pt}
\begin{tabular}{l|lll}
& $\D_{\mu\nu}$ & $\D$ & $\D_{\mu}$   \\ \hline
\# of components & $10$ & $1$ &\ $5$ \hspace{20pt}total 16 
\end{tabular}
\end{center}
If one performs a
naive dimensional reduction with
the following identifications
\begin{eqnarray}
\D_{+5} &\equiv& -W,\hspace{30pt}
\D_{\mu 5} \ \equiv\  -\tilde{\D}_{\mu}, \label{DR1}
\\[5pt]
\D_{-5} &\equiv& +F, \hspace{38pt}
\D_{5}\ \equiv\ -\tilde{\D}, \label{DR2}
\end{eqnarray}
the 
five dimensional constraints 
(\ref{c1})-(\ref{c3})
turn out to be the following 
$D=N=4$ SYM constraints,
\begin{eqnarray}
\{\D,\D_{\mu}\} &=& -i\D_{+\mu},\\[2pt]
\{\D_{\rho\sigma},\D_{\mu}\} &=& +i\delta_{\rho\sigma\mu\nu}\D_{-\nu},\\[2pt]
\{\D_{\rho\sigma},\tilde{\D}_{\mu}\} &=& -i\epsilon_{\rho\sigma\mu\nu}
\D_{+\nu},\\[2pt]
\{\tilde{\D},\tilde{\D}_{\mu}\} &=& -i\D_{-\mu},
\end{eqnarray}
\begin{eqnarray}
\{\D,\tilde{\D}\} &=& -iW,\hspace{30pt}
\{\D_{\mu\nu},\D_{\rho\sigma}\} \ =\ +i\epsilon_{\mu\nu\rho\sigma} W,\\
\{\D_{\mu},\tilde{\D}_{\nu}\} &=& -i\delta_{\mu\nu}F,
\end{eqnarray}
where $\mu,\nu,..$ runs from 1 to 4 and again $\D_{\pm\mu}=\D_{\underline{\mu}}
\pm\phi^{(\mu)}$. One can show that $(W,F,\D_{\underline{\mu}},\phi^{(\mu)})$ 
have appropriate $J$ and $R$ rotational properties and they actually consist   
the bosonic part of $N=D=4$ Dirac-K\"ahler twisted gauge  multiplet.
Therefore,
the constraints (\ref{c1})-(\ref{c3}) now can 
be understood as a
five dimensional lift-up of the above  $D=N=4$ SYM constraints.

\subsection{1st step 
Jacobi identities
and definitions of Fermions}
The Jacobi identities resulting from (\ref{c1})-(\ref{c3}) are
given as follows,
\begin{eqnarray}
[\D_{\mu},\D_{+\nu}]+[\D_{\nu},\D_{+\mu}] &=&0 \\[2pt]
[\D_{\rho\sigma},\D_{+\mu}] - \delta_{\rho\sigma\mu\nu}[\D,\D_{-\nu}] &=& 0\\[2pt]
\delta_{\rho\sigma\nu\tau}[\D_{\mu},\D_{-\tau}]
+\delta_{\rho\sigma\mu\tau}[\D_{\nu},\D_{-\tau}] &=& 0 \\[2pt]
\delta_{\alpha\beta\mu\nu}[\D_{\rho\sigma},\D_{-\nu}]
+\delta_{\rho\sigma\mu\nu}[\D_{\alpha\beta},\D_{-\nu}]
-\epsilon_{\alpha\beta\rho\sigma\lambda}[\D_{\mu},\D_{+\lambda}] &=& 0 \\[2pt]
\epsilon_{\mu\nu\alpha\beta\lambda}[\D_{\rho\sigma},\D_{+\lambda}]
+\epsilon_{\alpha\beta\rho\sigma\lambda}[\D_{\mu\nu},\D_{+\lambda}]
+\epsilon_{\rho\sigma\mu\nu\lambda}[\D_{\alpha\beta},\D_{+\lambda}] &=& 0
\end{eqnarray}
from which we can define
\begin{eqnarray}
[\D_{\mu},\D_{+\nu}] &\equiv& +\rho_{\mu\nu} \label{f1} \\[2pt]
[\D_{\mu},\D_{-\nu}] &\equiv& +\delta_{\mu\nu}\rho \label{f2} \\[2pt]
[\D,\D_{-\mu}] &\equiv& +\lambda_{\mu} \label{f3} \\[2pt]
[\D_{\rho\sigma},\D_{+\mu}] &=& +\delta_{\rho\sigma\mu\nu}\lambda_{\nu} 
\label{f4} \\[2pt]
[\D_{\rho\sigma},\D_{-\mu}] &=& -\frac{1}{2}
\epsilon_{\rho\sigma\mu\alpha\beta}\rho_{\alpha\beta} \label{f5} 
\end{eqnarray}
where $\rho_{\mu\nu}$ is anti-symmetric and also
note that $[\D,\D_{+\mu}]=0$ obeys from the starting constraints (\ref{c1}).
In the following, for notational simplicity, we denote the same symbols to represent 
the corresponding lowest component of the superfields. 

\begin{center}
\renewcommand{\arraystretch}{1.4}
\renewcommand{\tabcolsep}{10pt}
\begin{tabular}{l|lll}
& $\rho_{\mu\nu}$ & $\rho$ & $\lambda_{\mu}$   \\ \hline
\# of components & $10$ & $1$ &\ $5$ \hspace{20pt}total 16 
\end{tabular}
\end{center}

\subsection{2nd step 
Jacobi identities
and SUSY trans. laws}
Further operations of $\D_{A}$ on (\ref{f1})-(\ref{f5}) give,
\begin{eqnarray}
\{\D,\rho_{\mu\nu}\} + i[\D_{+\mu},\D_{+\nu}] &=& 0 \\[5pt]
\{\D_{\mu},\rho_{\rho\nu}\} + \{\D_{\rho},\rho_{\mu\nu}\}&=& 0 \\[5pt]
\{\D_{\rho\sigma},\rho_{\mu\nu}\} 
+ i\delta_{\rho\sigma\mu\lambda}[\D_{+\nu},\D_{-\lambda}]
+ \delta_{\rho\sigma\nu\lambda}\{\D_{\mu},\lambda_{\lambda}\} &=& 0 \\[5pt]
\delta_{\mu\nu}\{\D,\rho\} - i[\D_{-\nu},\D_{+\mu}]
+\{\D_{\mu},\lambda_{\nu}\} &=& 0 \\[5pt]
\{\D_{\mu},\rho\} &=& 0 \\[5pt]
\delta_{\mu\nu}\{\D_{\rho\sigma},\rho\}
+i\delta_{\rho\sigma\mu\lambda}[\D_{-\nu},\D_{-\lambda}]
-\frac{1}{2}\epsilon_{\rho\sigma\nu\alpha\beta}
\{\D_{\mu},\rho_{\alpha\beta}\} &=& 0 \\[5pt] 
\{\D,\lambda_{\mu}\} &=& 0 \\[5pt]
\frac{1}{2}\epsilon_{\rho\sigma\nu\alpha\beta}\{\D,\rho_{\alpha\beta}\}
-\{\D_{\rho\sigma},\lambda_{\nu}\} &=& 0\\[5pt]
\delta_{\alpha\beta\mu\nu}\{\D_{\rho\sigma},\lambda_{\nu}\}
-i\epsilon_{\rho\sigma\alpha\beta\lambda}[\D_{+\mu},\D_{+\lambda}]
+\delta_{\rho\sigma\mu\nu}\{\D_{\alpha\beta},\lambda_{\nu}\} &=& 0 \\[5pt]
\frac{1}{2}\epsilon_{\gamma\delta\mu\alpha\beta}\{\D_{\rho\sigma},\D_{\alpha\beta}\}
+i\epsilon_{\rho\sigma\gamma\delta\lambda}[\D_{-\mu},\D_{+\lambda}]
+\frac{1}{2}\epsilon_{\rho\sigma\mu\alpha\beta}
\{\D_{\gamma\delta},\rho_{\alpha\beta}\} &=& 0 
\end{eqnarray}
from which we obtain,
\begin{eqnarray}
\{\D,\lambda_{\mu}\} &=& 0 \\[2pt]
\{\D_{\mu},\lambda_{\nu}\} &=& -i[\D_{+\mu},\D_{-\nu}]
+\frac{i}{2}\delta_{\mu\nu}[\D_{+\rho},\D_{-\rho}] \\[2pt]
\{\D_{\rho\sigma},\lambda_{\mu}\} &=&
-\frac{i}{2}\epsilon_{\rho\sigma\mu\alpha\beta}[\D_{+\alpha},\D_{+\beta}] \\[10pt]
\{\D,\rho\} &=& -\frac{i}{2}[\D_{+\mu},\D_{-\mu}] \\[2pt]
\{\D_{\mu},\rho\} &=& 0 \\[2pt]
\{\D_{\rho\sigma},\rho\} &=& -i[\D_{-\rho},\D_{-\sigma}] \\[15pt]
\{\D,\rho_{\mu\nu}\} &=& -i[\D_{+\mu},\D_{+\nu}] \\[2pt]
\{\D_{\rho},\rho_{\mu\nu}\} &=& -\frac{i}{2}\epsilon_{\rho\mu\nu\alpha\beta}
[\D_{-\alpha},\D_{-\beta}] \\[5pt]
\{\D_{\rho\sigma},\rho_{\mu\nu}\} &=& 
-i\delta_{\rho\sigma\mu\lambda}[\D_{+\nu},\D_{-\lambda}]
+i\delta_{\rho\sigma\nu\lambda}[\D_{+\mu},\D_{-\lambda}]\\
&&+\frac{i}{2}\delta_{\rho\sigma\mu\nu}[\D_{+\lambda},\D_{-\lambda}].
\end{eqnarray}
The resulting SUSY transformation laws for the $N=4\ D=5$ twisted SYM multiplet 
$(\D_{\pm\mu},\lambda_{\mu},\rho,\rho_{\rho\sigma})$
can be read off from, as in the case of $N=D=4$,
\begin{eqnarray}
s_{A} \varphi &\equiv& [\D_{A},\varphi\}|_{\theta's=0}
\end{eqnarray}
where $\D_{A}$ represents any of fermionic supercovariant derivatives 
$(\D,\D_{\mu},\D_{\rho\sigma})$ 
and $s_{A}$ denotes the corresponding supercharge which operates
on the component fields, while
$\varphi$ represents any of the component fields 
$(\D_{\pm\mu},\lambda_{\mu},\rho,\rho_{\rho\sigma})$.
The results are summarized 
in Table \ref{trans}.

\begin{table}
\begin{center}
\renewcommand{\arraystretch}{1.8}
\renewcommand{\tabcolsep}{5pt}
\begin{tabular}{|c||c|c|c|}
\hline
& $s$ & $s_{\rho}$ & $s_{\rho\sigma}$
 \\ \hline
$\D_{+\mu}$ & $0$ & $\rho_{\rho\mu}$ &\ $+\delta_{\rho\sigma\mu\nu}\lambda_{\nu}$ \\ 
$\D_{-\mu}$ & $\lambda_{\mu}$ & $\delta_{\rho\mu}\rho$ 
&$-\frac{1}{2}\epsilon_{\rho\sigma\mu\alpha\beta}\rho_{\alpha\beta}$ \\
$\lambda_{\mu}$ & $0$ & $-i[\D_{+\rho},\D_{-\mu}]
+\frac{i}{2}\delta_{\rho\mu}[\D_{+\lambda},\D_{-\lambda}]$
& $-\frac{i}{2}\epsilon_{\rho\sigma\mu\alpha\beta}[\D_{+\alpha},\D_{+\beta}]$ \\
$\rho$ & $-\frac{i}{2}[\D_{+\lambda},\D_{-\lambda}]$ & $0$
& $-i[\D_{-\rho},\D_{-\sigma}]$ \\
$\rho_{\mu\nu}$ & $-i[\D_{+\mu},\D_{+\nu}]$ 
& $-\frac{i}{2}\epsilon_{\rho\mu\nu\alpha\beta}[\D_{-\alpha},\D_{-\beta}]$
& $-i\delta_{\rho\sigma\mu\lambda}[\D_{+\nu},\D_{-\lambda}]
+i\delta_{\rho\sigma\nu\lambda}[\D_{+\mu},\D_{-\lambda}]$ \\[-5pt]
 & & & $+\frac{i}{2}\delta_{\rho\sigma\mu\nu}[\D_{+\lambda},\D_{-\lambda}]$ \\ \hline
\end{tabular}
\caption{SUSY transformation laws for $(\D_{\pm\mu},\lambda_{\mu},\rho,\rho_{\rho\sigma})$}
\label{trans}
\end{center}
\end{table}

\subsection{On-shell closure of resulting algebra}

One can check that the resulting SUSY algebra for the bosonic components,
 $\D_{\pm\mu}$,
closes without any use of additional constraints,
\begin{eqnarray}
\{s,s_{\rho}\}
\D_{\pm\mu} 
&=& -i[\D_{+\rho},\D_{\pm\mu}] \\[5pt]
\{s_{\rho\sigma},s_{\mu}\}
\D_{\pm\nu}&=& +i\delta_{\rho\sigma\mu\lambda}
[\D_{-\lambda},\D_{\pm\nu}] \\[5pt]
\{s_{\alpha\beta},s_{\rho\sigma}\}
\D_{\pm\mu}&=& -i\epsilon_{\alpha\beta\rho\sigma\lambda}[\D_{+\lambda},\D_{\pm\mu}]\\[5pt]
\{others\}\D_{\pm\mu} &=& 0.
\end{eqnarray}
On the other hand,
for fermions $(\lambda_{\mu},\rho,\rho_{\rho\sigma})$,
SUSY algebra closes, provided some additional constraints
which can be interpreted as equations of motion hold.

More explicitly,
for  $\lambda_{\mu}$, one  obtains
after some calculations,
\begin{eqnarray}
\{s,s_{\rho}\}
\lambda_{\mu} &=&
-i[\D_{+\rho},\lambda_{\mu}] 
\ + \frac{i}{2}\delta_{\rho\mu}[\D_{+\lambda},\lambda_{\lambda}]\\[5pt]
\{s_{\rho\sigma},s_{\mu}\}
\lambda_{\nu} &=&
+\delta_{\rho\sigma\mu\lambda}[\D_{-\lambda},\lambda_{\nu}] \nonumber \\
&& -i\delta_{\rho\sigma\mu\lambda}
([\D_{-\lambda},\lambda_{\nu}]-[\D_{-\nu},\lambda_{\lambda}]
-\frac{1}{2}\epsilon_{\lambda\nu\alpha\beta\gamma}
[\D_{+\alpha},\rho_{\beta\gamma}]) \nonumber \\
&& -\frac{i}{2}\delta_{\mu\nu}
([\D_{-\rho},\lambda_{\sigma}]-[\D_{-\sigma},\lambda_{\rho}]
-\frac{1}{2}\epsilon_{\rho\sigma\alpha\beta\gamma}
[\D_{+\alpha},\rho_{\beta\gamma}])
\\[5pt]
\{s_{\alpha\beta},s_{\rho\sigma}\}
\lambda_{\mu} &=&
-i\epsilon_{\alpha\beta\rho\sigma\lambda}[\D_{+\lambda},\lambda_{\mu}]
\ +i\epsilon_{\alpha\beta\rho\sigma\mu}[\D_{+\tau},\lambda_{\tau}] \\[10pt]
\{s,s\}
\lambda_{\mu} &=&0 \\[5pt]
\{s,s_{\rho\sigma}\} 
\lambda_{\mu} &=& 0 \\[2pt]
\{s_{\rho},s_{\sigma}\} 
\lambda_{\mu}
&=& 
-\frac{i}{2}\delta_{\rho\mu}([\D_{+\sigma},\rho]+[\D_{-\lambda},\rho_{\sigma\lambda}]) \nonumber \\
&&-\frac{i}{2}\delta_{\sigma\mu}([\D_{+\rho},\rho]+[\D_{-\lambda},\rho_{\rho\lambda}]),
\end{eqnarray}
while as for $\rho$,
\begin{eqnarray}
\{s,s_{\rho}\}
\rho &=& -i[\D_{+\rho},\rho]
\ +\frac{i}{2}([\D_{+\rho},\rho]+[\D_{-\mu},\rho_{\rho\mu}])\\[5pt]
\{s_{\rho\sigma},s_{\mu}\}
\rho &=& 
+i\delta_{\rho\sigma\mu\nu}[\D_{-\nu},\rho] \\[5pt]
\{s_{\alpha\beta},s_{\rho\sigma}\}
\rho &=&
-i\epsilon_{\alpha\beta\rho\sigma\lambda}[\D_{+\lambda},\rho]
\ +i\epsilon_{\alpha\beta\rho\sigma\lambda}
([\D_{+\lambda},\rho]+[\D_{-\mu},\rho_{\lambda\mu}]) \\[10pt]
\{s,s\}
\rho &=& -i[\D_{+\mu},\lambda_{\mu}] \\[5pt]
\{s,s_{\rho\sigma}\}
\rho &=& 
-\frac{i}{2}([\D_{-\rho},\lambda_{\sigma}]-[\D_{-\sigma},\lambda_{\rho}]
-\frac{1}{2}\epsilon_{\rho\sigma\mu\alpha\beta}[\D_{+\mu},\rho_{\alpha\beta}])\\[5pt]
\{s_{\rho},s_{\sigma}\}
\rho &=& 0.
\end{eqnarray}

As for
$\rho_{\mu\nu}$,
\begin{eqnarray}
\{s,s_{\rho}\}
\rho_{\mu\nu} &=& -i[\D_{+\rho},\rho_{\mu\nu}] \nonumber \\
&&-\frac{i}{2}\epsilon_{\rho\mu\nu\alpha\beta}
([\D_{-\alpha},\lambda_{\beta}]-[\D_{-\beta},\lambda_{\alpha}]
-\frac{1}{2}\epsilon_{\alpha\beta\lambda\gamma\delta}[\D_{+\lambda},\rho_{\gamma\delta}])
\\[5pt]
\{s_{\rho\sigma}, s_{\mu}\}
\rho_{\alpha\beta} &=&
+i\delta_{\rho\sigma\mu\nu}[\D_{-\nu},\rho_{\alpha\beta}] \nonumber \\
&&+\frac{i}{2}\delta_{\rho\sigma\alpha\beta}
([\D_{-\lambda},\rho_{\mu\lambda}]+[\D_{+\mu},\rho]) \nonumber \\[5pt]
&&+i\delta_{\rho\sigma\mu\alpha}
([\D_{-\lambda},\rho_{\beta\lambda}]+[\D_{+\beta},\rho]) \nonumber \\[5pt]
&&-i\delta_{\rho\sigma\mu\beta}
([\D_{-\lambda},\rho_{\alpha\lambda}]+[\D_{+\alpha},\rho])\\[8pt]
\{s_{\alpha\beta},s_{\rho\sigma}\}
\rho_{\mu\nu} &=&
-i\epsilon_{\alpha\beta\rho\sigma\lambda}[\D_{+\lambda},\rho_{\mu\nu}] \nonumber \\
&&-\frac{i}{2}\delta_{\alpha\beta\mu\nu}
([\D_{-\rho},\lambda_{\sigma}]-[\D_{-\sigma},\lambda_{\rho}]
-\frac{1}{2}\epsilon_{\rho\sigma\lambda\tau\kappa}
[\D_{+\lambda},\rho_{\tau\kappa}]) \nonumber \\
&&-\frac{i}{2}\delta_{\rho\sigma\mu\nu}
([\D_{-\alpha},\lambda_{\beta}]-[\D_{-\beta},\lambda_{\alpha}]
-\frac{1}{2}\epsilon_{\alpha\beta\lambda\tau\kappa}
[\D_{+\lambda},\rho_{\tau\kappa}]) \nonumber \\
&&+i\delta_{\alpha\rho\mu\nu}
([\D_{-\beta},\lambda_{\sigma}]-[\D_{-\sigma},\lambda_{\beta}]
-\frac{1}{2}\epsilon_{\beta\sigma\lambda\tau\kappa}
[\D_{+\lambda},\rho_{\tau\kappa}]) \nonumber \\
&&-i\delta_{\beta\rho\mu\nu}
([\D_{-\alpha},\lambda_{\sigma}]-[\D_{-\sigma},\lambda_{\alpha}]
-\frac{1}{2}\epsilon_{\alpha\sigma\lambda\tau\kappa}
[\D_{+\lambda},\rho_{\tau\kappa}]) \nonumber \\
&&-i\delta_{\alpha\sigma\mu\nu}
([\D_{-\beta},\lambda_{\rho}]-[\D_{-\rho},\lambda_{\beta}]
-\frac{1}{2}\epsilon_{\beta\rho\lambda\tau\kappa}
[\D_{+\lambda},\rho_{\tau\kappa}]) \nonumber \\
&&+i\delta_{\beta\sigma\mu\nu}
([\D_{-\alpha},\lambda_{\rho}]-[\D_{-\rho},\lambda_{\alpha}]
-\frac{1}{2}\epsilon_{\alpha\rho\lambda\tau\kappa}
[\D_{+\lambda},\rho_{\tau\kappa}])\\[5pt]
\{s,s\}
\rho_{\mu\nu} &=& 0 \\[2pt]
\{s,s_{\alpha\beta}\}
\rho_{\mu\nu} &=&
+\frac{i}{2}\delta_{\alpha\beta\mu\nu}[\D_{+\lambda},
\lambda_{\lambda}
]\\[5pt]
\{s_{\rho},s_{\sigma}\}
\rho_{\mu\nu} &=& 0.
\end{eqnarray}
All of the above relations form closed SUSY
algebra for the fermionic components
$\varphi = (\lambda_{\mu},\rho,\rho_{\mu\nu})$,
\begin{eqnarray}
\{s,s_{\rho}\}
\varphi &=& -i[\D_{+\rho},\varphi] \\[5pt]
\{s_{\rho\sigma},s_{\mu}
\}\varphi &=&
+i\delta_{\rho\sigma\mu\nu}[\D_{-\nu},\varphi]\\[5pt]
\{s_{\alpha\beta},s_{\rho\sigma}\}
\varphi &=&
-\epsilon_{\alpha\beta\rho\sigma\lambda}
[\D_{+\lambda},\varphi]
\end{eqnarray}
provided the following additional constraints holds,
\begin{eqnarray}
[\D_{+\mu},\lambda_{\mu}] &=& 0 \label{eom1} \\[5pt]
[\D_{+\mu},\rho] + [\D_{-\nu},\rho_{\mu\nu}] &=& 0 \label{eom2} \\[5pt]
[\D_{-\mu},\lambda_{\nu}]-[\D_{-\nu},\lambda_{\mu}]
-\frac{1}{2}\epsilon_{\mu\nu\alpha\beta\gamma}
[\D_{+\alpha},\rho_{\beta\gamma}] &=& 0, \label{eom3}
\end{eqnarray}
which will turn out to be the equations of motion for the resulting action
(\ref{5D_action_cont}).

It is worthwhile to note that in \cite{Catterall:2013roa}, twisted SUSY transformations of the Dirac-K\"ahler twisted $N=4\ D=5$ SYM multiplet in the continuum spacetime have been worked out by paying a particular attention to R symmetries. 
However, here in the present formulation, the whole algebraic structure of Dirac-K\"ahler twisted SUSY of $N=4\ D=5$ SYM
has clearly and entirely been revealed due to our manifestly covariant formulation.

\subsection{Construction of the Action
in the continuum spacetime
}

Regarding the relations (\ref{eom1})-(\ref{eom3}) as
the equations of motion of the system,
one can determine the fermion kinetic terms of the action
as
\begin{eqnarray}
S_{F} =
\int d^{5}x \ {\rm tr}\ \biggl[
 i\lambda_{\mu}[\D_{+\mu},\rho]
+i \lambda_{\mu}[\D_{-\nu},\rho_{\mu\nu}]
+\frac{i}{8}\epsilon_{\mu\nu\lambda\rho\sigma}
\rho_{\mu\nu}[\D_{+\lambda},\rho_{\rho\sigma}]
\biggr],
\end{eqnarray}
and further observing the 
$s$ transformation of $\mathcal{L}_{F}$,
one can also uniquely determine the bosonic part of the action
in such a way that the sum of those vanishes under the operation of $s$.
Namely,
\begin{eqnarray}
S_{B}
=
\int d^{5}x \ {\rm tr}\ \biggl[
\frac{1}{2}[\D_{+\mu},\D_{+\nu}][\D_{-\mu},\D_{-\nu}]
-\frac{1}{4}[\D_{+\mu},\D_{-\mu}][\D_{+\nu},\D_{-\nu}]
\biggr],
\end{eqnarray}
\begin{eqnarray}
S^{N=4\, D=5}_{TSYM}
&=&
S_{B}+S_{F}
\\
&=& 
\int d^{5}x \ {\rm tr}\ \biggl[
\frac{1}{2}[\D_{+\mu},\D_{+\nu}][\D_{-\mu},\D_{-\nu}]
-\frac{1}{4}[\D_{+\mu},\D_{-\mu}][\D_{+\nu},\D_{-\nu}] \nonumber \\
&&+i\lambda_{\mu}[\D_{+\mu},\rho]
+i \lambda_{\mu}[\D_{-\nu},\rho_{\mu\nu}]
+\frac{i}{8}\epsilon_{\mu\nu\lambda\rho\sigma}
\rho_{\mu\nu}[\D_{+\lambda},\rho_{\rho\sigma}]
\biggr].
\label{5D_action_cont}
\end{eqnarray}
One can show the total action (\ref{5D_action_cont})
is invariant 
under all the supercharges $s_{A}=(s,  s_{\mu},s_{\mu\nu})$,
with use of trace properties and Jacobi identities, namely
\begin{eqnarray}
\delta_{A} S^{N=4\,D=5}_{TSYM} (\varphi) &\equiv&
S^{N=4\,D=5}_{TSYM} (\varphi+\delta_{A}\varphi) - S^{N=4\,D=5}_{TSYM} (\varphi) \\[5pt]
&=&0,
\end{eqnarray}
where $\varphi$ represents any of the component fields 
$(\D_{\pm\mu},\rho,\lambda_{\mu},\rho_{\rho\sigma})$,
and,
as in the case of $N=D=4$, 
we define the variation of the component fields as
\begin{eqnarray}
\delta_{A}\varphi &\equiv&
\xi_{A}
(s_{A}\varphi) \\[5pt]
&=&
\xi_{A}
[\D_{A},\varphi\},
\end{eqnarray}
where we introduced
a set of Grassmann parameters $\xi_{A} = (\xi, \xi_{\mu}, \xi_{\mu\nu})$,
each of which anti-commutes with any of the fermionic covariant derivatives $\nabla_{B}=(\D,\D_{\mu},\D_{\mu\nu})$,
\begin{eqnarray}
\{ \xi_{A}, \nabla_{B} \}
 &=& 0.
\end{eqnarray}
As in the case of $N=D=4$, the invariance of the action under the operations of twisted supercharges can be shown without use of equations of motion, although the closure of the algebra holds up to equations of motion.

\subsection{$N=4\ D=5$ SYM constraints on a lattice}
A lattice implementation of the starting constraints (\ref{c1})-(\ref{c3})
can be done, as in the case of $N=D=4$, 
 through replacing the fermionic covariant derivatives $\D_{A}=(\D,\D_{\mu},\D_{\mu\nu})$ by fermionic link covariant derivatives,
\begin{eqnarray}
\D_{A} \ \rightarrow\  (\D_{A})_{x+a_{A},x},
\end{eqnarray}
and replacing the bosonic covariant derivatives $\D_{\pm\mu}$ by 
bosonic gauge link variables,
\begin{eqnarray}
\D_{\pm\mu} \ \rightarrow\ \mp (\U_{\pm\mu})_{x\pm n_{\mu},x} \ = \ 
\mp (e^{\pm i (A_{\mu}\pm i\phi^{(\mu)})})_{x\pm n_{\mu},x}.
\end{eqnarray}
We then have the following $N=5\ D=4$ Twisted SYM constraints on a lattice,
\begin{eqnarray}
\{\D,\D_{\mu}\}_{x+a+a_{\mu},x} &=& +i(\U_{+\mu})_{x+n_{\mu},x} \label{c1_lat}\\[2pt]
\{\D_{\rho\sigma},\D_{\mu}\}_{x+a_{\rho\sigma}+a_{\mu},x}
 &=& +i\delta_{\rho\sigma\mu\nu}(\U_{-\nu})_{x-n_{\nu},x}\label{c2_lat} \\[2pt]
\{\D_{\rho\sigma},\D_{\mu\nu}\}_{x+a_{\mu\nu}+a_{\rho\sigma}} &=& + i\epsilon_{\rho\sigma\mu\nu\lambda}(\U_{+\lambda})_{x+n_{\lambda},x}.
\label{c3_lat}
\end{eqnarray}
Lattice Leibniz rule conditions, or equivalently, gauge covariant conditions for
the lattice SYM constraints
(\ref{c1_lat})-(\ref{c3_lat}) requires 
\begin{eqnarray}
a+a_{\mu} &=& + n_{\mu} \label{l1} \\
a_{\mu\nu}+a_{\mu} &=& - n_{\nu} \label{l2}\\
a_{\mu\nu} + a_{\rho\sigma} &=& + |\epsilon_{\mu\nu\rho\sigma\lambda}|
n_{\lambda}, \hspace{20pt} for\ 
\mu, \nu, \rho, \sigma\  {\it all\ different\ each\ other}
\label{l3}.
\end{eqnarray} 
From (\ref{l1}) one gets, 
$a_{\mu} = +n_{\mu} - a$ and
further from (\ref{l2}),
$ a_{\mu\nu} = -n_{\mu} -n_{\nu} + a $.
Then (\ref{l3}) requires
$-n_{\mu}-n_{\nu}-n_{\rho}-n_{\sigma}+2a
=+|\epsilon_{\mu\nu\rho\sigma\lambda}|n_{\lambda}$
which leads 
\begin{eqnarray}
a &=& \frac{1}{2}\sum_{\lambda=1}^{5} n_{\lambda}.
\end{eqnarray}
Namely, for (\ref{c1_lat})-(\ref{c3_lat}), the $N=4\ D=5$ twisted 
SYM constraints with 16 supercharges, 
only the  following
symmetric choice 
is allowed to be consistent with
the five dimensional lattice Leibniz rule.
Note that, in the symmetric choice, $a_{A}$'s correspond to 16 vertices symmetically 
selected from 32 vertices of a 5-cube (5 dimensional hypercube).  
\begin{center}
\underline{Symmetric choice ($N=4\ D=5$)}
\end{center}
\begin{eqnarray}
a &=& (+\frac{1}{2},+\frac{1}{2},+\frac{1}{2},+\frac{1}{2},+\frac{1}{2}), \hspace{20pt}
a_{1}\ =\ (+\frac{1}{2},-\frac{1}{2},-\frac{1}{2},-\frac{1}{2},-\frac{1}{2}), \label{5D_Symm1} \\[2pt]
a_{12} &=& (-\frac{1}{2},-\frac{1}{2},+\frac{1}{2},+\frac{1}{2},+\frac{1}{2}), \hspace{20pt}
a_{2}\ =\ (-\frac{1}{2},+\frac{1}{2},-\frac{1}{2},-\frac{1}{2},-\frac{1}{2}), \\[2pt] 
a_{13} &=& (-\frac{1}{2},+\frac{1}{2},-\frac{1}{2},+\frac{1}{2},+\frac{1}{2}), \hspace{20pt}
a_{3}\ =\ (-\frac{1}{2},-\frac{1}{2},+\frac{1}{2},-\frac{1}{2},-\frac{1}{2}), \\[2pt] 
a_{14} &=& (-\frac{1}{2},+\frac{1}{2},+\frac{1}{2},-\frac{1}{2},+\frac{1}{2}), \hspace{20pt}
a_{4}\ =\ (-\frac{1}{2},-\frac{1}{2},-\frac{1}{2},+\frac{1}{2},-\frac{1}{2}), \\[2pt] 
a_{15} &=& (-\frac{1}{2},+\frac{1}{2},+\frac{1}{2},+\frac{1}{2},-\frac{1}{2}), \hspace{20pt}
a_{5}\ =\ (-\frac{1}{2},-\frac{1}{2},-\frac{1}{2},-\frac{1}{2},+\frac{1}{2}), \\[2pt] 
a_{23} &=& (+\frac{1}{2},-\frac{1}{2},-\frac{1}{2},+\frac{1}{2},+\frac{1}{2}), \hspace{20pt}
a_{45}\ =\ (+\frac{1}{2},+\frac{1}{2},+\frac{1}{2},-\frac{1}{2},-\frac{1}{2}), \\[2pt] 
a_{24} &=& (+\frac{1}{2},-\frac{1}{2},+\frac{1}{2},-\frac{1}{2},+\frac{1}{2}), \hspace{20pt}
a_{35}\ =\ (+\frac{1}{2},+\frac{1}{2},-\frac{1}{2},+\frac{1}{2},-\frac{1}{2}), \\[2pt]
a_{25} &=& (+\frac{1}{2},-\frac{1}{2},+\frac{1}{2},+\frac{1}{2},-\frac{1}{2}), \hspace{20pt}
a_{34}\ =\ (+\frac{1}{2},+\frac{1}{2},-\frac{1}{2}, \label{5D_Symm8}
-
\frac{1}{2},
+
\frac{1}{2}). 
\end{eqnarray}

We can repeat basically the same calculations with the continuum case,
with the understanding of the link (anti)commutators, and derive the twisted SUSY transformation laws on a lattice which are summarized in Table \ref{trans_lat}
where the link indices are omitted for notational simplicity.

As in the case of $N=D=4$ super Yang-Mills multiplet, 
one can also show that
link properties of the non-vanishing fermionic components 
$(\rho,\lambda_{\mu},\rho_{\mu\nu})$
are just opposite to those of 
$(\D,\D_{\mu},\D_{\mu\nu})$,
which are summarized in Table \ref{D=5N=4shift_fermions}.
\textcolor{blue}{
\begin{table}
\begin{center}
\renewcommand{\arraystretch}{1.4}
\renewcommand{\tabcolsep}{6pt}
\begin{tabular}{c|c|ccc|ccc}
& $\U_{\pm\mu}$ & $\rho$ & $\lambda_{\mu}$ & $\rho_{\mu\nu}$ 
& $\D$ & $\D_{\mu}$ & $\D_{\mu\nu}$  \\ \hline
shift & $\pm n_{\mu}$ & 
$-a$ & $-a_{\mu}$ & $-a_{\mu\nu}$ 
& $+a$ & $+a_{\mu}$ & $+a_{\mu\nu}$ 
\end{tabular}
\caption{Shift nature for  Dirac-K\"ahler twisted $D=5\ N=4$ lattice SYM multiplet}
\label{D=5N=4shift_fermions}
\end{center}
\end{table}
}

\begin{table}[htbp]
\begin{center}
\renewcommand{\arraystretch}{1.8}
\renewcommand{\tabcolsep}{5pt}
\begin{tabular}{|c||c|c|c|}
\hline
& $s$ & \ $s_{\rho}$ & $s_{\rho\sigma}$
 \\ \hline
$\U_{+\mu}$ & $0$ & $-\rho_{\rho\mu}$ &\ $-\delta_{\rho\sigma\mu\nu}\lambda_{\nu}$ \\ 
$\U_{-\mu}$ & $\lambda_{\mu}$ & $\delta_{\rho\mu}\rho$ 
&$-\frac{1}{2}\epsilon_{\rho\sigma\mu\alpha\beta}\rho_{\alpha\beta}$ \\
$\lambda_{\mu}$ & $0$ & $+i[\U_{+\rho},\U_{-\mu}]
-\frac{i}{2}\delta_{\rho\mu}[\U_{+\lambda},\U_{-\lambda}]$
& $-\frac{i}{2}\epsilon_{\rho\sigma\mu\alpha\beta}[\U_{+\alpha},\U_{+\beta}]$ \\
$\rho$ & $+\frac{i}{2}[\U_{+\lambda},\U_{-\lambda}]$ & $0$
& $-i[\U_{-\rho},\U_{-\sigma}]$ \\
$\rho_{\mu\nu}$ & $-i[\U_{+\mu},\U_{+\nu}]$ 
& $-\frac{i}{2}\epsilon_{\rho\mu\nu\alpha\beta}[\U_{-\alpha},\U_{-\beta}]$
& $+i\delta_{\rho\sigma\mu\lambda}[\U_{+\nu},\U_{-\lambda}]
-i\delta_{\rho\sigma\nu\lambda}[\U_{+\mu},\U_{-\lambda}]$ \\[-5pt]
 & & & $-\frac{i}{2}\delta_{\rho\sigma\mu\nu}[\U_{+\lambda},\U_{-\lambda}]$ \\ \hline
\end{tabular}
\caption{
Lattice SUSY transformation laws for $(\U_{\pm\mu},\lambda_{\mu},\rho,\rho_{\rho\sigma})$
}
\label{trans_lat}
\end{center}
\end{table}

One can also show that the resulting twisted $N=4\ D=5$ SUSY algebra for the component
fields closes on-shell,
\begin{eqnarray}
\{s,s_{\mu}\}(\varphi)_{x+a_{\varphi},x} 
&\dot{=}& +i[\U_{+\mu},\varphi]_{x+n_{\mu}+a_{\varphi},x},
\label{5DSYMlatalgf1}\\[2pt]
\{s_{\rho\sigma},s_{\mu}\}(\varphi)_{x+a_{\varphi},x} 
&\dot{=}& +i\delta_{\rho\sigma\mu\nu}[\U_{-\nu},\varphi]_{x-n_{\nu}+a_{\varphi},x},
\label{5DSYMlatalgf2}\\[2pt]
\{s_{\rho\sigma},s_{\mu\nu}\}(\varphi)_{x+a_{\varphi},x} 
&\dot{=}& +i\epsilon_{\rho\sigma\mu\nu\lambda}
[\U_{+\lambda},\varphi]_{x+n_{\lambda}+a_{\varphi},x},\label{5DSYMlatalgf3} \\[2pt]
\{others\}(\varphi)_{x+a_{\varphi},x} 
&\dot{=}& 0,
\label{5DSYMlatalgf4}
\end{eqnarray}
namely, up to the equations of motion,
\begin{eqnarray}
[\U_{+\mu},\lambda_{\mu}] &=& 0 \label{eom1_lat} \\[5pt]
[\U_{+\mu},\rho] - [\U_{-\nu},\rho_{\mu\nu}] &=& 0 \label{eom2_lat} \\[5pt]
[\U_{-\mu},\lambda_{\nu}]-[\U_{-\nu},\lambda_{\mu}]
+\frac{1}{2}\epsilon_{\mu\nu\alpha\beta\gamma}
[\U_{+\alpha},\rho_{\beta\gamma}] &=& 0, \label{eom3_lat}
\end{eqnarray}
where all the link indices are omitted for notational simplicity.

The corresponding $N=4\ D=5$ twisted SYM action on a lattice can be derived as
\begin{eqnarray}
S^{N=4\, D=5}_{lat.\ TSYM}
&=&
\sum_{x}  \ {\rm tr}\ \biggl[
\frac{1}{2}[\U_{+\mu},\U_{+\nu}]_{x,x-n_{\mu}-n_{\nu}}[\U_{-\mu},\U_{-\nu}]_{x-n_{\mu}-n_{\nu},x}
\nonumber  \\[5pt]
&&-\frac{1}{4}[\U_{+\mu},\U_{-\mu}]_{x,x}[\U_{+\nu},\U_{-\nu}]_{x,x} 
-i(\lambda_{\mu})_{x,x+a_{\mu}}[\U_{+\mu},\rho]_{x+a_{\mu},x} \nonumber \\[5pt]
&&+i (\lambda_{\mu})_{x,x+a_{\mu}}[\U_{-\nu},\rho_{\mu\nu}]_{x+a_{\mu},x}
-\frac{i}{8}\epsilon_{\mu\nu\lambda\rho\sigma}
(\rho_{\mu\nu})_{x,x+a_{\mu\nu}}[\U_{+\lambda},\rho_{\rho\sigma}]_{x+a_{\mu\nu},x}
\biggr]. \nonumber \\
\label{5D_action_lat}
\end{eqnarray}
One can show the total action (\ref{5D_action_lat})
is invariant 
under all the supercharges $s_{A}=(s,  s_{\mu},s_{\mu\nu})$,
with use of trace properties and Jacobi identities, 
\begin{eqnarray}
\delta_{A} S^{N=4\,D=5}_{lat.\ TSYM} (\varphi) &\equiv&
S^{N=4\,D=5}_{lat.\ TSYM} (\varphi+\delta_{A}\varphi) - S^{N=4\,D=5}_{lat.\ TSYM} (\varphi) \\[5pt]
&=&0,
\end{eqnarray}
or more explicitly,
\begin{eqnarray}
\delta\, S^{N=D=4}_{lat.\ TSYM} & = &
\delta_{\mu}\, S^{N=D=4}_{lat.\ TSYM} \ = \ 
\delta_{\mu\nu}\, S^{N=D=4}_{lat.\ TSYM} \ =\ 
0, 
\end{eqnarray}
where $\varphi$ represents any of the component fields 
$(\U_{\pm\mu},\rho,\lambda_{\mu},\rho_{\rho\sigma})$,
and,
as in the case of $N=D=4$, 
we define the variation of the component fields on a lattice as
\begin{eqnarray}
(\delta_{A}\varphi)_{x+a_{\varphi},x} &\equiv&
(\xi_{A})_{x+a_{\varphi},x+a_{\varphi}+a_A}
(s_{A}\varphi)_{x+a_{\varphi}+a_A,x}
\label{deltaphi-def} \\
&=& 
(\xi_{A})_{x+a_{\varphi},x+a_{\varphi}+a_A}
s_{A}(\varphi)_{x+a_{\varphi},x} \\
&=&
(\xi_{A})_{x+a_{\varphi},x+a_{\varphi}+a_A}
[\D_{A},\varphi\}_{x+a_{\varphi}+a_A,x},
\end{eqnarray}
where $(\varphi)_{x+a_{\varphi},x}$ again denotes any components of the lattice multiplet
with explicit link indices.
In the above definition of the variation of the component fields, we introduced 
a set of Grassmann parameters $\xi_{A} = (\xi, \xi_{\mu}, \xi_{\mu\nu})$
with link nature from $x$ to $x-a_{A}$ 
which is an opposite link nature of the corresponding fermionic covariant derivative 
$(\nabla_{A})_{x+a_{A},x}$.
We also assume that, as in the case of $N=D=4$, the Grassmann parameter $(\xi_{A})_{x-a_{A},x}$ 
anti-commutes with any of the fermionic covariant derivatives $\nabla_{B}=(\D,\D_{\mu},\D_{\mu\nu})$
in the sense of link (anti)commutators,
\begin{eqnarray}
\{ \xi_{A}, \nabla_{B} \}_{x+a_{B}-a_{A},x}
 &=&
(\xi_{A})_{x+a_{B}-a_{A},x+a_{B}} (\nabla_{B})_{x+a_{B},x} \nonumber \\[5pt] 
&& +
 (\nabla_{B})_{x+a_{B}-a_{A},x-a_{A}} (\xi_{A})_{x-a_{A},x} \nonumber \label{NC_xi_nabla1} \\[5pt]
 &=& 0.
\end{eqnarray}
In order to support the above calculations, 
the summation over $x$ in  the lattice action (\ref{5D_action_lat})
needs to appropriately cover the coresponding lattice  structure
which may be expressed as both of the integer sites and half-integer sites, 
\begin{eqnarray}
\sum_{x} &=& \sum_{(m_1,m_2,m_3,m_4,m_5)}
+\sum_{(m_1+\frac{1}{2},m_2+\frac{1}{2},m_3+\frac{1}{2},m_4+\frac{1}{2},,m_5+\frac{1}{2})}, 
\label{Sum_x_1}
\end{eqnarray}
where each of $m_1,m_2,m_3,m_4,m_5$ denotes any integer.

Regarding the five-dimensional formulation, it is important to mention the 
aspect of fermion 
doubling.
In four dimension, as we have seen, 
the number of 
vertices of 
dual
hypercube is 16 which coincides with
the number of 
species doubler
lattice fermions, and the $N=D=4$  
super Yang-Mills can be
formulated via Dirac-K\"ahler mechanism
without chiral fermion problems
since the doubler fermions can be identified as physical supermultiplet.
Here, in five-dimensions, the number of vertices of five-dimensional 
dual
hypercube is 32, 
while the half of the vertices are assigned to the lattice supercharges or lattice fermions
as explicitly indicated in (\ref{5D_Symm1})-(\ref{5D_Symm8}).
It is non-trivial to ask how the fermion doubilng problem is 
circumvented 
in five dimensions.
In this regard, in \cite{Joseph_D=5},
bosonic and fermionic propagators of the lattice theory have been explicitly calculated 
in the context of twisted SUSY,
and it was shown that the theory does not suﬀer from spectrum doublers.
Therefore, as a future study, it is important to look at these aspects explicitly
in  our link formulation.

\section{Our reply to the criticism}

\indent
The group and algebraic interpretation of the link approach
given in the subsection \ref{GAI}
 systematically resolves the ordering ambiguity \cite{Bruckmann,BKC} mentioned in the Introduction. Let us illustrate the essence of the resolution with some explicit computations below.

Firstly, as shown  in the subsection \ref{GAI}, 
promoting bosonic supercovariant derivatives to their exponentials (\ref{Nabla2-1})-(\ref{Nabla2-2})
consistently with the lattice Leibniz rule
naturally gives rise to the notion of gauge covariant link (anti-)commutators (\ref{Link_comm1})-(\ref{Link_comm2}).
This provides an explanation of how and why the two fundamental concepts of 
non-commutativity and link nature are tightly related each other.
From this point of view, the non-commutativity relations
(\ref{non-com-rel1}) are naturally understood as link commutative relations between
$(Q \phi_{1})_{x+a_{Q},x}$ and $(\phi_{2})_{x,x}$, and  between
$(Q \phi_{2})_{x+a_{Q},x}$ and $(\phi_{1})_{x,x}$,
\begin{eqnarray}
(Q \phi_{1})_{x+a_{Q},x}(\phi_{2})_{x,x} - (\phi_{2})_{x+a_{Q},x+a_{Q}}(Q \phi_{1})_{x+a_{Q},x} &=& 0, \\[5pt]
(Q \phi_{2})_{x+a_{Q},x}(\phi_{1})_{x,x} - (\phi_{1})_{x+a_{Q},x+a_{Q}}(Q \phi_{2})_{x+a_{Q},x} &=& 0.
\end{eqnarray}
The relations
(\ref{ord-amb}) are also naturally understood as link equations as follows:
\begin{eqnarray}
\bigl(Q (\phi_{1}\phi_{2})\bigr)_{x+a_{Q},x}
&=& (Q \phi_{1})_{x+a_{Q},x}(\phi_{2})_{x,x} + (\phi_{1})_{x+a_{Q},x+a_{Q}}(Q \phi_{2})_{x+a_{Q},x}, 
\label{chap7_link1}
\\[5pt]
\bigl(Q (\phi_{2}\phi_{1})\bigr)_{x+a_{Q},x}
&=& (Q \phi_{2})_{x+a_{Q},x}(\phi_{1})_{x,x} + (\phi_{2})_{x+a_{Q},x+a_{Q}}(Q \phi_{1})_{x+a_{Q},x}.
\label{chap7_link2}
\end{eqnarray}
The r.h.s.\ of the former equation (\ref{chap7_link1}) can be written with using (\ref{ops}) as
\begin{equation}
\begin{aligned}
&\bigl((Q)_{x+a_Q,x}(\phi_1)_{x,x} - (\phi_1)_{x+a_Q,x+a_Q}(Q)_{x+a_Q,x}\bigr)(\phi_2)_{x,x} \\
&+
(\phi_1)_{x+a_Q,x+a_Q}\bigl((Q)_{x+a_Q,x}(\phi_2)_{x,x} - (\phi_2)_{x+a_Q,x+a_Q}(Q)_{x+a_Q,x}\bigr) \\[5pt]
& =
(Q)_{x+a_Q,x}(\phi_1)_{x,x}
(\phi_2)_{x,x}
-
(\phi_1)
_{x+a_Q,x+a_Q}
(\phi_2)
_{x+a_Q,x+a_Q}
(Q)_{x+a_Q,x}.
\end{aligned}
\end{equation}
Similar manipulation can be delivered to the latter equation  (\ref{chap7_link2}).
We thus obtain
\begin{eqnarray}
\bigl(Q (\phi_{1}\phi_{2})\bigr)_{x+a_{Q},x}
&=& (Q)_{x+a_Q,x}(\phi_1)_{x,x}(\phi_2)_{x,x}  \nonumber \\
&& 
- (\phi_1)
_{x+a_Q,x+a_Q}
(\phi_2)
_{x+a_Q,x+a_Q}
(Q)_{x+a_Q,x},   \\[5pt]
\bigl(Q (\phi_{2}\phi_{1})\bigr)_{x+a_{Q},x}
&=& (Q)_{x+a_Q,x}(\phi_2)_{x,x}(\phi_1)_{x,x} \nonumber \\
&&
 - (\phi_2)
_{x+a_Q,x+a_Q}
(\phi_1)
_{x+a_Q,x+a_Q}
(Q)_{x+a_Q,x}.
\end{eqnarray}
It is clear that, if the fields $\phi_1$ and $\phi_2$ commute with each other as in a non-gauge or Abelian gauge theory, the r.h.s.\ of these equations as well as the l.h.s.\ are identical,
which do not bring about any inconsitency.
Therefore, the ordering ambiguity posed in \cite{Bruckmann} is not applied.

Secondly, we might consider more general non-Abelian gauge extension of (\ref{ord-amb}):
\begin{eqnarray}
\delta_A\tr \bigl[(\varphi_{1})_{x+a_\varphi,x}(\varphi_{2})_{x,x+a_\varphi}\bigr]
&=& \tr \bigl[(\delta_A\varphi_{1})_{x+a_\varphi,x}(\varphi_{2})_{x,x+a_\varphi} \nonumber \\
&& \phantom{\tr\bigl[}
+ (\varphi_{1})_{x+a_\varphi,x}(\delta_A\varphi_{2})_{x,x+a_\varphi}, 
\label{chap7_gauge_link1} \\[5pt]
\delta_A\tr \bigl[(\varphi_{2})_{x,x+a_\varphi}(\varphi_{1})_{x+a_\varphi,x}\bigr]
&=& \tr \bigl[(\delta_A\varphi_{2})_{x,x+a_\varphi}(\varphi_{1})_{x+a_\varphi,x} \nonumber \\
&& \phantom{\tr\bigl[}
+ (\varphi_{2})_{x,x+a_\varphi}(\delta_A\varphi_{1})_{x+a_\varphi,x},
\label{chap7_gauge_link2}
\end{eqnarray}
similar equations to which were considered in \cite{BKC} to claim an ordering ambiguity in a gauge theory.
Note that we have taken the fields $\varphi_1$ and $\varphi_2$ whose shifts are $a_{\varphi_1}=a_{\varphi}$ and $a_{\varphi_2} = -a_{\varphi}$, respectively, so that the product becomes shiftless and thus gauge invariant.
These two equations are identical up to the factor $(-1)^{|\varphi_1||\varphi_2|}$ since we can (cyclically) exchange the order of fields under the trace.

Using the definition (\ref{deltaphi-def}), we can write the r.h.s.\ of the former equation 
(\ref{chap7_gauge_link1}) as
\begin{equation}
\begin{aligned}
&\tr\bigl[
(\xi_A)_{x+a_\varphi,x+a_\varphi+a_A}
(s_A\varphi_1)_{x+a_\varphi+a_A,x}(\varphi_2)_{x,x+a_\varphi} \\
&\phantom{\tr\bigl[}
+
(\varphi_1)_{x+a_\varphi,x}(\xi_A)_{x,x+a_A}(s_A\varphi_2)_{x+a_A,x+a_\varphi}
\bigr] \\[5pt]
&=\tr\Bigl[
(\xi_A)_{x+a_\varphi,x+a_\varphi+a_A}
\Bigl(
(s_A\varphi_1)_{x+a_\varphi+a_A,x}(\varphi_2)_{x,x+a_\varphi} \\
&\phantom{=\tr\bigl[}
+(-1)^{|\varphi_1|}
(\varphi_1)_{x+a_\varphi+a_A,x+a_A}(s_A\varphi_2)_{x+a_A,x+a_\varphi}
\Bigr)
\Bigr],
\label{ord-gauge1}
\end{aligned}
\end{equation}
where we have used (\ref{non-comm_xi1}) for exchanging $\xi_A$ and $\varphi_1$ in the second term.
Notice that the supertransformation of the product $\varphi_1\varphi_2$ w.r.t.\ $s_A$ obeys the lattice Leibniz rule as can be seen in the terms without the Grassmann parameter $\xi_A$.
Similar computation leads the r.h.s\ of the latter equation (\ref{chap7_gauge_link2}) to
\begin{equation}
\begin{aligned}
&\tr\Bigl[
(\xi_A)_{x,x+a_A}
\Bigl(
(s_A\varphi_2)_{x+a_A,x+a_\varphi}(\varphi_1)_{x+a_\varphi,x} \\
&\phantom{=\tr\bigl[}
+(-1)^{|\varphi_2|}
(\varphi_2)_{x+a_A,x+a_\varphi+a_A}(s_A\varphi_1)_{x+a_\varphi+a_A,x}
\Bigr)
\Bigr].
\label{ord-gauge2}
\end{aligned}
\end{equation}
As was pointed out earlier, this expression (\ref{ord-gauge2}) should coincide with 
(\ref{ord-gauge1})
 up to the Grassmann sign factor because of the permutation of fields under the trace. However, we have to be cautious in exchanging the order of the relevant terms since the Grassmann parameter follows the link (anti)commutation relations with fields. In fact, using the (anti)commutation relations
and moving the parameter towards the rightmost end in (\ref{ord-gauge2}), we have
\begin{equation}
\begin{aligned}
&-(-1)^{|\varphi_1|+|\varphi_2|}\tr\Bigl[
\Bigl(
(s_A\varphi_2)_{x,x+a_\varphi-a_A}(\varphi_1)_{x+a_\varphi-a_A,x-a_A} \\
&\phantom{=-(-1)^{|\varphi_1|+|\varphi_2|}\tr\bigl[}
+(-1)^{|\varphi_2|}
(\varphi_2)_{x,x+a_\varphi}(s_A\varphi_1)_{x+a_\varphi,x-a_A}
\Bigr)
(\xi_A)_{x-a_A,x}
\Bigr].
\end{aligned}
\end{equation}
Using in turn the permutation property under the trace, we can further transform this to
\begin{equation}
\begin{aligned}
&(-1)^{|\varphi_1||\varphi_2|}\tr\Bigl[
\Bigl(
(\varphi_1)_{x+a_\varphi-a_A,x-a_A}(\xi_A)_{x-a_A,x}(s_A\varphi_2)_{x,x+a_\varphi-a_A}\\
&\phantom{=(-1)^{|\varphi_1||\varphi_2|}\tr\bigl[}
-(-1)^{|\varphi_1|}
(s_A\varphi_1)_{x+a_\varphi,x-a_A}(\xi_A)_{x-a_A,x}(\varphi_2)_{x,x+a_\varphi}
\Bigr)
\Bigr] \\[5pt]
&=
(-1)^{|\varphi_1||\varphi_2|}\tr\Bigl[
(\xi_A)_{x+a_\varphi-a_A,x+a_\varphi}
\Bigl(
(-1)^{|\varphi_1|}
(\varphi_1)_{x+a_\varphi,x}(s_A\varphi_2)_{x,x+a_\varphi-a_A}
\Bigr)\\
&\phantom{=(-1)^{|\varphi_1||\varphi_2|}\tr\Bigl[}
+
(\xi_A)_{x+a_\varphi,x+a_\varphi+a_A}
\Bigl(
(s_A\varphi_1)_{x+a_\varphi+a_A,x}(\varphi_2)_{x,x+a_\varphi}
\Bigr)
\Bigr].
\end{aligned}
\end{equation}
Taking into account  
the sign factor $(-1)^{|\varphi_{1}||\varphi_{2}|}$, we can see that
this last form is identical to (\ref{ord-gauge1}), 
up to the total difference:
\begin{equation}
(-1)^{|\varphi_1|(|\varphi_2|+1)}\tr\Bigl[
(\xi_A)_{x+a_\varphi,x+a_\varphi+a_A}(\varphi_1)_{x+a_\varphi+a_A,x+a_A}(s_A\varphi_2)_{x+a_A,x+a_\varphi}
- (x\rightarrow x-a_A)
\Bigr].
\end{equation}
We therefore confirm that the SUSY transformation of gauge invariant operators is well-defined if we take summation over both the original and dual lattice coordinates. 
The most important example of such a gauge invariant is the action of the theory. The potential ambiguity discussed in \cite{BKC} can essentially be avoided within the current approach.

Another aspect worth clarifying explicitly is the gauge invariance of the SUSY transformation of the action. As we have just observed in the previous paragraphs, the SUSY transformation of the gauge invariant action can be schematically represented as
\begin{equation}
\delta_A S = \sum_x\tr\bigl[(\xi_A)_{x,x+a_A}(s_A\mathcal{L})_{x+a_A,x}\bigr].
\label{Sum_x_2}
\end{equation}
Its gauge transformation can then be expressed as
\begin{equation}
\begin{aligned}
\delta_A S
&\rightarrow
\sum_x\tr\bigl[(\xi_A)_{x,x+a_A}G^{-1}(x+a_A)(s_A\mathcal{L})_{x+a_A,x}G(x)\bigr] \\[5pt]
&
=
\sum_x\tr\bigl[G(x)(\xi_A)_{x,x+a_A}G^{-1}(x+a_A)(s_A\mathcal{L})_{x+a_A,x}\bigr],
\end{aligned}
\end{equation}
where we have used the trace property for moving the factor $G(x)$.
Since, by (\ref{gauge_inv_xi1}),
\begin{equation}
G(x)(\xi_A)_{x,x+a_A}G^{-1}(x+a_A) = (\xi_A)_{x,x+a_A}G(x+a_A)G^{-1}(x+a_A) = (\xi_A)_{x,x+a_A},
\end{equation}
it is clear that the transformation of the action $\delta_A S$ is gauge invariant.
As mentioned in \cite{BKC}, this invariance is crucial to construct a physically well-defined quantum theory based on the partition function
\begin{equation}
Z = \int\mathcal{D}\varphi\, e^{-S}.
\end{equation}
The SUSY transformation of the partition function is
\begin{equation}
\delta_A Z = \int\mathcal{D}\varphi\,(-\delta_A S)e^{-S} = -Z\langle\delta_A S\rangle,
\label{deltaZ}
\end{equation}
where $\langle\ \rangle$ denotes a vacuum expectation value.
According to this equation, if the SUSY transformation of the action were gauge covariant, the transformation of the partition function would vanish trivially on the gauge invariant vacuum, regardless of the SUSY invariance or the actual content of the action. In our formulation, the Grassmann parameter is inevitable to assure the gauge invariance of the SUSY transformation of the action, without which we would end up with an unphysical result where any scalar form could be used to construct a SUSY invariant partition function \cite{BKC}.
In this sense, the parameter plays a role of the ``discriminator'' for the physical partition function; by including it in the SUSY transformation, we can confirm that (\ref{deltaZ}) vanishes if and only if the action is SUSY invariant, as it should be.
The gauge invariance issue raised in \cite{BKC} can be thus resolved with the introduction of the Grassmann parameter.

\section{Summary and Discussions}
\indent

We have constructed $N=D=4$ 
and $N=4\ D=5$
SYM models on a lattice
by means of the Jacobi identity method.
As we have seen, Dirac-K\"ahler twisted SUSY
serves as a promising framework
to be realized exactly on the lattice.

After considering the relation between $N=2\ D=4$ twisting and $N=D=4$
twisting, it was explicitly shown that the Dirac-K\"ahler twisted SUSY algebra
 can satisfy the corresponding lattice Leibniz rule requirements.

As a direct consequence of lattice Leibniz rule considerations,
Dirac-K\"ahler twisted supercharges should be generically assigned on links,
which is one of the important features proposed in this formulation.
The manifest gauge covariant method based on the
Jacobi identities gives the systematic 
formulations of $N=D=4$ 
and $N=4\ D=5$
lattice SYM multiplets.

Employing Grassmann parameters with link nature,
we have explicitly 
shown that the resulting super Yang-Mills action is  invariant under all the supercharges 
on a lattice.

Furthermore, as a group and algebraic interpretation of the link approach, we have shown 
that promoting bosonic supercovariant derivatives to their exponentials (\ref{Nabla2-1})-(\ref{Nabla2-2})
consistently with the lattice Leibniz rule
naturally gives rise to the notion of gauge covariant link (anti-)commutators (\ref{Link_comm1})-(\ref{Link_comm2}).
This provides an explanation of how and why the two fundamental concepts of 
non-commutativity and link nature are tightly related with each other.
The approach also systematically resolves the ordering ambiguity \cite{Bruckmann,BKC} mentioned in the Introduction. 

We have also seen that the exponentiation of the bosonic supercovariant derivatives 
provides us with a deep insight among gauge group, SUSY algebra, and finite translations.
These sorts of insight can be obtained only when we stand on the geometrical understanding of the spacetime
through the Dirac-K\"ahler mechanism.
Having seen that, in association with the above mentioned exponentiation, the scale operator $\bf{D}$ superficially splits to the coordinates  $x_{\mu}$ as symmetry operators, while the Weyl-'tHooft algebra (\ref{NC_a_nabla})-(\ref{NC_a_nabla2}) between $\D_{\pm\mu}$ and $x_{\nu}$ naturally emerges, one may expect a rich algebraic and group structure behind this formulation.  
In particular, investigations from the viewpoints of superspace and matrix model 
which accommodate the above structure would be necessary.

It is also worthwhile to remind of the ``no-go theorem" claiming that the simultaneous 
    preservation of full SUSY and locality in lattice spacetime is forbidden \cite{No-Go}.  
Since our formulation utilizes the notion of gauge covariant link (anti-)commutators, it is not immediately clear that the 
``no-go theorem'' may or may not be applied to the current framework.
In particular,     we observe 
    that our claim of exact lattice SUSY statement in this paper and the ``no-go theorem" 
    may not be compatible with each other. 
    As it is explicitly pointed out in Eq. (
    \ref{Sum_x_1}) in Section 6 and in Eq. (
    \ref{Sum_x_2}) 
    in Section 7, original lattice sites 
    and dual lattice sites should be summed over 
    the 
    coordinate phase space to keep exact 
    lattice SUSY and gauge invariance. If we keep only the sum of original sites to keep locality,
    exact lattice SUSY and gauge symmetry are lost. Fields on the dual lattice have nonlocal 
    nature with respect to the fields on the original lattice. 
We should also remember that the above-explained group and algebraic structure lies behind the nonlocal  nature.
    It could possibly be this semi-nonlocal
     nature of the fields on the dual lattice sites
associated with the underlying group and algebraic structure,
    which may avoid the ``no-go theorem". 
    Link approach with Dirac-K\"{a}hler twisting current formulation can accommodate these 
   semi locally scattered nonlocal fields on the dual lattice. 
   We should, however, admit that we may need numerical investigations to clarify  
the above semi-nonlocal nature of ﬁelds leading to a local ﬁeld theory in the  
continuum limit. 
   Finally, 
   from the aspect of practical applications, it is also important to ask how the Grassmann link 
   parameter $\xi_A$ could be implemented in numerical calculations.

A part of the study of these aspects is in progress and the result will be given elsewhere.

\subsection*{Acknowledgments}
\indent

Since the first paper of our link approach on lattice Super Yang-Mills in 
2005,
we had 
numerous fruitful and constructive and useful discussions and criticisms from many 
friends over so many years. They include K. Asaka, G. Bergner, F. Bruckmann, S. Catterall, 
P. H. Damgaard, A. Feo, J. Giedt, D. Kadoh, I. Kanamori, D. B. Kaplan, 
J. Kato, M. Kato, Y. Kondo, S. Matsuura, A. Miyake, H. B. Nielsen,  H. So, F. Sugino,  
H. Suzuki, M. \"Unsal.      
 
 This work had been supported in part by the funds of Japanese Ministry of Education,
Science, Sports and Culture and also by INFN research funds.

This paper is dedicated to Alessandro D'Adda who passed away on April 23, 2022.


\end{document}